\documentclass[usenatbib,usegraphicx,letterpaper]{mn2e}
\usepackage[totalwidth=480pt,totalheight=680pt]{geometry}

\usepackage{amssymb}
\usepackage{epsfig}
\usepackage{amsmath}
\usepackage{color}
\usepackage[dvipsnames]{xcolor}
\usepackage{yfonts}

\newcommand{\apjl}{ApJL~}
\newcommand{\apjs}{ApJS~}

\newcommand{\Omegam}{\Omega_{m}}
\newcommand{\Omegab}{\Omega_{b}}

\newcommand{\tilt}{n_{\mathrm{s}}}
\newcommand{\rhocrit}{\rho_{\mathrm{crit}}}

\newcommand{\mvir}{M_\mathrm{vir}} 
 
\newcommand{\deltavir}{\Delta_{\mathrm{vir}}}

\newcommand{\unit}[1]{\mathrm{#1}}

\newcommand{\mpc}{\unit{Mpc}} 
\newcommand{\kpc}{\unit{kpc}}

\newcommand{\kms}{\unit{km \ s^{-1}}}

\newcommand{\mstar}{M_{\ast}}
\newcommand{\rvir}{R_\mathrm{vir}}

\newcommand{\Msun}{\mathrm{M}_{\odot}}
\newcommand{\msun}{\mathrm{M}_{\odot}}

\newcommand{\ngal}{N_{\mathrm{g}}}
\newcommand{\nsat}{N_\mathrm{s}}
\newcommand{\ncen}{N_\mathrm{c}}

\newcommand{\deltancen}{\delta N_{\mathrm{c}}}
\newcommand{\deltansatone}{\delta N_{\mathrm{s},1}}
\newcommand{\deltansattwo}{\delta N_{\mathrm{s},2}}
\newcommand{\deltancenone}{\delta N_{\mathrm{c},1}}
\newcommand{\deltancentwo}{\delta N_{\mathrm{c},2}}

\newcommand{\deltangalone}{\delta N_{\mathrm{g},1}}

\newcommand{\Abias}{\mathcal{A}_{\mathrm{bias}}}

\newcommand{\pnm}[2]{P(#1|#2)}
\newcommand{\mean}[2]{\left\langle#1 \vert {#2}\right\rangle}
\newcommand{\avg}[1]{\left\langle #1 \right\rangle}

\newcommand{\pstd}[2]{P_{\rm std}(#1|#2)}
\newcommand{\pdec}[2]{P_{\rm dec}(#1|#2)}

\newcommand{\sigmalogm}{\sigma_{\log{}\mathrm{M}}}

\newcommand{\vmax}{V_{\rm max}}

\newcommand{\abias}{\mathcal{A}_{\rm bias}}

\newcommand{\fshm}{\bar{M}_{\ast}}
\newcommand{\mstarzero}{M_{\ast, 0}}

\newcommand{\mstarscatter}{\sigma_{\log\mstar}}

\newcommand{\mstarthresh}{\mstar^{\rm thresh}}

\newcommand{\mhalo}{M_{\rm h}}
\newcommand{\csmf}{\phi_{\rm c}(\mstar|\mhalo)}

\newcommand{\dndmvir}{\frac{\dd n}{\dd\mvir}}
\newcommand{\dndmhalo}{\frac{\dd n}{\dd\mhalo}}

\newcommand{\mcoll}{M_{\rm coll}}

\newcommand{\xigg}{\xi_{\mathrm{gg}}}
\newcommand{\xiggr}{\xi_{\mathrm{gg}}(r)}
\newcommand{\xiggroneh}{\xi^{1h}_{\mathrm{gg}}(r)}
\newcommand{\xiggrtwoh}{\xi^{2h}_{\mathrm{gg}}(r)}

\newcommand{\xigm}{\xi_{\mathrm{gm}}}

\newcommand{\ngalaxy}{\bar{n}_{\mathrm{g}}}

\newcommand{\dd}{\mathrm{d}}

\newcommand{\beq}{\begin{equation}}
\newcommand{\eeq}{\end{equation}}
\newcommand{\beqray}{\begin{eqnarray}}
\newcommand{\eeqray}{\end{eqnarray}}

\newcommand{\ben}{\begin{enumerate}}
\newcommand{\een}{\end{enumerate}}
\newcommand{\bit}{\begin{itemize}}
\newcommand{\eit}{\end{itemize}}

\usepackage{epsfig}  \usepackage{graphicx}   \usepackage{rotating}

\begin{document}

%-----------------------------------------------------------------------------------
\title[Introducing Decorated HODs]
{Introducing Decorated HODs: modeling assembly bias in the galaxy-halo connection}

%-----------------------------------------------------------------------------------

\author[Hearin et al.]
{Andrew P. Hearin$^{1}$, Andrew R. Zentner$^{2}$, Frank C. van den Bosch$^{3},$ \newauthor
Duncan Campbell$^{3},$ Erik Tollerud$^{3, 4}$ \\
$^1$Yale Center for Astronomy \& Astrophysics, Yale University, New Haven, CT\\
$^2$Department of Physics and Astronomy \& Pittsburgh Particle Physics, Astrophysics, and Cosmology Center (PITT PACC),\\ University of Pittsburgh, Pittsburgh, PA 15260\\
$^3$Department of Astronomy, Yale University, P.O. Box 208101, New Haven, CT \\
$^4$ Space Telescope Science Institute, 3700 San Martin Dr, Baltimore, MD 21218
}

\maketitle

%%%%%%%%%%%%%%%%%%%%%%%%%%%% ABSTRACT %%%%%%%%%%%%%%%%%%%%%%%%%%%%
\begin{abstract}

  The connection between galaxies and dark matter halos is often
  inferred from data using probabilistic models, such as the Halo Occupation
  Distribution (HOD). Conventional HOD formulations 
  assume that {\em only} halo mass governs the galaxy-halo connection. 
  Violations of this assumption, known as {\em galaxy assembly bias},
  threaten the HOD program. We introduce {\em decorated HODs,} a
  new, flexible class of models designed to account for 
  assembly bias. Decorated HODs minimally expand the 
  parameter space and maximize the independence between 
  traditional and novel HOD parameters. 
  We use decorated HODs to quantify the influence of assembly bias 
  on clustering and lensing statistics. For SDSS-like samples, 
  the impact of assembly bias on galaxy clustering 
  can be as large as a factor of two on $r \sim 200$~kpc
  scales and $\sim15\%$ in the linear regime. Assembly bias can 
  either enhance or diminish clustering on large scales, but generally 
  increases clustering on scales $r \lesssim 1$~Mpc. 
  We performed our calculations with {\tt Halotools,} 
  an open-source, community-driven python package for 
  studying the galaxy-halo connection ({\tt http://halotools.readthedocs.org}). 
  We conclude by describing the use of decorated HODs to treat 
  assembly bias in otherwise conventional likelihood analyses.
  
\end{abstract}
%%%%%%%%%%%%%%%%%%%%%%%%%%%%%%%%%%%%%%%%%%%%%%%%%%%%%%%%%%%%%%%%%%


%%%%%%%%%%%%%%%%%%%%%%%%%%%%%% INTRO %%%%%%%%%%%%%%%%%%%%%%%%%%%%%%

\section{INTRODUCTION}
\label{sec:intro}

%%%%%%%%%%%%%%%%%%%%%%%%%%%%%%%%%%%%%%%%%%%%%%%%%%%%%%%%%%%%%%%%%%%%%

The last decade has seen the development of a new, powerful technique
to inform models of galaxy formation, interpret large-scale structure,
and constrain cosmological parameters. This technique, called halo
occupation modeling, establishes a statistical connection between
galaxies and dark matter halos.  Using clustering and lensing data
from various galaxy redshift surveys, halo occupation modeling has
been used to put tight constraints on the stellar mass-to-halo mass
relation of galaxies from redshift $z=0$ to $z \sim 2$ \citep{yang03,
  tinker05, vdBosch07, Zheng07, wake_etal11, leauthaud_etal12}, which
in turn has put tight constraints on the co-evolution of galaxies and
their host halos across cosmic time \citep{conroy_wechsler09, yang09b,
  yang11a, yang12, behroozi13, moster13}.  In addition, this technique
is routinely used to infer the masses of dark matter halos that host
extreme populations, such as quasars \citep{porciani04, porciani06},
Lyman-break galaxies \citep{bullock01}, or luminous red galaxies
\citep{wake_etal08}, while applications to star forming and quenched
galaxies separately has proven instrumental for learning about the
demographics of galaxy quenching \citep{vdBosch03a, zehavi05a,
  zehavi11, collister05, skibba_sheth09, rod_puebla11, tinker_etal13,
  guo_etal11b, guo_etal14, zu_mandelbaum15b}. 
  And finally, halo occupation modeling is
used to test cosmology on small scales \citep[e.g.,][]{vdBosch03b,
  tinker05, cacciato_etal13, more_etal13}.

The key concept that underlies halo occupation modeling is that all
galaxies reside in dark matter halos, and that these halos themselves
are biased tracers of the dark matter density field.  If the so-called
``halo bias" $b_{\rm h}$ were only a function of halo mass, then
knowledge of the manner by which galaxies reside in halos as a
function of halo mass would be sufficient to make predictions for the
large-scale galaxy clustering that we observe.

However, it is now well-established that halo clustering depends upon
attributes other than halo mass, including halo formation time and
concentration \citep[e.g.,][]{gao_etal05, wechsler06, gao_white07, zentner07, 
  dalal_etal08, lacerna11}. Such dependence of the spatial
distribution of dark matter halos upon properties besides mass is
generically referred to as {\em halo assembly bias,} and on large
scales it can be quantified as $b_{\rm h} = b_{\rm h}(M,X)$, where $X$
is the set of halo properties other than halo mass that influence halo
bias.  {\em If} the relationship between galaxies and halos depends
upon any of these additional halo properties $X,$ then conventional
occupation models will fail.

Despite this threat of assembly bias, virtually all studies of halo
occupation statistics published to date are predicated upon the
assumption that the mass of a dark matter halo, $\mhalo,$ {\it
  completely} determines the statistical properties of its resident
galaxy population.\footnote{Some models replace halo mass by a single
  other halo property, such as the maximum circular velocity of the
  halo.  The upshot remains the same; it is implicitly assumed that
  the occupation statistics depend on that property alone.} Under this
`Mass-Only ansatz', the properties of galaxies that reside in halos of
fixed mass are uncorrelated with any other halo property, and {\it
  any} environmental dependence of galaxies is merely a manifestation
of different environments probing different halo masses. 

In light of our standard paradigm for galaxy formation 
\citep[see][for an overview and extensive list of references]{mo_vdb_white10}, 
in which galaxy growth is governed by halo growth,
galaxy size is linked to halo spin, galaxy morphology is related to
halo merger history, and star formation quenching is triggered by
local environmental processes, such an ansatz is suspect.
Indeed, semi-analytical models for galaxy formation predict
significant correlations between galaxy properties and halo properties
other than mass \citep[e.g.,][]{zhu_etal06, croton_etal07}; such
correlations are also present in hydrodynamical simulations 
of galaxy formation and evolution
\citep{feldmann_mayer14, bray_etal15}.

The Mass-Only ansatz was not proposed with any theoretical prejudice
regarding galaxy formation; rather, it is the simplest implementation
that has proven to be successful in fitting the clustering properties
of the available data. Consequently, it is generally argued that the
impact of assembly bias, if present, is too weak to have a significant
impact, and that there is thus no need for more sophisticated models
\citep[e.g.,][]{tinker08b}. This argument is strengthened by several
studies that have demonstrated that halo mass is clearly the dominant
parameter governing the environmental demographics of galaxies
\citep[e.g.,][]{mo_etal04, kauffmann_etal04, blanton06}.

\cite{zentner_etal14} demonstrated explicitly that ignoring assembly
bias in halo occupation modeling yields constraints on the galaxy-dark
matter connection that may be plagued by significant, systematic
errors. The magnitude of this error is especially large for extreme
populations, such as star forming or quenched galaxies, and calls into
question many of the inferences that have been made using halo
occupation modeling. Furthermore, some level of galaxy assembly bias
seems difficult to avoid, given that the abundance of
subhalos (which is directly related to the occupation number of
satellite galaxies) depends strongly on the assembly time of the host
halo \citep[][]{zentner05, vdBosch05, mao_etal15, Jiang_vdB15}; hence,
one expects that earlier-forming, more strongly-clustered host halos
have fewer satellite galaxies.

Additionally, a number of studies have presented compelling evidence
that observational data is affected by galaxy assembly
bias. \cite{yang_etal06a} showed that, at fixed halo mass, the
clustering strength increases as the star formation rate of the
central galaxy decreases \citep[see also][]{wang_etal08, wang_etal13,
  lin_mandelbaum_etal15}.  In \cite{blanton_berlind07} it was shown
that shuffling the colors and luminosities of galaxies among groups of
similar mass modifies the clustering at the $\sim 5-10\%$ level. More
recently, \cite{miyatake_etal15} reported a strong detection of
assembly bias in an SDSS sample of galaxy clusters. Another
manifestation of assembly bias was presented by \cite{weinmann06b},
who showed that, at fixed group (halo) mass, both the color and star
formation rate (SFR) of satellite galaxies depend on the color/SFR of
the group's central galaxy.  \cite{kauffmann_etal13} demonstrated that
this `galactic conformity' (i.e., spatial correlations in specific
star formation rates) persists out to several Mpc. As emphasized in
\cite{hearin_etal14}, if halo mass has been properly controlled for in the 
observational measurements, then conformity manifestly violates the Mass-Only
ansatz, and is a smoking-gun signature of galaxy assembly bias.

To summarize, without some remedy for assembly bias, halo occupation
modeling is doomed to hit a systematic error ``ceiling" before
realizing its potential to exploit the wealth of extant and
forthcoming observational data in an optimal and unbiased way.  To
address this challenge, in this paper we introduce a natural extension
of the standard Halo Occupation Distribution (HOD) formalism with
parametric freedom that allows for galaxy assembly bias. This {\em
  decorated HOD} describes the halo occupation statistics in terms of
two halo properties rather than one. In this pilot study we describe
the formalism and demonstrate how the extra degree of freedom can
cause dramatic changes in the clustering and lensing of galaxies,
further advocating a proper treatment of assembly bias. 

All calculations in this paper are performed using {\tt Halotools}, a new
open-source, community-driven python package for studying the
galaxy-halo connection ({\tt http://halotools.readthedocs.org}).  {\tt
  Halotools} is built upon the Astropy package-template\footnote{See
  {\tt https://github.com/astropy/package-template}.} and is being
developed to be an Astropy-affiliated package
\citep{astropy}.\footnote{See {\tt http://www.astropy.org/affiliated}.}
Since its inception, {\tt Halotools} has been developed in public on 
{\tt https://github.com/astropy/halotools}, in the spirit of open science. 

The remainder of this paper is organized as follows.  We give a
pedagogical review of standard halo occupation methods in
\S\ref{sec:halomodel}.  In \S\ref{sec:Halotools} we describe how we
use the {\tt Halotools} package to conduct all of our halo-occupation
and large-scale structure calculations.  We develop the analytical
formalism of the decorated HOD in \S\ref{sec:hodgen}.  In
\S\ref{sec:results} we use a simple implementation of the decorated
HOD to estimate the magnitude and scale-dependence of the effects
galaxy assembly bias can have on galaxy clustering and galaxy-galaxy
lensing.  We conclude in \S\ref{sec:discussion} by comparing our model
to previous formulations of the galaxy-halo connection, and by
describing the significance of our results for the precision cosmology
program. We summarize our primary findings in \S\ref{sec:summary}.

%%%%%%%%%%%%%%%%%%%%%%%%%%%%%% STANDARD HALO MODEL %%%%%%%%%%%%%%%%%%%%%%%%%%%%%%

\section{STANDARD HOD MODEL}
\label{sec:halomodel}

In halo models of the galaxy distribution, it is assumed that every
galaxy resides in some dark matter halo.  Under this assumption,
knowledge of how galaxies populate, and are distributed within, dark
matter halos is sufficient to describe the statistics of the observed
galaxy distribution
\citep[e.g.][]{seljak00,mafry00,scoccimarro01a,berlind02}. In this
section, we discuss the standard halo models of galaxy clustering with
an eye toward pedagogy in order to enable qualitative understanding of
results that follow later in the paper.

\subsection{HOD formulation of Galaxy Clustering}
\label{subsec:hodclustering}

One of the most well-studied statistics of the galaxy distribution is
the galaxy-galaxy correlation function, $\xiggr,$ which expresses the
probability in excess of random of finding a pair of galaxies
separated by three-dimensional distance $r.$ By dividing galaxy
pair-counts into terms containing either pairs residing in the same
dark matter halo or pairs of galaxies residing in distinct halos,
$\xigg$ can be decomposed into a ``one-halo term" and a ``two-halo
term'', $$\xiggr=1+\xiggroneh+\xiggrtwoh.$$

As mentioned briefly in the introduction, one of the leading
approaches to characterizing the galaxy-halo connection is via the
Halo Occupation Distribution (HOD). In the HOD approach, the central
quantity is $\pnm{\ngal}{\mhalo},$ the probability that a halo of mass
$\mhalo$ hosts $\ngal$ galaxies. Given a specific HOD, the one- and
two-halo terms of the galaxy correlation function can be computed in
terms of the first two moments of $\pnm{\ngal}{\mhalo}$:

\beqray
\label{eq:onehaloterm}
1+\xiggroneh & \simeq &
\frac{1}{4\pi{}r^{2}\ngalaxy^{2}}\int\dd\mhalo\dndmhalo\Xi_{\rm
  gg}(r|\mhalo) \nonumber \\ & \times & \mean{\ngal(\ngal-1)}{\mhalo},
\eeqray and

\beqray
\label{eq:twohaloterm}
\xiggrtwoh & \simeq & \xi_{\mathrm{mm}}(r) \times \\ \nonumber & &
\left[\frac{1}{\ngalaxy}\int\dd\mhalo\dndmhalo\mean{\ngal}{\mhalo}b_{\mathrm{h}}(\mhalo)\right]^{2}
\eeqray

In Eqs.~(\ref{eq:onehaloterm}) and (\ref{eq:twohaloterm}), $\ngalaxy$
is the cosmic mean number density of the galaxy sample,
$\dd\mathrm{n}/\dd\mhalo$ is the halo mass function,
$b_{\mathrm{h}}(\mhalo)$ the spatial bias of dark matter halos, and
$\xi_{\rm mm}$ is the dark matter two-point correlation function.  If
the spatial distribution of galaxies within a halo is represented by a
spherically symmetric, unit-normalized number density $n_{\rm g}(r),$
then the quantity $\Xi_{\rm gg}(r)$ is the convolution of $n_{\rm
  g}(r)$ with itself. Note that we have employed several simplifying
assumptions in these expressions for the correlation function; in
particular, halos and their associated galaxy distributions are
assumed to be spherically symmetric, the halo bias is assumed to have
no radial dependence, and halo exclusion is not taken into account
\citep[see e.g.,][and references
  therein]{cooray02,mo_vdb_white10,vdBosch13}.  As we describe in
\S\ref{sec:Halotools}, the {\tt Halotools} methodology used in the
present work has the distinct advantage that it is immune to errors
and uncertainties arising from these approximations.

Under the same simplifying assumptions as above, one can use the same
formalism to also express the one- and two-halo terms of the
galaxy-matter cross-correlation function, $\xigm$, as
\beqray
\label{eq:onehalotermlensing}
1+ \xigm^{\rm 1h} & \simeq &
\frac{1}{4\pi{}r^{2}\ngalaxy\bar{\rho}_{\rm
    m}}\int\dd\mhalo\dndmvir\Xi_{\rm gm}(r|\mhalo) \nonumber \\ &
\times & \mean{\ngal}{\mhalo}, \eeqray
and
\beqray
\label{eq:twohalotermlensing}
\xigm^{\rm 2h} & \simeq & \xi_{\mathrm{mm}}(r) \times \\ \nonumber & &
\left[\frac{1}{\ngalaxy}\int\dd\mhalo\dndmhalo\mean{\ngal}{\mhalo}b_{\mathrm{h}}(\mhalo)\right]
\\ \nonumber & & \times \left[\frac{1}{\bar{\rho}_{\rm
      m}}\int\dd\mhalo\dndmhalo b_{\mathrm{h}}(\mhalo)\right] \eeqray
Here, $\bar{\rho}_{\rm m}$ is the mean matter density in the universe,
and $\Xi_{\rm gm}$ is the convolution of the normalized dark matter
halo density profile with $n_{\rm g}(r)$. Note that $\xigm$ is 
the fundamental two-point function underlying the excess surface densities 
probed by the galaxy-galaxy lensing
signal \citep[e.g.,][]{mandelbaum05,seljak05,yoo06,cacciato_etal09}.

\subsection{Central-Satellite Decomposition}
\label{subsec:censat}

Conventionally, occupation statistics of central galaxies are modeled
separately from satellites, so that
$\mean{\ngal}{\mhalo}=\mean{\ncen}{\mhalo}+\mean{\nsat}{\mhalo}.$ The
starting point for any HOD-style model is then choosing an analytical
form for $\avg{\ncen}$ and $\avg{\nsat}.$ Central galaxies are
commonly assumed to reside at the center of the halo, so that
$n_{\mathrm{cen}}(r)=\delta(r)$, while satellite galaxies are
typically assumed to follow a radial number density distribution,
$n_{\rm sat}(r)$, that traces the NFW density distribution
\citep{nfw97} of the underlying dark matter halo, albeit with a
concentration parameter that is sometimes allowed to differ from that
of the dark matter.

Consider Eq.~(\ref{eq:onehaloterm}) in light of this
decomposition. The one-halo term receives a contribution from the
second satellite moment, $\mean{\nsat(\nsat-1)}{\mhalo};$ computing
this contribution requires additional assumptions beyond a model for
$\mean{\nsat}{\mhalo}.$ Motivated by the occupation statistics of
subhalos in high-resolution N-body simulations \citep{kravtsov04a},
the PDF of satellite occupation is commonly assumed to be Poissonian,
so that $\mean{\nsat(\nsat-1)}{\mhalo} = \mean{\nsat}{\mhalo}^{2}.$

\subsection{Central-Satellite Correlations}
\label{subsec:censatcorrelations}

The one-halo term in Eq.~(\ref{eq:onehaloterm}) also contains a
contribution from $\mean{\ncen\nsat}{\mhalo},$ which itself requires
an additional assumption to compute.  In the vast majority of HOD
studies, it is assumed that the satellite and central HODs are
completely uncorrelated; however, we are unaware of any data
indicating that this should be the case despite theoretical prejudices
that $\ncen$ and $\nsat$ may well be correlated. If satellites have no
knowledge of the central galaxy occupation of their host halos, then
$\mean{\ncen\nsat}{\mhalo}=\mean{\ncen}{\mhalo}\mean{\nsat}{\mhalo}$.
On the other hand, if the presence of a central in a halo is required
for one or more satellites to occupy that halo, then
$\mean{\ncen\nsat}{\mhalo}=\mean{\nsat}{\mhalo}$. It may also be
possible that centrals and satellites ``repel" each other so that they
never reside in the same halo, in which case $\mean{\ncen
  \nsat}{\mhalo}=0$. These possibilities represent the extremes of no
correlation, complete correlation, and complete
anticorrelation. Galaxy samples selected from the observed universe
likely exhibit correlations somewhere between these extremes.

The means by which $\mean{\ncen\nsat}{\mhalo}$ is computed is not
simply academic: both \citet{reid_etal14} and \citet{guo_etal14} have
shown that the satellite fraction inferred from HOD analyses of BOSS
galaxies is impacted at the $\sim 50\%$ level depending on whether one
assumes maximal or zero central-satellite correlations \citep[see
  also][for analogous findings for color-selected
  samples]{ross_brunner09}. \citet{zentner_etal14} demonstrated 
 that identical HODs with distinct central-satellite
correlations motivated by physical considerations can lead to large
(as much as a factor of $\sim 2$), scale-dependent differences in
correlation functions for separations $r \lesssim 1\, \mpc$.

The degree of central-satellite correlation is more than simply a
nuisance systematic. Such correlations are induced by astrophysics
that is interesting in its own right and which we aim to learn about
through the analysis of statistically-large galaxy samples.  This
correlation encodes the extent to which the properties of a satellite
galaxy (stellar mass, color, etc.) may be correlated with the
properties of its central galaxy at fixed halo mass. Such a
correlation could easily arise, for example, from galactic
cannibalism: if a satellite galaxy merges with the central galaxy, the
latter's mass increases while $\nsat$ decreases
\citep[e.g.,][]{purcell_etal07}; in addition, the one-halo galactic
conformity detected by \citet{weinmann06b} is an example of a
correlation between central and satellite properties.  As we will see,
the decorated HOD formalism introduced in \S\ref{sec:hodgen} naturally
permits an analytical means to explore parametrically cases of
intermediary correlation between the two extremes sketched above, so
that quantitative constraints can be placed on the effects giving rise
to central-satellite correlations.

%------------------------------------------------------------------------------------------------
\subsection{Baseline HOD Parameterization}
\label{subsec:baselineHOD}

In order to understand the potential influence of assembly bias, we
perturb the galaxy-halo connection about a baseline HOD model that has
no assembly bias. For the remainder of this paper, unless otherwise
specified, we use the HOD parameterization introduced in
\citet{leauthaud11b} as our baseline model. In this model, the first
occupation moment of central galaxies $\mean{\ncen}{\mhalo}$ is
defined in terms of the conditional stellar mass function (CSMF),
$\csmf;$ the CSMF is the probability distribution for the stellar mass
of a central galaxy residing in a halo of mass $\mhalo$
\citep[][]{yang03,yang09a}. For a volume-limited sample of galaxies
more massive than $M_{\ast}^{\rm thresh},$ the relationship between
$\langle\ncen\rangle$ and $\csmf$ is given by:
\beqray
\mean{\ncen}{\mhalo} = \int_{\mstarthresh}^{\infty}\dd\mstar\csmf
\eeqray
The function $\csmf$ is assumed to be a log-normal distribution with a
first moment that varies with halo mass according to the
stellar-to-halo-mass relation $\fshm(\mhalo).$ For $\fshm(\mhalo)$ we
use the model developed in \citet{behroozi10}, in which the
stellar-to-halo-mass relation is defined by the inverse relation
$\bar{M}_{\rm h}(M_{\ast}),$
\beq
\begin{split}
\label{eq:shmdef}
\log_{10}[\bar{M}_{\rm h}(\mstar)] = \\  \log_{10}(M_1) &+ \beta\log_{10}\left(\frac{\mstar}{M_{\ast, 0}}\right) \\ &+ 
\frac{(\mstar / M_{\ast, 0})^{\delta}}{1 + (\mstar / M_{\ast, 0})^{-\gamma}} - \frac{1}{2}. 
\end{split}
\eeq
One can then compute the stellar-to-halo-mass relation $\fshm(\mhalo)$
by numerically inverting Eq.~(\ref{eq:shmdef}). The model for
$\fshm(\mhalo)$ at redshift-zero therefore has five parameters: $M_1$
is the characteristic halo mass, $\mstarzero$ is the characteristic
stellar mass, $\beta$ is the low-mass slope, $\delta$ the high-mass
slope, and $\gamma$ controls the transition region.

We will model redshift-dependence in the HOD by allowing the
parameters of $\fshm(\mhalo)$ to vary linearly with the scale factor,
as in \citet{behroozi10}, so that
\beqray
\label{eq:zdepsmhmparams}
\mstarzero(a) & = & \mstarzero - (1 - a)M_{\ast,a}\nonumber \\
M_{1}(a) & = & M_{1} - (1 - a)M_{1, a} \nonumber \\
\beta(a) & = & \beta - (1 - a)\beta_{a}  \\ 
\gamma(a) & = & \gamma - (1 - a)\gamma_{a} \nonumber \\ 
\delta(a) & = & \delta - (1 - a)\delta_{a} \nonumber 
\eeqray
The model for $\fshm(\mhalo, z)$ across redshift then has a total of
$5\times2=10$ parameters.  Throughout this paper, the values of all
ten of these parameters are taken directly from column 1 of Table 2 of
\citet{behroozi10}.

Throughout this paper, we assume that $\csmf$ has constant scatter
$\mstarscatter,$ which will serve as the $11^{\rm th}$ parameter
governing central occupations statistics across redshift. Under the
log-normal assumption the first occupation moment of central galaxies
can be computed analytically as:
\begin{align}
\label{eq:ncendef}
 \mean{\ncen}{\mhalo} & = & \\ \nonumber \frac{1}{2} - & \frac{1}{2}{\rm erf}\left(\frac{\log_{10}(\mstar^{\rm thresh}) - \log_{10}[\fshm(\mhalo)]}{\sqrt{2}\sigma_{\rm log\mstar}}\right).  
\end{align}
We will explicitly study how the impact of assembly bias is influenced
by the level of scatter, but unless otherwise stated we will use
$\mstarscatter=0.4$ as our fiducial value.

The first occupation moment of satellites is modeled as
\beqray
\label{eq:nsatdef}
\mean{\nsat}{\mhalo} = \mean{\ncen}{\mhalo} \left(\mhalo/M_{\rm sat}\right)^{\alpha_{\rm sat}}e^{-M_{\rm cut}/\mhalo}. 
\eeqray
The parameter $\alpha_{\rm sat}$ controls the power-law increase in
satellite number with halo mass; $M_{\rm sat}$ defines the amplitude
of the power law; and $M_{\rm cut}$ sets the scale of an exponential
cutoff that guarantees that halos with masses $\mhalo \ll M_{\rm cut}$
are extremely unlikely to host a satellite galaxy. In light of the
discussion in \S\ref{subsec:censat}, note that $\mean{\nsat}{\mhalo}$
depends on $\mean{\ncen}{\mhalo}$, indicating that in this particular
model the occupation statistics of centrals and satellites are
correlated. The satellite amplitude is parameterized as
\beqray M_{\rm sat} = 10^{12}\msun B_{\rm sat}\left[\frac{\bar{M}_{\rm
    h}(\mstarthresh)}{10^{12}\msun}\right]^{\beta_{\rm sat}}, \eeqray
while the cutoff is parameterized as  
\beqray M_{\rm cut} = 10^{12}\msun B_{\rm cut}\left[\frac{\bar{M}_{\rm
    h}(\mstarthresh)}{10^{12}\msun}\right]^{\beta_{\rm cut}}.  \eeqray
Satellite occupation statistics therefore have five parameters, namely
$\alpha_{\rm sat},$ $B_{\rm sat},$ $\beta_{\rm sat},$ $B_{\rm sat}$
and $\beta_{\rm cut}.$ We set the values of these parameters in our
fiducial baseline model to be those listed in the ``SIG\_MOD1" values
of Table 5 in \citet{leauthaud_etal12}, to which we refer the reader for a
full discussion of this model.

By using this model, we are working with an HOD parameterization that
has been used to describe survey data with success. This model has
sufficient complexity to be relevant to the interpretation of
observations. This adds to the realism and relevance of the
calculations that follow. However, we should note that if assembly
bias is a non-negligible effect in the universe, these HODs may be
systematically in error \cite[e.g.,][]{zentner_etal14} and we may not
be perturbing about the true, underlying, mass-only baseline HOD
realized in nature.

%------------------------------------------------------------------------------------------------
\section{Halo Occupation Modeling with {Halotools}}
\label{sec:Halotools}

The conventional analytical methods for calculating $\xigg$ and
$\xigm$ described in \S\ref{sec:halomodel} rely on a large number of
assumptions and resrictions that limit the accuracy of these
commonly-used techniques. Halos are typically assumed to be spherical,
virialized matter distributions, characterized by an NFW profile
\citep{nfw97}, while fitting functions for the halo mass function $\dd
n/\dd\mhalo$ and large-scale halo bias $b_{\rm h}(\mhalo)$ are only
accurate to the $\sim 5\%$ level \citep{tinker08a, tinker10}. In
addition, halo exclusion and scale-dependence of the halo bias are
difficult to treat properly, resulting in additional uncertainties and
inaccuracies \citep{vdBosch13}. Details regarding the implementations
of these assumptions vary from author to author, and can give rise to
systematic uncertainties that easily exceed $10\%$.  Hence, if the
demands for sub-percent accuracy of the precision-cosmology program
are to be taken seriously, it is clear that conventional analytical
methods face a serious problem.

The {\tt Halotools} package is designed to remedy these and other
shortcomings of conventional large-scale structure analyses by
directly populating dark matter halos in numerical simulations with
mock galaxies. {\tt Halotools} therefore makes no appeal whatsoever to
fitting functions for the abundance or spatial distribution of dark
matter halos, while automatically taking halo exclusion into
account. In addition, large-scale structure observables such as
clustering and lensing are computed in {\tt Halotools} directly from
the mock galaxy distributions, using exactly the same method as used
for the actual observational measurements. Finally, we stress that all
calculations have been heavily optimized for MCMC-type applications.

Written exclusively in Python,\footnote{Some specific performance-critical elements 
are implemented with Cython \citep{cython}, a tool that compiles a python-like code into C code 
({\tt http://cython.org}).} {\tt Halotools} provides a
highly modular, object-oriented platform for building halo occupation
models, so that individual modeling features can easily be swapped in
and out. Beyond the conveniences of readability and
ease-of-development that comes with using contemporary design patterns
in a modern programming language, this modularity facilitates
rigorous, systematic study of each and every component that makes up a
halo occupation model. Of particular relevance to the present work is
that {\tt Halotools} has been designed from the ground-up with
assembly bias applications in mind.

Ultimately, in order for any cosmological likelihood analysis to
proceed it will be necessary to calibrate new fitting functions or
emulators \citep[e.g.,][]{heitmann_etal08, heitmann_etal10} so that
predictions can be made for un-simulated sets of cosmological
parameters. Either {\tt Halotools}, or a package very much like it,
will need to go hand-in-hand with such an effort, as
direct-mock-population provides the gold standard of precision for
large-scale structure predictions.

Although the equations in \S\ref{sec:halomodel} are indeed only rough
approximations, these equations are and will remain useful to gain
physical insight into the connection between occupation statistics and
two-point clustering. Throughout the paper, however, for all our
results concerning galaxy clustering and lensing, we do not perform
our calculations using
Eqs.~(\ref{eq:onehaloterm})-(\ref{eq:twohalotermlensing}). Instead, we
calculate $\xigg$ and $\xigm$ by using {\tt Halotools} to populate
dark matter halos with mock galaxies and then apply the
\citet{landyszalay93} estimator on the resulting set of point data. In
\S\ref{subsec:montecarlo} we provide a brief sketch of the Monte Carlo
techniques used in this mock population; we refer the reader to {\tt
  http://halotools.readthedocs.org} for comprehensive documentation of these
methods. We describe the simulation and halo catalogs we use in
\S\ref{subsec:sim}.

\subsection{Monte Carlo Methods}
\label{subsec:montecarlo}

To populate halos with central galaxies, first we calculate the value
of $\langle\ncen\rangle$ for every halo in the simulation. For
standard HOD models we use Eq.~(\ref{eq:ncendef}), whereas for
assembly-biased models we use the analytical expressions derived in
\S\ref{sec:hodgen} below. For every halo in the simulation, we then
draw a random number $r$ from $\mathcal{U}(0, 1),$ a uniform
distribution between zero and unity; for all halos with
$r\leq\avg{\ncen},$ we place a central galaxy at the halo center,
leaving all other halos devoid of centrals.

Populating satellites is more complicated because the spatial
distributions are non-trivial. The first step is the same: we compute
$\avg{\nsat}$ for every halo, using either Eq.~(\ref{eq:nsatdef}) or
the methods of \S\ref{sec:hodgen}, whichever is appropriate for the
model in question. For each halo, we then determine the number of
satellites that will be assigned to the halo by drawing an integer
from the assumed satellite occupation distribution,
$P(\nsat\vert\mhalo)$ (or $P(\nsat\vert\mhalo, x),$ see
\S\ref{sec:hodgen}).

In this paper, we use the {\tt Halotools} framework to model
satellites as being isotropically distributed within their halos
according to a NFW profile with concentration given by the value in
the halo catalog.\footnote{The modular design of {\tt Halotools} of
  course permits alternative modeling choices for the intra-halo
  distribution, e.g., simulated subhalo positions can be used
  directly.} We generate Monte Carlo realizations of both radial and
angular positions via the method of inverse transformation sampling,
which we sketch in the following paragraph. Briefly, first we generate
realizations of points uniformly distributed on the unit sphere; we
then multiply these halo-centric $x, y, z$ coordinates by the
corresponding realization of the radial position.

To realize points on the unit sphere, we draw random numbers $\phi$
and $t\equiv\cos(\theta)$ from $\mathcal{U}(0, 1),$ computing
$\sin(\theta) = \sqrt{1 - t^{2}}.$ The $x, y, z$ positions on the unit
sphere are then computed as $x = \sin(\theta)\cos(\phi),$ $y =
\sin(\theta)\sin(\phi),$ and $z = \cos(\theta).$ For the radial
positions, first we calculate $P_{\rm NFW}(<{\tilde{r}}\vert c),$ the
cumulative probability function of the mass profile of an NFW halo
with concentration $c:$
\beq
\label{eq:nfwmasspdf}
P_{\rm NFW}(<{\tilde{r}}\vert c) \equiv \frac{M_{\rm NFW}(<\tilde{r}\vert c)}{M_{\rm tot}} =
\frac{g(c\,\tilde{r})}{g(c)},
\eeq
where $g(x) \equiv \ln(1+x) - \frac{x}{1+x},$ and $\tilde{r}\equiv
r/R_{\rm vir}.$ For a halo with concentration $c$ populated by $\nsat$
satellites, we draw $\nsat$ random numbers $p$ from $\mathcal{U}(0,
1).$ Each value $p$ is interpreted as a probability where the
corresponding value for the scaled radius $\tilde{r}$ comes from
numerically inverting $p=P_{\rm NFW}(<{\tilde{r}}\vert c).$ Scaling
the $x, y, z$ points on the unit sphere by the value $r$ gives the
halo-centric position of the satellites.

%%%%%%%%%%%%%%%%%%%%%%%%%%%%%% SIMULATION %%%%%%%%%%%%%%%%%%%%%%%%%%%%%%

\subsection{Simulation and Halo Catalogs}
\label{subsec:sim}

Throughout this paper, the foundation of our results is the
collisionless N-body Bolshoi simulation \citep{bolshoi_11}. The
$\Lambda{\rm CDM}$ cosmological parameters of Bolshoi are
$\Omegam=0.27,$ $\Omega_{\Lambda}=0.73,$ $\Omegab=0.042,$
$\tilt=0.95,$ $\sigma_8 = 0.82,$ and $H_0=70$ $\kms\mpc^{-1}.$ The
gravity-solver of the simulation is the Adaptive Refinement Tree code
\citep[ART;][]{kravtsov_eta97,gottloeber_klypin08}, run on $2048^3$
particles in a $250\mpc$ periodic box. Bolshoi particles have a mass
of $\mathrm{m_p}\approx1.35\times10^{8}\msun;$ the force resolution
of the simulation is $\epsilon\approx1\kpc.$ Here and throughout the paper, 
all numerical values of length and mass will be understood to be in $h=1$ units. 
Snapshot data and halo
catalogs for Bolshoi are part of the Multidark Database
\citep{riebe_etal11}, accessible at {\tt http://www.multidark.org}.

We use the {\tt ROCKSTAR} halo finder
\citep{behroozi_rockstar11,behroozi_trees13} to identify host halos in
the $z=0$ Bolshoi snapshot. Halo catalogs based on {\tt ROCKSTAR} are
publicly available at {\tt
  http://hipacc.ucsc.edu/Bolshoi/MergerTrees.html}. Halos in these
catalogs are defined to be spherical regions centered on a local
density peak, such that the average density inside the sphere is
$\deltavir\approx360$ times the mean matter density of the simulation
box. The radius of each such sphere defines the {virial radius}
$\rvir$ of the halo; the mass enclosed inside this sphere is the
so-called ``virial mass"
$\mvir=\frac{4}{3}\pi\rvir^3\deltavir\Omegam\rhocrit,$ where
$\rhocrit=3H_{0}^{2}/8\pi\mathrm{G}$ is the critical energy density of
the universe. In the model developed in this work, it will be useful
to refer to a more generic halo mass $\mhalo$ that is defined
according to some density threshold that may deviate from $\deltavir,$
but it will be understood that all our numerical computations were
carried out on halos defined according $\deltavir.$

%%%%%%%%%%%%%%%%%%%%%%%%%%%%%% GENERALIZING THE HALO MODEL %%%%%%%%%%%%%%%%%%%%%%%%%%%%%%

\section{GENERALIZING THE HOD}
\label{sec:hodgen}

%-----------------------------------------------------------------------
\subsection{Basic Considerations}
\label{sub:basic}
%-----------------------------------------------------------------------

The standard HOD formalism described in \S~\ref{sec:halomodel}
presumes that halo mass is the sole property that influences the
probability that a (host) dark matter halo contains one or more
galaxies in some sample. If there exists some halo property $x$, such
that the HOD depends upon both $x$ and $\mhalo$, $P(\ngal \vert
\mhalo, x) \ne P(\ngal \vert \mhalo)$, {\em and} the clustering of
halos depends upon $x$, then the dependence of the HOD on $x$ must be
accounted for in order to faithfully model the clustering of
galaxies. {\em Galaxy assembly bias} refers to the situation when
there exists such a property $x$.

In this section, we discuss a simple and convenient generalization of
the HOD formalism to account for galaxy assembly bias. The formalism
that we describe is general in the sense that it can, in principle, be
applied to any halo property in addition to halo mass and it can be
straightforwardly extended to describe HODs that depend upon numerous
additional halo properties.

%-----------------------------------------------------------------------
\subsubsection{The principle of HOD conservation}
%-----------------------------------------------------------------------

We will differentiate between the occupation statistics of a standard
HOD model and a {\em decorated} HOD model by denoting these two
distributions as $P_{\rm std}$ and $P_{\rm dec},$ respectively, so
that $$\pstd{\ngal}{\mhalo, x} = \pstd{\ngal}{\mhalo},$$
but $$\pdec{\ngal}{\mhalo, x}\neq\pdec{\ngal}{\mhalo}.$$ In the
decorated HOD, we will refer to the {\em conditional moments of order
  k} as
\beq
\label{eq:conditionalmomentdef}
\mean{\ngal^k}{\mhalo, x}_{\rm dec} \equiv \sum_{\ngal=0}^{\infty} \ngal^{k} \pdec{\ngal}{\mhalo, x}, 
\eeq
and the corresponding {\em marginalized moments} as 
\beq
\label{eq:marginalizedmoments}
\mean{\ngal^k}{\mhalo}_{\rm dec} \equiv \int\, \mean{\ngal^k}{\mhalo, x}_{\rm dec} P(x\vert\mhalo)\, \dd x, 
\eeq
where $P( x \vert \mhalo)$ is the normalized probability distribution
that a halo of mass $\mhalo$ has a particular value of $x$.

Beginning with any standard HOD, our goal is to identify the
conditions under which the marginalized moments of a new, decorated
model are equal to the moments of the standard model:
\beq
\label{eq:momentpreservation}
\mean{\ngal^k}{\mhalo}_{\rm dec} = \mean{\ngal^k}{\mhalo}_{\rm std}.
\eeq
Eq.~(\ref{eq:momentpreservation}) defines our notion of {\em HOD
  conservation}; we will say that any model $\pdec{\ngal}{\mhalo, x}$
with marginalized moments that respect
Eq.~(\ref{eq:momentpreservation}) {\em preserve the moments} of
$\pstd{\ngal}{\mhalo}.$ Conserving the HOD minimizes the modifications
that are needed to allow for assembly bias. By using decorated HODs
that are modeled off of existing HODs one can continue to reap the
benefits of the infrastructure that has been developed over the past
decade.

Now we will define a perturbation to the halo occupation induced by
the auxiliary variable $x$,
\begin{equation}
\label{eq:delta}
\delta \ngal^k(\mhalo,x) \equiv \mean{\ngal^k}{\mhalo, x}_{\mathrm{dec}} - \mean{\ngal^k}{\mhalo}_{\mathrm{std}}.
\end{equation}
We will refer to $\delta\ngal^k$ as the $k^{\rm th}-${\em order
  decoration function}. The $\delta \ngal^k(x,\mhalo)$ represent the
change in the moments of the decorated HOD with respect to a standard
HOD without assembly bias. Using this definition,
Eq.~(\ref{eq:momentpreservation}) becomes
\begin{equation}
\label{eq:deltangalconstraint}
0 = \int\, \delta \ngal^k(\mhalo, x)\, P(x\vert\mhalo) \dd x
\end{equation}
for all $k=1, 2, 3, \dots$ Whenever no confusion can arise, we will
drop the superscript and it will be understood that $\delta\ngal =
\delta\ngal^1.$

%%%%%%
%-----------------------------------------------------------------------
\subsubsection{Application to centrals}
%-----------------------------------------------------------------------

Let us proceed by considering the central and satellite populations
separately because different assumptions are typically made to
describe these distributions.  In particular, halos have either zero
or one central galaxy and the HOD (both decorated and standard)
therefore follows a nearest-integer distribution.  This implies that
\begin{equation}
\mean{\ngal^k}{\mhalo, x}_{\mathrm{dec}} = \mean{\ngal^1}{\mhalo, x}_{\mathrm{dec}}
\end{equation}
for all $k$. 
Hence, once Eq.~(\ref{eq:momentpreservation}) is satisfied for $k=1,$ 
it is trivially satisfied for all higher-order central moments. 
Therefore, in
order to construct decorated HOD models that preserve the full
$\pstd{\ncen}{\mhalo}$ we only need to ensure that the first-order
decoration function $\deltancen^1$ respects
$$0 = \int\, \delta \ncen^1(\mhalo, x)\, P(x\vert\mhalo) \dd x.$$

%-----------------------------------------------------------------------
\subsubsection{Application to satellites}
%-----------------------------------------------------------------------

Moving on, consider the more difficult case of satellite galaxies. Typically, 
satellite galaxies are considered to be drawn from a Poisson distribution. 
We begin with the natural assumption that both $\pdec{\nsat}{\mhalo}$ and 
$\pdec{\nsat}{\mhalo, x}$ are Poisson distributions. Under this Poisson assumption, 
as well as the assumption that the first moment of the HOD is preserved, 
one can show that  
\begin{equation}
\label{eq:second_moment}
0 = \int\, \left[\delta \nsat^1(\mhalo, x )\right]^2 \, P(x \vert \mhalo )\, \dd x. 
\end{equation}
Eq.~(\ref{eq:second_moment}), however, cannot be satisfied for a
non-trivial decoration function $\delta \nsat^1(x)$ because the
integrand is non-negative over its entire domain. {\em If we wish for
  both $\pnm{\nsat}{\mhalo}$ and $\pnm{\nsat}{\mhalo, x}$ to obey
  Poisson statistics then it is not possible to conserve the HOD.}  The
converse is also true: if we insist on conserving the HOD, one cannot
have both $\pnm{\nsat}{\mhalo}$ and $\pnm{\nsat}{\mhalo, x}$ obey
Poisson statistics.

There are numerous reasonable ways to proceed.  First, one could
assume that {\it both} $\pstd{\nsat}{\mhalo}$ and
$\pdec{\nsat}{\mhalo, x}$ are Poisson distributions.  This implies
that
$$\mean{\nsat(\nsat-1)}{\mhalo, x}_{\rm dec} = \mean{\nsat}{\mhalo, x}^2_{\rm dec}.$$ 
and thus that
\beqray
\label{eq:secondmomentfail}
\mean{\nsat(\nsat-1)}{\mhalo}_{\rm dec} &  =  & \int\, \mean{\nsat}{\mhalo, x}_{\rm dec}^2\, P(x\vert\mhalo)\, \dd x \nonumber \\
                                     & \neq & \mean{\nsat}{\mhalo}^2_{\rm dec},
\eeqray
Hence, in this case the {\it marginalized}, decorated HOD,
$\pdec{\ngal}{\mhalo}$, cannot be Poissonian and 
Eq.~(\ref{eq:momentpreservation}) cannot be satisfied for $k=2$
(i.e., one cannot conserve the second moment of the HOD). 

As a second alternative, one could insist upon the principle of HOD
conservation, $\pdec{\nsat}{\mhalo} = \pstd{\nsat}{\mhalo}$, and assume 
$\pdec{\nsat}{\mhalo,x}$ to be Poissonian. In that case
neither $\pdec{\nsat}{\mhalo}$ nor $\pstd{\nsat}{\mhalo}$ can be
Poissonian.

Third, still insisting on HOD conservation, one could choose
for $\pstd{\nsat}{\mhalo}$ to be Poissonian. Now one has that
$\pdec{\nsat}{\mhalo}$ is Poissonian as well, but then of course
$\pdec{\nsat}{\mhalo, x}$ cannot obey Poisson statistics for all $x$.
As a simple, explicit demonstration of this last case, suppose that
$\pdec{\nsat}{\mhalo,x}$ is such that
\begin{equation}
\mean{\nsat(\nsat-1)}{\mhalo, x}_{\rm dec} = \alpha^2\mean{\nsat}{\mhalo, x}_{\rm dec}^2, 
\end{equation}
where $\alpha$ is a constant. Note that $\alpha$ cannot be equal to
unity, which would correspond to $\pdec{\nsat}{\mhalo,x}$ being
Poissonian.  Eq.~(\ref{eq:deltangalconstraint}) now implies that
\begin{equation}
  \alpha^2 = \left[1+\int\, \frac{\left[\delta \nsat^{1}(\mhalo, x)\right]^2}{\mean{\nsat}{\mhalo, x}^2}\, P(x\vert\mhalo)\, \dd x
    \right]^{-1}.
\end{equation}
Since the integral in the above expressions is non-negative, this
implies that $\alpha^2 < 1$.  Hence, {\em in order to conserve the
  HOD, the distribution $\pdec{\nsat}{\mhalo, x}$ must have a
  distribution that is narrower than a Poisson distribution with mean
  $\mean{\nsat}{\mhalo,x}_{\rm dec}.$} The intuitive way of thinking
of this result is that there is an additional source of variance
associated with the allocation of satellites into sub-populations at a
given halo mass.  Of course, introduction of the constant $\alpha$ is
only one simple way in which to characterize $\pdec{\nsat}{\mhalo,
  x}$; in practice there is an infinite set of possible decorations
that satisfy Eq.~(\ref{eq:deltangalconstraint}).

%-----------------------------------------------------------------------
\subsubsection{The central-satellite term}
\label{subsub:censatterm}
%-----------------------------------------------------------------------

As discussed in \S\ref{subsec:censatcorrelations}, halo occupation models also
need to make an assumption regarding $\langle \ncen\nsat \rangle$,
which specifies the expectation value for the number of
central-satellite pairs.  Throughout this paper we assume that
\begin{equation}\label{eq:ncnsassumption}
  \langle\ncen\nsat \vert \mhalo,x \rangle_{\rm dec} =
  \langle\ncen \vert \mhalo,x \rangle_{\rm dec} \cdot
  \langle\nsat \vert \mhalo,x \rangle_{\rm dec}\,,
\end{equation}
and thus that the occupation numbers for centrals and satellites {\it
  at fixed $\mhalo$ and $x$} are independent. Note, though, that
because of the mutual covariance with the secondary halo property $x$,
this will generally {\it not} be the case for the corresponding
marginalized moment, i.e.,
\begin{eqnarray}
  \langle\ncen\nsat \vert \mhalo \rangle_{\rm dec} & = &
  \int\, \langle\ncen\nsat \vert \mhalo,x \rangle_{\rm dec} \, P(x \vert \mhalo )\, \dd x \nonumber \\ 
  & \ne & \langle\ncen \vert \mhalo \rangle_{\rm dec} \cdot \langle \nsat \vert\mhalo\rangle_{\rm dec}.
\end{eqnarray}
Assembly bias induces a non-trivial $\langle \ncen \nsat \vert \mhalo \rangle$ 
through the auxiliary property $x$. 
This assumption is not a necessary feature of the decorated HOD, 
and as described in \S\ref{sec:discussion} the {\tt Halotools} framework 
is written to accommodate alternative assumptions for central-satellite correlations.

As discussed in more detail in \S\ref{sub:toy} below, the decoration
of centrals is limited by the requirement that $0 \leq \langle \ncen
\rangle \leq 1$. Consequently, if $\langle\ncen \vert
\mhalo\rangle_{\rm dec} = 1$ (which is typically the case for massive
halos), one also has that $\langle\ncen \vert \mhalo, x\rangle_{\rm
  dec} = 1$ (see \S\ref{sub:toy} below), and thus
\begin{equation}
  \langle\ncen\nsat \vert \mhalo \rangle_{\rm dec} = \langle\nsat \vert \mhalo \rangle_{\rm dec}.
\end{equation}
Similarly, if $\langle\ncen \vert \mhalo\rangle_{\rm dec} = 0$ (which
is typically the case in low mass halos), then $\langle\ncen \vert
\mhalo, x\rangle_{\rm dec} = 0$ and thus $\langle \ncen\nsat \vert
\mhalo \rangle_{\rm dec} = 0$. Hence, under HOD conservation, the
average number of central-satellite pairs in massive halos for a
decorated model is identical to that of its standard baseline model,
except for the fairly narrow range in halo masses for which $0 <
\langle \ncen \vert \mhalo \rangle < 1$ (see also
\S\ref{subsubsec:smallscaleclustering} and
\S\ref{subsec:dependencebaseline}). 

%%%%%%%%%%%%%%%%%%%%%%%%% FIGURE %%%%%%%%%%%%%%%%%%%%%%%%%%%%%

\begin{figure}
\begin{center}
\includegraphics[width=8cm]{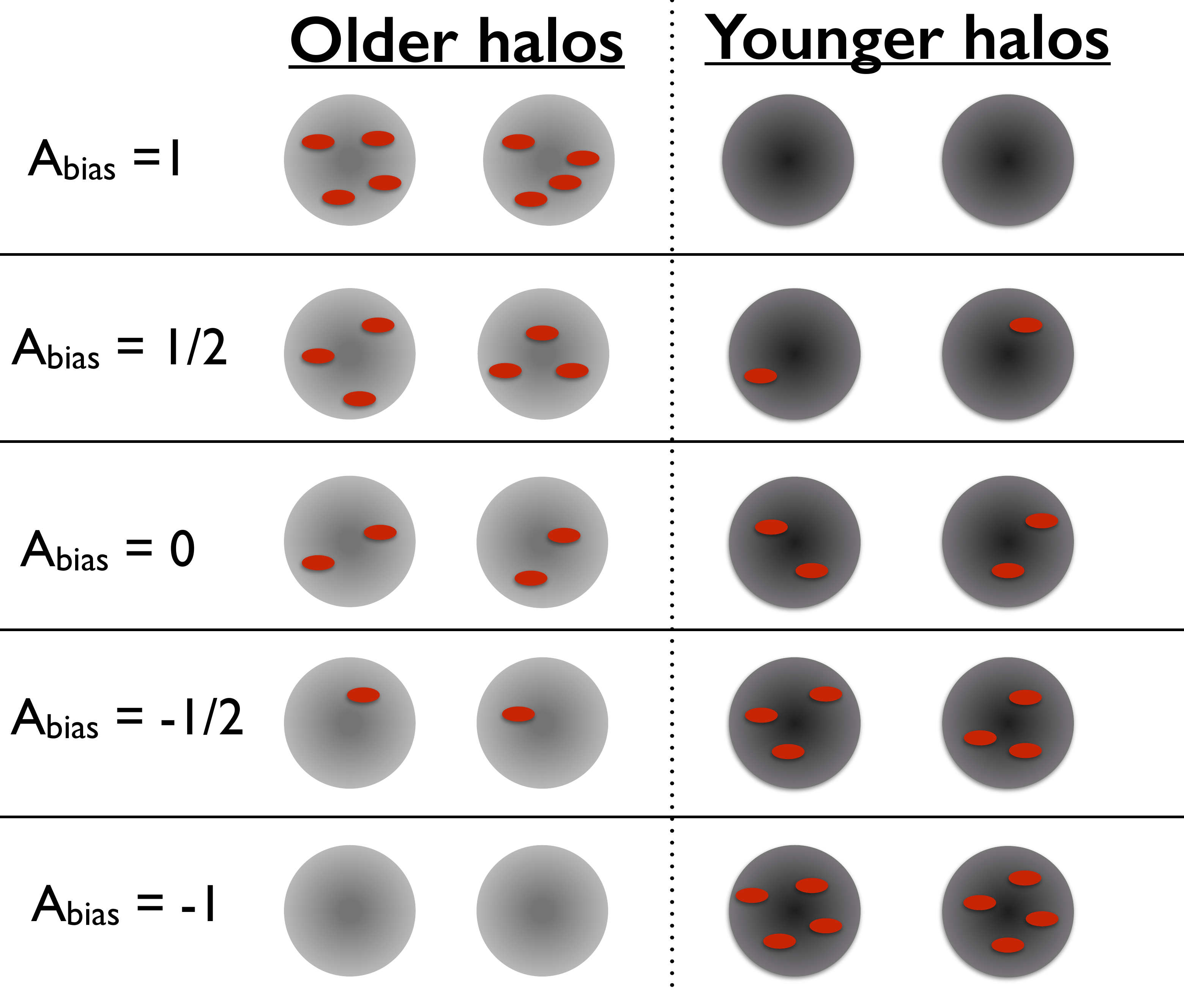}
\caption{ {\bf Cartoon illustration of the decorated HOD.} Each row of
  circles represents a population of halos of the same mass, divided
  evenly on the left and right into halos that are old and young for
  their mass, respectively. The number of galaxies in each halo is
  represented with the small, red ellipses. As described in
  \S\ref{sub:toy}, the $\abias$ parameter governs the strength of
  assembly bias in our two-population model. Each row gives a visual
  representation of a different value of $-1 \le \abias \le 1.$ More
  positive values of $\abias$ correspond to models in which
  later-forming halos host more galaxies relative to earlier-forming
  halos of the same mass, and conversely for $\abias<0.$ Note that
  changing values of $\abias$ does not change $\mean{\ngal}{\mhalo},$
  the mean number of galaxies averaged across all halos of a fixed
  mass; this is the defining feature of the decorated HOD, and the
  meaning of the principle of HOD conservation.  }
\label{fig:cartoon}
\end{center}
\end{figure}

%%%%%%%%%%%%%%%%%%%%%%%%%%%%%%%%%%%%%%%%%%%%%%%%%%%%%%%%%%%%%%%%%%%%

%%%%%%%%%%%%%%%%%%%%%%%%% FIGURE %%%%%%%%%%%%%%%%%%%%%%%%%%%%%

\begin{figure}
\begin{center}
\includegraphics[width=8cm]{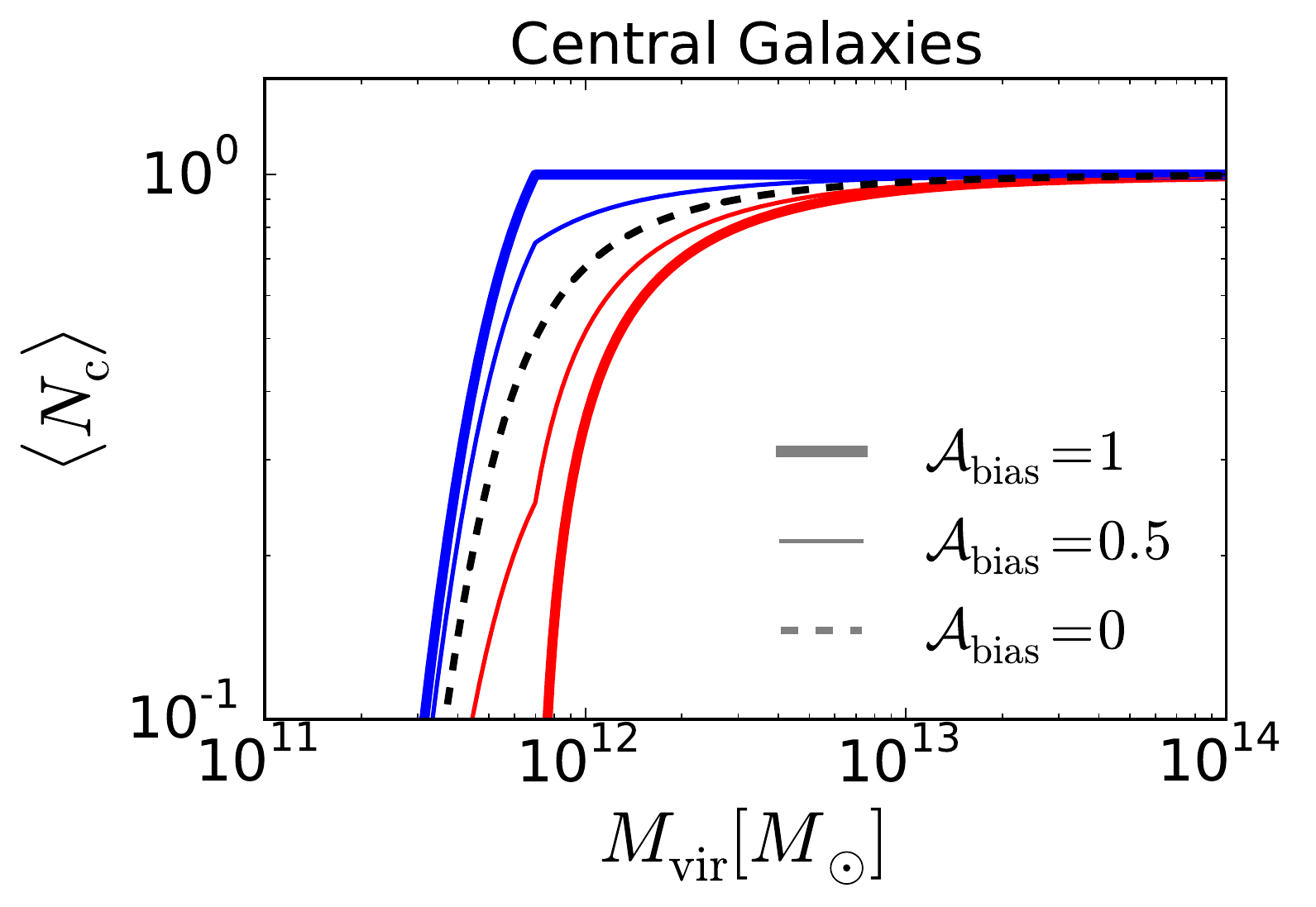}
\includegraphics[width=8cm]{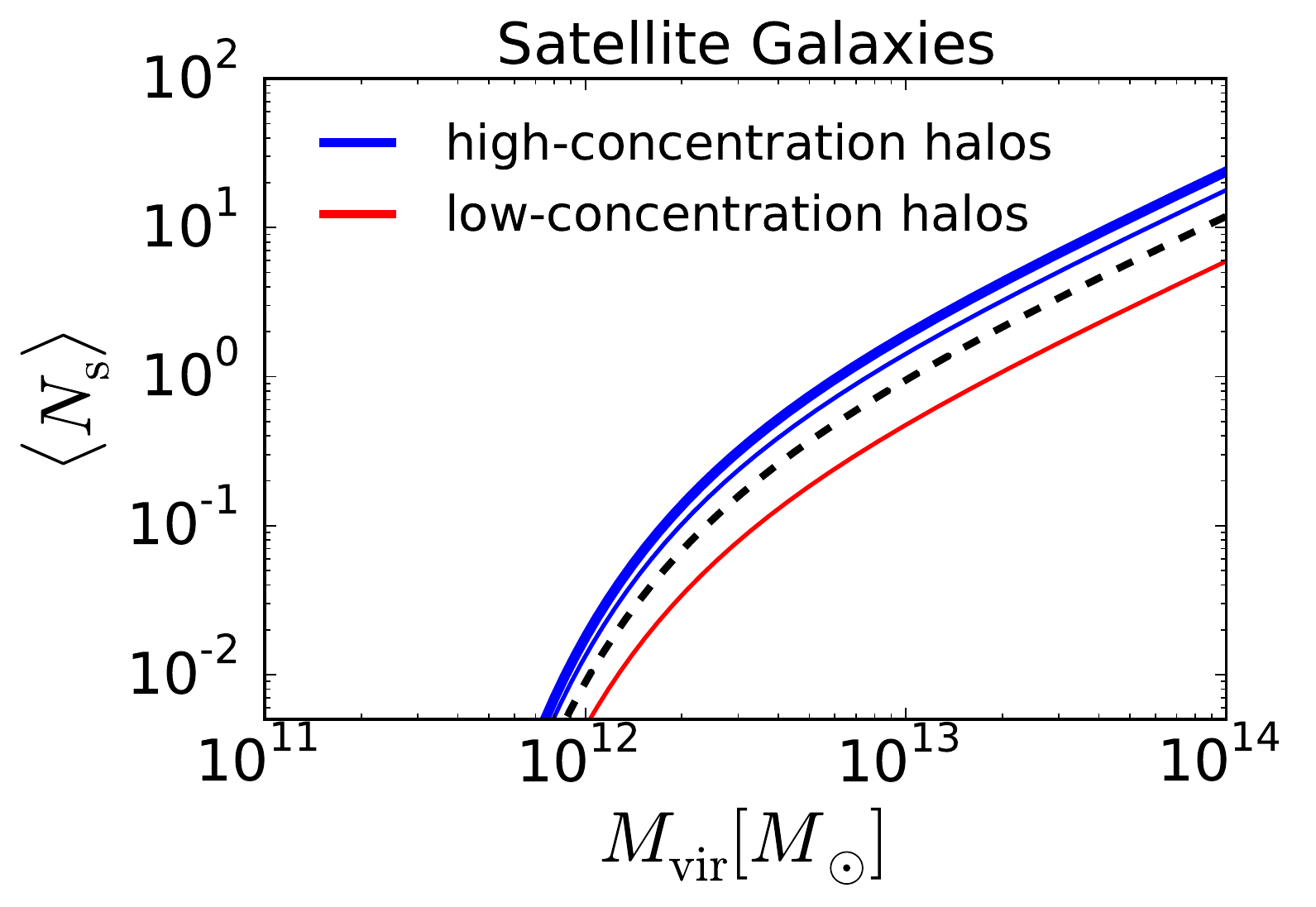}
\caption{ {\bf Decorated HOD occupation statistics.}  All curves
  pertain to the \citet{leauthaud11b} HOD (see \S\ref{subsec:baselineHOD}) with stellar mass threshold
  $M_{\ast}>10^{10.5}\msun$ with $\sigma_{\rm logM_{\ast}} = 0.4.$ The
  parameter $\abias$ controls the strength of assembly bias. The value
  $\abias=1$ ({\em thick blue and red curves}) shows the case of
  maximum assembly bias allowed by the HOD-conservation constraint;
  $\abias=0.5$ ({\em thin blue and red curves}) shows a model with
  $50\%$ allowable strength; the standard HOD result $\abias=0$ is
  shown with the {\em dashed black curve}.  Assembly bias can have no
  impact on central occupation statistics in the $\langle {\rm
    N_{cen}}\rangle = 1$ regime, whereas satellite occupations are
  unbounded and so can be biased at large halo mass. There is no thick
  red curve in the bottom panel because in the $\abias=1$ model there
  are exactly zero satellites in high-concentration halos.  }
\label{fig:occupations}
\end{center}
\end{figure}

%%%%%%%%%%%%%%%%%%%%%%%%%%%%%%%%%%%%%%%%%%%%%%%%%%%%%%%%%%%%%%%%%%%%

%----------------------------------------------------------------------------------------
\subsection{A Toy Model: Discrete Halo Sub-Populations}
\label{sub:toy}
%----------------------------------------------------------------------------------------

In this section we will develop a simple toy example of a decorated
HOD in which the halo population at fixed mass is split into two
sub-populations. For example, there is one sub-population that
contains a fraction $P_1$ of all halos (the ``type 1" halos) at fixed
mass for which $\delta \nsat(\mhalo, x) = \deltansatone$ (a constant)
and a second population containing $P_2=1-P_1$ of all halos at fixed
mass and for which $\delta \nsat(\mhalo, x) = \deltansattwo$. This
could be achieved by splitting the halo population into the $P_1$
percentile of highest-$x$ halos and assigning them a satellite galaxy
occupation enhancement of $\deltansatone$. Likewise, the remaining
$1-P_1$ percentile of lowest $x$ halos receive a satellite galaxy
occupation decrement of $\deltansattwo$ This is a simple case of two,
discrete halo sub-populations, with different occupation statistics,
at fixed mass.

In such a scenario, conserving the first moment of the HOD
(Eq.~[\ref{eq:deltangalconstraint}], for $k=1$) requires that
\begin{equation}
\label{eq:toy_mean_conservation}
0 = P_1 \deltansatone + P_2 \deltansattwo,
\end{equation}
and likewise for centrals.  As discussed in \S\ref{sub:basic}, the
second central occupation moment is automatically conserved. For the
satellites, we choose to assume that both $\pdec{\nsat}{\mhalo, x}$
and $\pstd{\nsat}{\mhalo}$ are Poisson, and thus that the second
occupation moment is not conserved. \citep[see][for the motivation for
  this choice based on subhalo occupation statistics]{mao_etal15}.
With these two assumptions, one need only specify the two first-order
decoration functions, $\deltancenone$ and $\deltansatone,$ and then
$\pdec{\ngal}{\mhalo, x}$ is completely determined.

In order to guarantee that the mean number of galaxies is always
non-negative and that the mean number of centrals is never greater
than one, there is a restricted set of values which the decoration
functions $\deltansatone$ and $\deltancenone$ may take on. It is easy
to show that the maximum strength of assembly bias for satellites in
this two-population toy model is
\begin{equation}
\label{eq:sat_max}
\deltansatone(\mhalo, x) \le \deltansatone^{\mathrm{max}}(\mhalo) =
\frac{1-P_1}{P_1}\, \mean{\nsat}{\mhalo}_{\rm std}.
\end{equation}
Eq.~(\ref{eq:sat_max}) ensures that $\avg{\nsat} \ge 0$ in type-2
halos ($x < \bar{x}(\mhalo)$).  The constraint that $\avg{\nsat} \ge
0$ in type-1 halos ($x > \bar{x}(\mhalo)$) can be written as
\begin{equation}
\label{eq:sat_min}
\deltansatone(\mhalo, x) \ge \deltansatone^{\mathrm{min}}(\mhalo) = -
\mean{\nsat}{\mhalo}_{\rm std}.
\end{equation}
Notice that the range of possible values for $\deltansatone$ depends
upon both $P_1$ and the value of the first moment of the ``baseline"
model, $\mean{\nsat}{\mhalo}_{\rm std}.$ We return to this important
point in the discussion of our results.

Central galaxies are subject to the same positivity constraint that
$\mean{\ncen}{\mhalo, x} \ge 0$ as well as the additional constraint
that the number of central galaxies in any halo can never exceed
one, $\mean{\ncen}{\mhalo, x} \le 1.$ 
As a result, the maximum strength of assembly bias for centrals
is given by
\begin{equation}
\label{eq:cen_max}
\deltancenone^{\mathrm{max}} = \mathrm{min}\left\{ 1-\mean{\ncen}{\mhalo}_{\rm std}, 
\frac{1-P_1}{P_1}\, \mean{\ncen}{\mhalo}_{\rm std} \right\}.
\end{equation}
The minimum in Eq.~(\ref{eq:cen_max}) selects that constraint on the
central population which is most restrictive. Ensuring that centrals
in type-2 halos also respect $0 \le \mean{\ncen}{\mhalo, x} \le 1$ 
places the
following constraint on centrals in type-1 halos
\begin{equation}
\label{eq:cen_min}
\deltancenone^{\mathrm{min}} = \mathrm{max}\left\{-\mean{\ncen}{\mhalo}_{\rm std}, 
\frac{1-P_1}{P_1}\, (\mean{\ncen}{\mhalo}_{\rm std}-1) \right\}.
\end{equation}
The quantities $\deltancentwo$ and $\deltansattwo$ are subject to the
same constraints described by
Eqs.~(\ref{eq:sat_max})-(\ref{eq:cen_min}). Note that when one
population attains its maximum (minimum) allowed value,
Eq.~(\ref{eq:toy_mean_conservation}) guarantees that the other
population automatically attains its minimum (maximum).

%%%%%%%%%%%%%%%%%%%%%%%%% FIGURE %%%%%%%%%%%%%%%%%%%%%%%%%%%%%
\begin{figure*}
\begin{center}
\includegraphics[width=8.3cm]{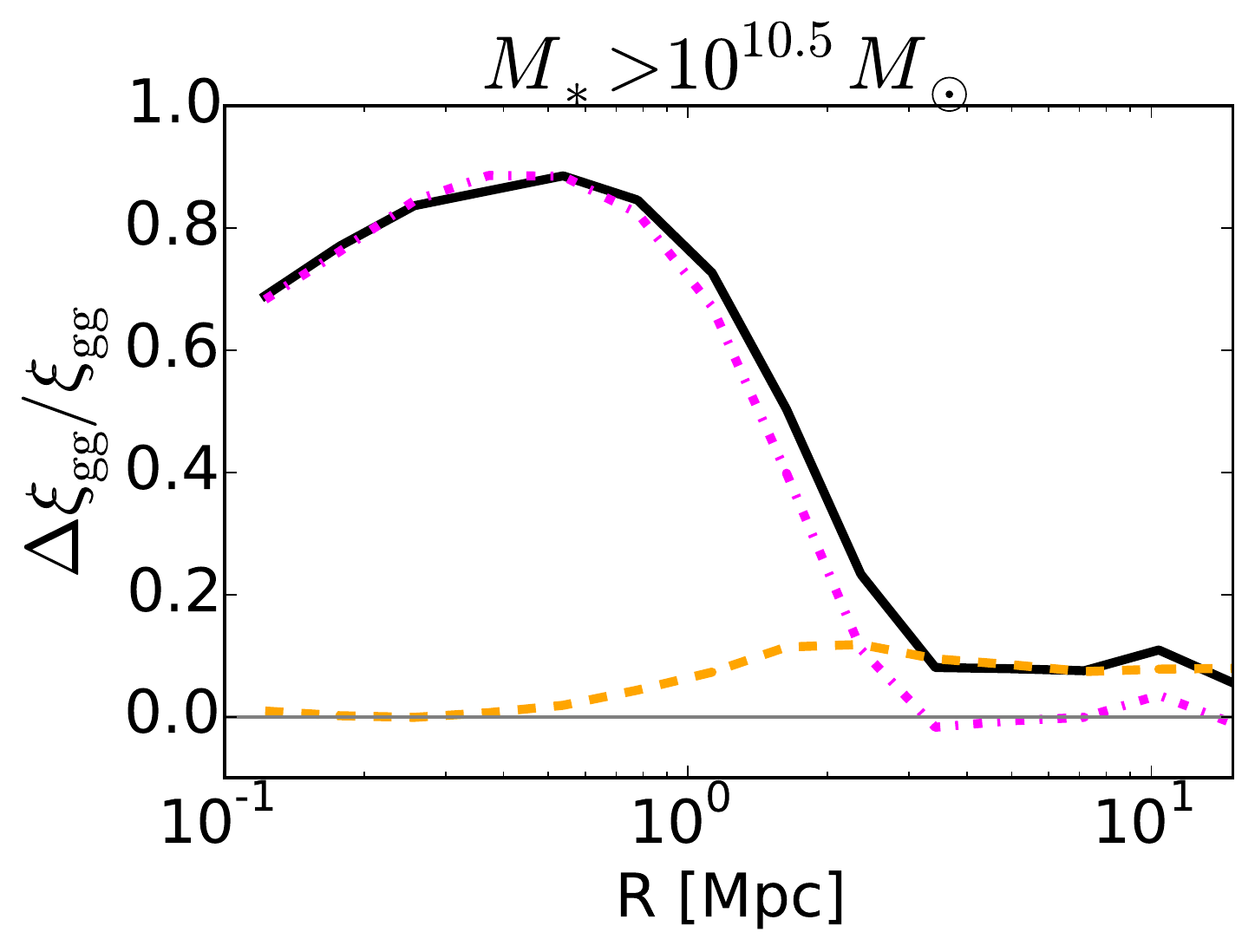}
\includegraphics[width=8.3cm]{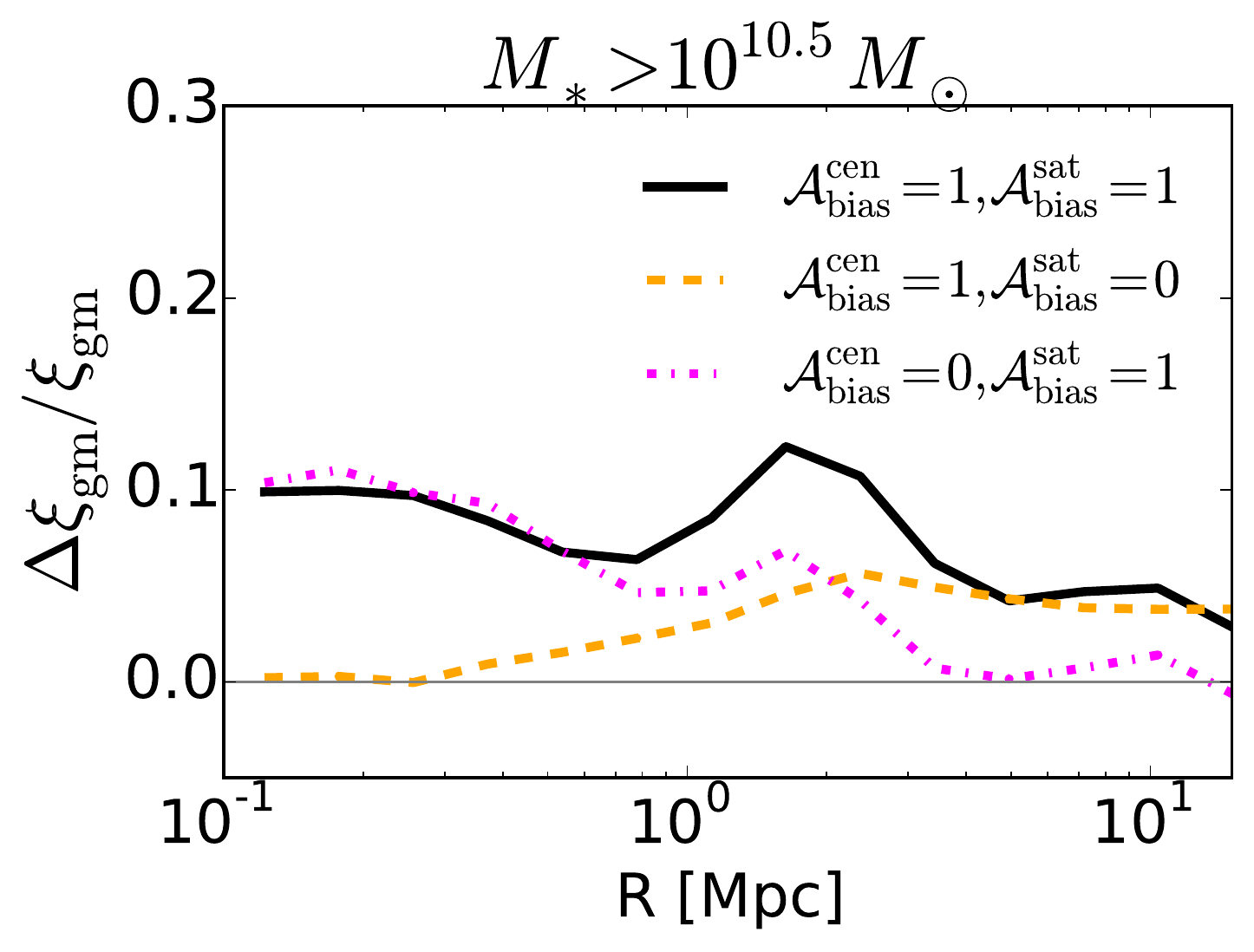}
\caption{ {\bf Parsing central and satellite assembly bias.} The {\em
    left} panel shows the effect of assembly bias on galaxy
  clustering. The curves show the fractional change to $\xigg$ as a
  function of spatial separation for three different assembly bias
  models.  The {\em solid, black} line shows a model in which both
  centrals and satellites have maximum assembly bias of the same sign,
  $\abias^{\rm cen} = \abias^{\rm sat} = 1.$ The {\em dashed, orange}
  lines shows the case of maximum assembly bias in the centrals, zero
  assembly bias in the satellites; conversely, the {\em dot-dashed,
    purple} line shows the effect of standard central occupations but
  maximum assembly bias in the satellites. The {\em right} panel is
  similar, but shows the fractional effect of these models on the
  galaxy-mass cross correlation.  }
\label{fig:satcen}
\end{center}
\end{figure*}
%%%%%%%%%%%%%%%%%%%%%%%%%%%%%%%%%%%%%%%%%%%%%%%%%%%%%%%%%%%%%%%%%%%%

%--------------- Defining Assembly Bias
\subsubsection{Defining the assembly bias parameter $\abias$}

In the following sections, we give examples of the plausible strength
of assembly bias effects based upon this simple, but illustrative,
two-population model. In those sections, we refer to the strength of
assembly bias using the parameter $-1 \le \Abias(\mhalo) \le 1.$ The $\abias$
parameter governs assembly bias in this two-population, decorated HOD
as
\begin{displaymath}
   \deltangalone(\mhalo) = \left\{
     \begin{array}{lr}
       \vert\abias(\mhalo)\vert\deltangalone^{\rm max}(\mhalo) & : \abias > 0  \\
       \vert\abias(\mhalo)\vert\deltangalone^{\rm min}(\mhalo) & : \abias < 0  
     \end{array}
   \right.
\end{displaymath}
Thus in our nomenclature, the dimensionless $\abias$ parameter
specifies the strength of assembly bias as a fraction of its maximum
allowable effect. The maximum and minimum assembly bias strengths,
$\deltancenone^{\rm max}, \deltancenone^{\rm min}, \deltansatone^{\rm
  max}$ and $\deltansatone^{\rm min},$ are given in
Eqs.~(\ref{eq:sat_max})-(\ref{eq:cen_min}). Our sign convention is
always to choose the ``type-1" halos to be those in the
upper-percentile of the secondary property (e.g., $x \ge
\bar{x}(\mhalo)$). Fig.~\ref{fig:cartoon} is a cartoon illustration of
a two-population scenario such as this one.

As a specific example, suppose we choose the secondary halo property
$x$ to be halo concentration, that we split our halo population in
half at each halo mass, and that we wish for both centrals and
satellites in halos with above-average concentration to receive a
boost to their mean occupations by $50\%$ of the maximum allowable
strength at each mass.  Then we have $P_1=0.5,$ $\Abias = 0.5,$ and
thus
\beqray
\deltansatone(\mhalo) & = & 0.5\, \deltansatone^{\mathrm{max}} = 0.5\mean{\nsat}{\mhalo}_{\rm std} \nonumber \\
\deltansattwo(\mhalo) & =  & 0.5\, \deltansattwo^{\mathrm{min}} = -0.5\mean{\nsat}{\mhalo}_{\rm std} \nonumber,
\eeqray
with directly analogous equations holding for centrals. 

Figure~\ref{fig:occupations} shows examples of HODs constructed by
splitting the host halo population in the Bolshoi simulation on halo
concentration using $P_1=0.5.$ Fig.~\ref{fig:occupations} shows
examples of galaxy assembly bias of strength $\Abias=0$ (no assembly
bias, {\em dashed black curves}), $\Abias=1$ (maximum allowable
assembly bias in a two-population model, {\em thick red and blue
  curves}), and an intermediate value of $\Abias=0.5$ ({\em thin red
  and blue curves}).

The top panel of Fig.~\ref{fig:occupations} exhibits the HODs of
central galaxies in this model. In the case of central galaxies,
assembly bias is restricted to the regime $0< \avg{\ncen} < 1.$ Thus
the scatter in the stellar-to-halo-mass relation controls the
operative halo mass range for central galaxy assembly bias; we will
return to this point in \S\ref{subsec:dependencebaseline}.

The lower panel of Fig.~\ref{fig:occupations} shows the effect of
galaxy assembly bias on the satellites within the two host halo
sub-populations. Unlike the case of central galaxies, satellite galaxy
assembly bias is relevant at all host halo masses for which
$\mean{\nsat}{\mhalo}_{\rm std}>0.$ In particular, note that there is
no thick red curve in the lower panel of Fig.~\ref{fig:occupations}.
This is because in the $\abias^{\rm sat}=1$ case, there are exactly
zero satellites in low-concentration halos.  This is different for the
$\abias^{\rm cen}=1$ case due to the constraint that a halo cannot be
occupied by more than one central.

%%%%%%%%%%%%%%%%%%%%%%%%%%%%%%%%%%%%%%%%%%%%%%%%%
% Results on clustering
%%%%%%%%%%%%%%%%%%%%%%%%%%%%%%%%%%%%%%%%%%%%%%%%%

\section{IMPACT OF ASSEMBLY BIAS ON CLUSTERING}
\label{sec:results}

In order to demonstrate the potential impact of assembly bias, we
 use {\tt Halotools} to populate dark matter
halos in the Bolshoi simulation with galaxies with a stellar mass
$\mstar>10^{10.5}\msun$. We use the standard HOD model described in
\S\ref{subsec:baselineHOD} as our `no-assembly-bias' baseline model,
which we decorate using a simply two-population model as follows. At
each halo mass, we split the halo population into low- and
high-concentration halos; those within the top $50\%$ of concentration
at fixed $\mhalo$ are assigned to the first sub-population, and the
remaining to the second population (so $P_1=P_2=0.5$). With this
choice, a {\em positive} value for $\Abias$ implies that halos with
above-average concentration have {\em boosted} galaxy
occupations.\footnote{This is opposite the intuitive expectation that
  high-concentration halos should have fewer satellites
  \citep{zentner05}. This more natural expectation simply corresponds
  to a negative value of $\Abias$, and will be addressed in the next
  subsection.} We assume that both $\pstd{\nsat}{\mhalo}$ and
$\pdec{\nsat}{\mhalo, x}$ are Poisson distributions, so that the
decorated HOD is entirely specified by two free parameters:
$\abias^{\rm cen}$ and $\abias^{\rm sat}$. When both of these
parameters equal zero, the model is formally equivalent to the baseline
`no-assembly-bias' model of \cite{leauthaud11b}.

Once we have populated the Bolshoi simulation volume with mock
galaxies, we compute the corresponding galaxy-galaxy and galaxy-matter
correlation functions using the fast pair-counting facilities of {\tt
  Halotools} combined with the Landy \& Szalay (1993) correlation
function estimator. Our figures can be reproduced in quantitative detail 
using the annotated IPython Notebook in the repository stored at {\tt https://github.com/aphearin/decorated-hod-paper}. 
This repository also contains a frozen copy of the exact version of {\tt Halotools} that generated our results.

%---------------------------------------------------------------------------
\subsection{Central vs. Satellite Assembly Bias}
\label{subsec:basicfeatures}

Figure~\ref{fig:satcen} exhibits our first demonstration of the
sense and size of assembly bias effects on two-point functions of
galaxy samples in our simple two-population toy model. We explore
three different models for the decoration that only differ in their
treatment of centrals and satellites, and compare the resulting
correlation functions to the corresponding signal in the baseline
model:
\ben
\item $\abias^{\rm cen} = 1, \abias^{\rm sat} = 1,$ maximum assembly
  bias in both populations ({\em black curves}).
\item $\abias^{\rm cen} = 1, \abias^{\rm sat} = 0,$ maximum central
  assembly bias only ({\em dashed, yellow curves}).
\item $\abias^{\rm cen} = 0, \abias^{\rm sat} = 1,$ maximum satellite
  assembly bias only ({\em dot-dashed, magenta curves}).
\een
Note how assembly bias in satellites vs. centrals imprints 
a distinct signature on galaxy clustering as well as lensing, as we discuss in
more detail what follows.

%-----------------------------------------------------
\subsubsection{Small-scale ($R \lesssim 1 \mathrm{Mpc}$) clustering}
\label{subsubsec:smallscaleclustering}

Interestingly, the small- and large-scale influences of assembly bias
on galaxy clustering can be qualitatively different. From the close
agreement between the solid, black and dot-dashed, magenta curves in
Fig.~\ref{fig:satcen}, it is evident that for large values of
$\abias,$ the influence of satellite assembly bias dominates that of
centrals on small scales.

We can understand this through the analytical halo-model expression,
Eq.~(\ref{eq:onehaloterm}).  The one-halo term has two contributions:
one from $\langle\ncen\nsat\rangle,$ and another from $\langle
\nsat(\nsat-1) \rangle.$ The central-satellite term may only be
altered by a small amount due to assembly bias. This is because we
have assumed that the occupation statistics of centrals and satellites
obey Eq.~(\ref{eq:ncnsassumption}). Consequently, as discussed in
\S\ref{subsub:censatterm}, the halo mass range over which assembly
bias can influence central galaxy occupation at all is limited to host
halo masses where $0 < \mean{\ncen}{\mhalo}_{\rm std} < 1$.
This range is typically fairly narrow, as is evident in the top
panel of Fig.~\ref{fig:occupations}, and because halos of these masses
typically have zero satellite galaxies, the expectation value $\langle
\ncen \nsat \rangle$ is left nearly unaffected by assembly bias.

This is the behavior that is typical of most $\mstar-$threshold HODs explored in
the literature as reasonable descriptions of observational data;
however, exceptions to this reasoning may be realized, particularly
for samples selected such that $\mean{\ncen}{\mhalo}_{\rm std} < 1$
while $\mean{\nsat}{\mhalo}_{\rm std} > 0$ over a broad range of halo
masses. For a commonly encountered example, 
see the blue-selected galaxy HODs inferred by
\citet{zehavi11} or discussed in the mock catalogs of
\citet{zentner_etal14}.

The boost in small-scale clustering due to assembly bias is nearly
entirely due to an increase in the number of satellite-satellite pairs
(an increase in the $\langle \nsat^2 \rangle$ term in the halo model)
for relatively high-mass host halos. This effect is simple to
understand. Packing the same number of satellites into fewer hosts
increases the average number of pairs per host and decreases the
number of relevant hosts. This leads to a significant boost in
small-scale clustering \citep[see][]{watson_powerlaw11} as the same
number of galaxies are apportioned among fewer, richer groups. This
dependence should be fairly generic.

Finally we point out that since we have assumed that satellite
galaxies trace the dark matter, i.e., $n_{\rm sat}(r)$ is modeled as a NFW
profile with the same concentration parameter as for the dark matter
host halo, implementing assembly bias via a split on host halo
concentration modifies the average, satellite-number-weighted, number
density profile of satellite galaxies. However, this has only a very
small impact on the clustering signal, which is completely dominated
by changes in the typical number of satellite-satellite pairs.  In
other words, the shift in clustering on small scales is dominated by
the changes in $\langle \nsat^2 \rangle$, and would be present even if
we had divided our host halo sample into two random sub-populations at
fixed mass \citep[see][section 2, for a discussion of this same point
  in a different context]{paranjape15}.

%------------------------------------------------------
\subsubsection{Large-scale ($R > 1 \mathrm{Mpc}$) clustering}
\label{subsubsec:largescaleclustering}

On large scales, the distinct signatures of central vs. satellite assembly bias 
arise from different considerations.  For host halos whose mass is below the non-linear
collapse mass, $\mcoll \approx 10^{12.7}\msun$, high-concentration
halos exhibit a larger clustering bias relative to low-concentration
halos of the same mass.  This trend weakens with increasing halo mass
and may even change sign for halos with $\mhalo\gg\mcoll$
\citep{wechsler06}, such that for halos with masses greater than the
collapse mass it is the low-concentration halos that cluster more
strongly.  Large-scale galaxy clustering is a weighted average of halo
clustering (see Eq.~[\ref{eq:twohaloterm}]) and the role of {\em
  galaxy assembly bias} on large scales is in determining the
weighting of halo-halo pairs. Apportioning more galaxies into a subset
of halos that is more strongly clustered will weight those halos more
highly and boost clustering, whereas preferentially placing galaxies
into a subset of halos that clusters more weakly at fixed mass will
suppress clustering.

Figure \ref{fig:satcen} demonstrates that central galaxy assembly bias
has a far more important influence on the two-halo term relative to
satellite galaxy assembly bias.  This is due to two distinct effects.
First and foremost, centrals dominate satellites by number; for the
stellar mass threshold of $\mstar>10^{10.5}\msun$ shown here, more
than $75\%$ of galaxies are centrals. So central-central pairs are far
more common than central-satellite or satellite-satellite pairs in the
two-halo regime \citep[see][]{watson_powerlaw11} and altering central
galaxy occupations induces greater deviations in large-scale
clustering.

%%%%%%%%%%%%%%%%%%%%%%%%% FIGURE %%%%%%%%%%%%%%%%%%%%%%%%%%%%%

\begin{figure*}
\begin{center}
\includegraphics[width=8.3cm]{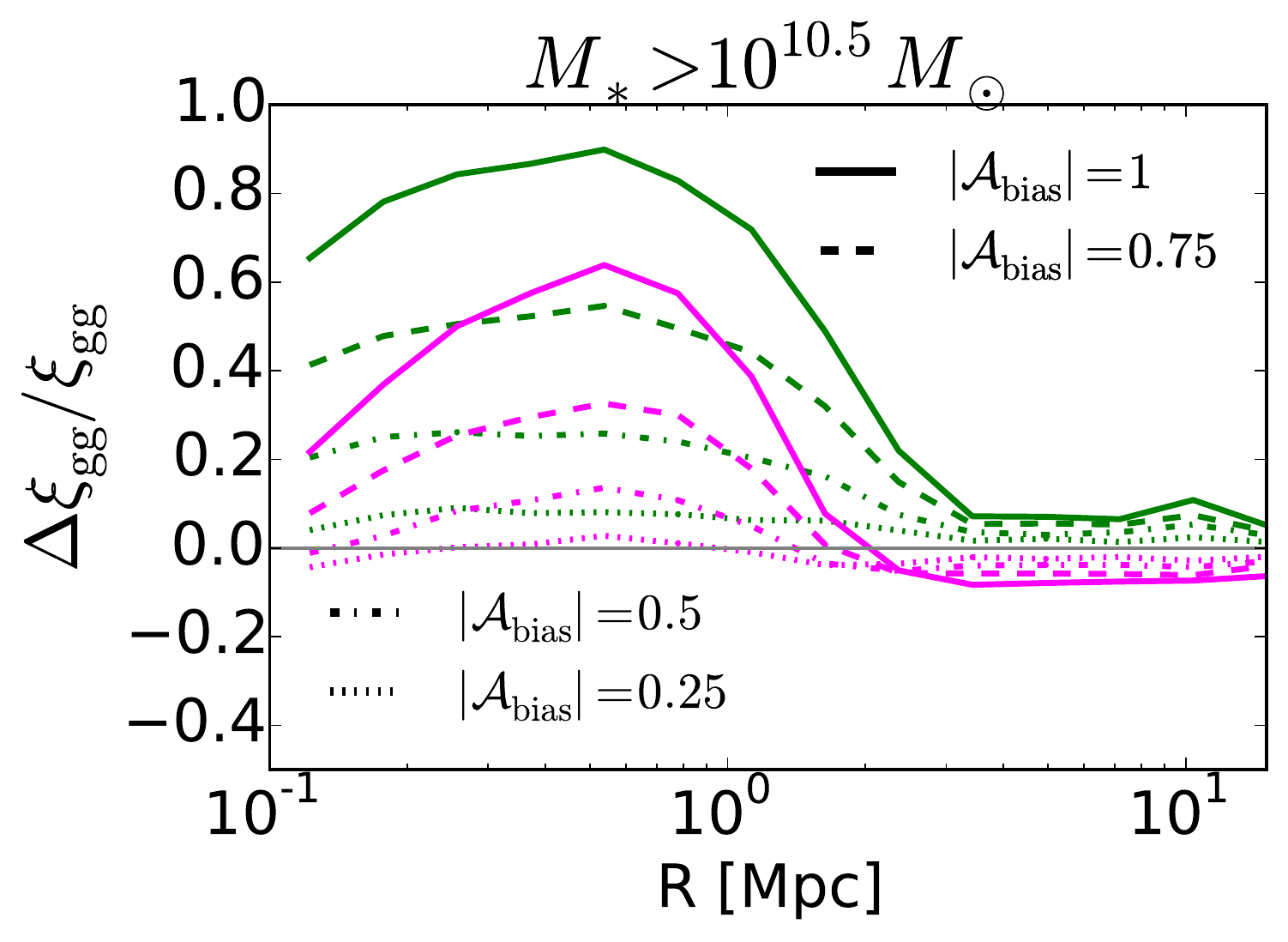}
\includegraphics[width=8.3cm]{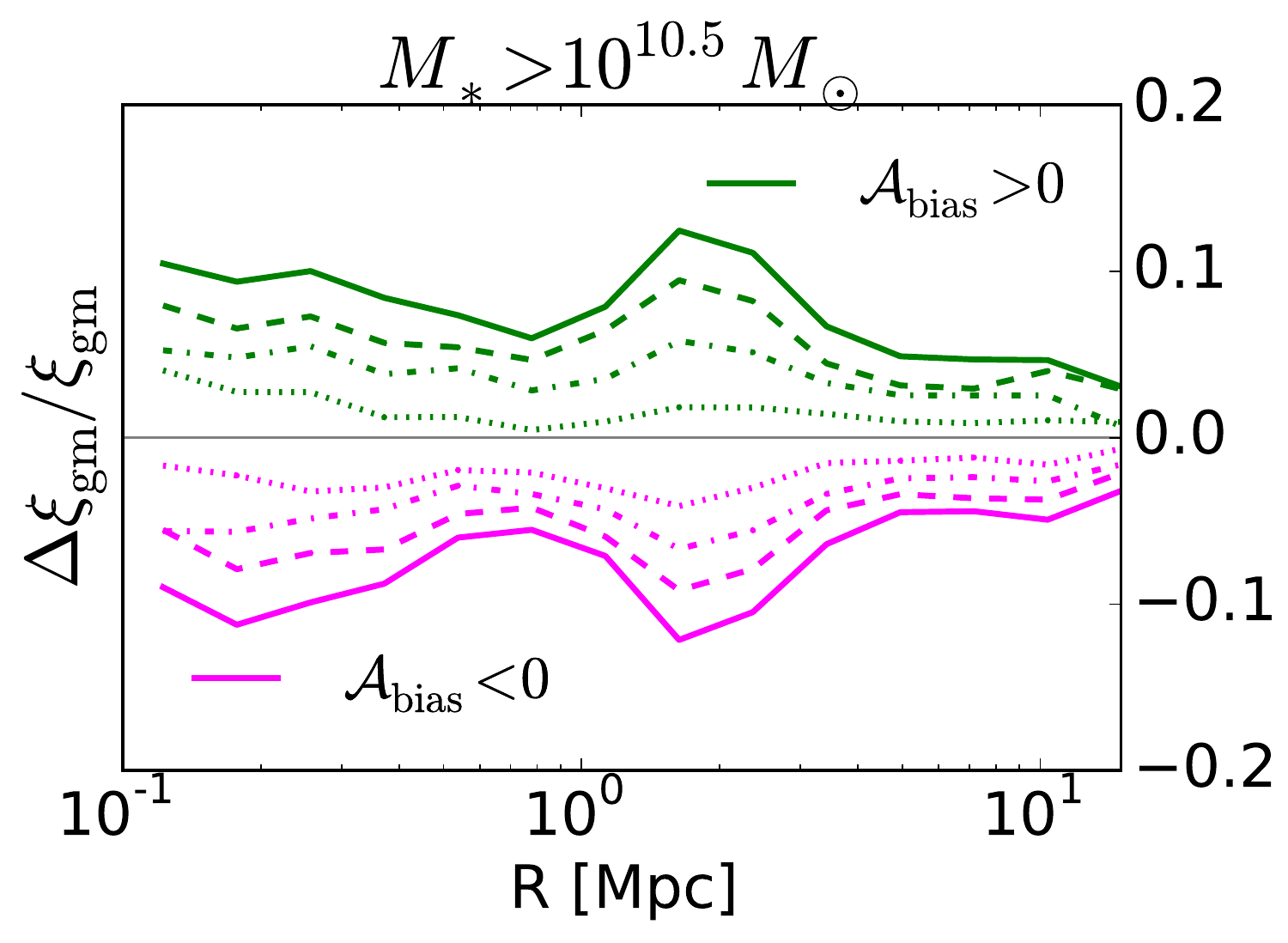}
\caption{ {\bf Varying the strength of assembly bias.} Impact of
  assembly bias on galaxy-galaxy clustering ({\rm left}) and the
  galaxy-mass cross correlation ({\rm right}).  The strength of the
  assembly bias is shown in the legend in the {\em left} panel. Green
  curves show results for $\Abias>0,$ in which case high-concentration
  halo occupations are boosted relative to low-concentration halos of
  the same mass; magenta curves show the opposite case of $\Abias<0.$
  All curves pertain to a baseline, mass-only HOD with stellar mass
  threshold $M_{\ast}>10^{10.5}\msun$ with $\sigma_{\rm logM_{\ast}} =
  0.4,$ as in Figures \ref{fig:occupations} \&
  \ref{fig:satcen}. Notice that the effects of assembly bias, even in
  this simple toy model, are potentially large and
  complex. Furthermore, note that under the assumptions of this model,
  in the pure one-halo regime assembly bias always boosts
  galaxy-galaxy clustering (but {\em not} necessarily the galaxy-mass
  cross correlation).}
\label{fig:clustering}
\end{center}
\end{figure*}

%%%%%%%%%%%%%%%%%%%%%%%%%%%%%%%%%%%%%%%%%%%%%%%%%%%%%%%%%%%%%%%%%%%%

%%%%%%%%%%%%%%%%%%%%%%%%% FIGURE %%%%%%%%%%%%%%%%%%%%%%%%%%%%%

\begin{figure*}
\begin{center}
\includegraphics[width=8.3cm]{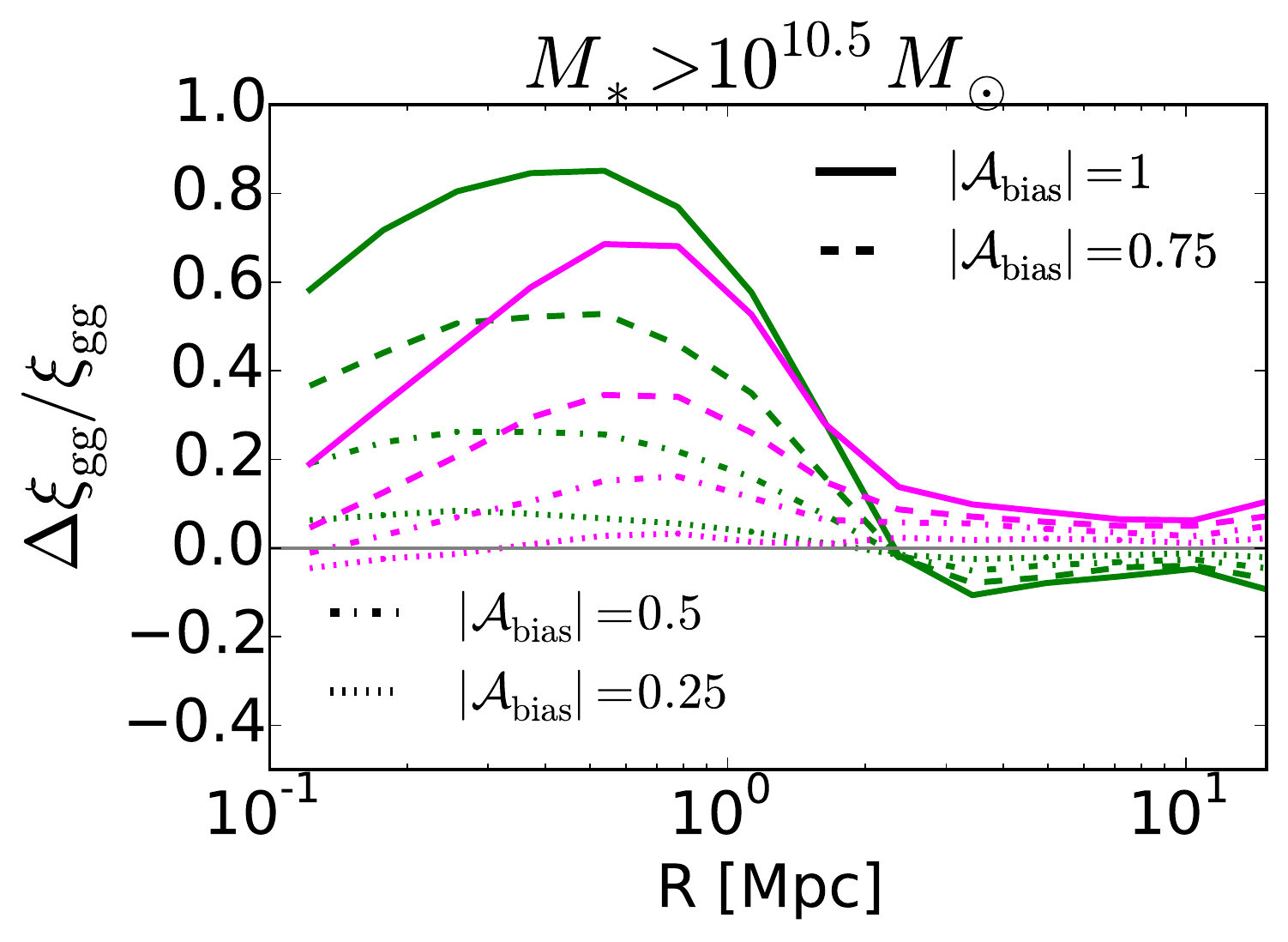}
\includegraphics[width=8.3cm]{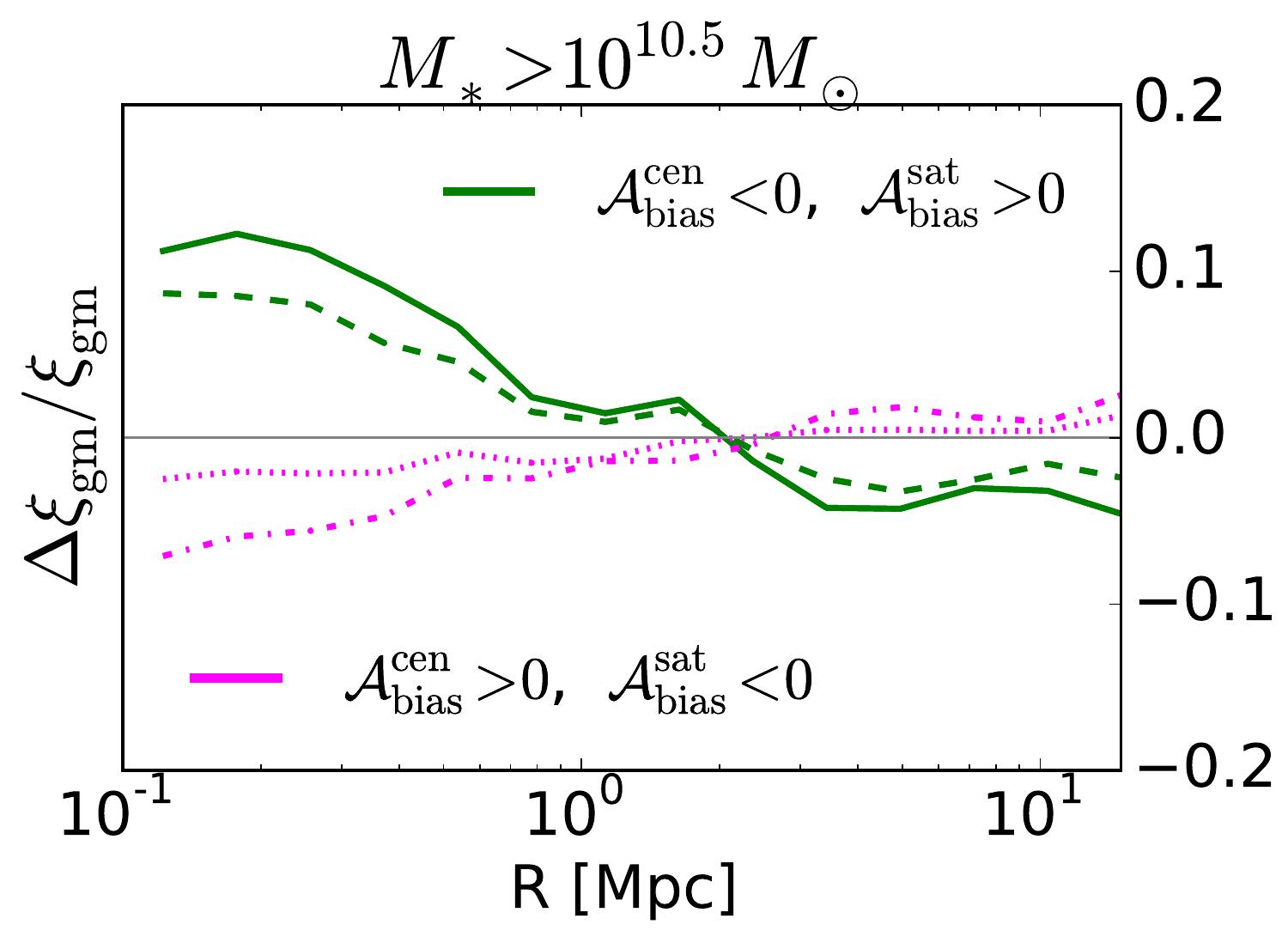}
\caption{ Same as Fig.~\ref{fig:clustering}, but central and satellite
  assembly bias have opposite sign. }
\label{fig:clustering_opposite_sign}
\end{center}
\end{figure*}

%%%%%%%%%%%%%%%%%%%%%%%%%%%%%%%%%%%%%%%%%%%%%%%%%%%%%%%%%%%%%%%%%%%%

Second, satellites preferentially occupy more massive halos than the
vast majority of central galaxies in most samples (see
Fig.~\ref{fig:occupations}).  For our particular example, which
represents galaxies with stellar mass $\mstar > 10^{10.5}\msun$, $\sim
75\%$ of satellites occupy host halos with masses $\mhalo >
10^{13}\msun$. The majority of satellites in this sample occupy hosts
with masses greater than the collapse mass, $\mcoll \approx
10^{12.7}\msun$, for which halo assembly bias is markedly weaker (and
perhaps of opposite sense) than the assembly bias of lower-mass host
halos. Consequently, altering the weighting of halos that are large
enough to host satellite galaxies produces only very modest changes in
large-scale clustering.  As with our discussion of the influence of
assembly bias on small-scale clustering, these considerations should
be fairly typical for the vast majority of HODs explored in the
literature.  However, assembly bias will lead to different effects if
the galaxy sample is selected in such a way that the satellite
fraction is particularly large and/or halos with masses $\mhalo \ll
\mcoll$ host satellites in significant numbers.

%--------------------------------------------------------
\subsubsection{Galaxy-galaxy lensing}
\label{subsubsec:gglensing}

Now consider the right-hand panel of Figure \ref{fig:satcen}. This
panel shows the influence of galaxy assembly bias on the galaxy-matter
cross correlation, the fundamental two-point statistic underlying the
galaxy-galaxy lensing signal. On large scales ($R \gg 1
\mathrm{Mpc}$), the physical picture discussed in
\S~\ref{subsubsec:largescaleclustering} is unchanged: the signal can
be understood entirely in terms of the relative weighting of halo-
halo pairs, and central galaxy assembly bias dominates the effect in
this regime.

On small scales $R \sim 100-400$ $\kpc$, the $\langle \nsat^{2}
\rangle$ boost discussed in \S~\ref{subsubsec:smallscaleclustering} is
not relevant because the galaxy-matter cross correlation is sensitive
to galaxy-matter ``pairs," not galaxy-galaxy pairs. Instead, in this
regime, there are two contributions to the clustering: one
proportional to $\mean{\ncen}{\mvir}$ and another proportional to
$\mean{\nsat}{\mvir}$. For both contributions, the one-halo term of
galaxy-matter clustering is a probe of the halo mass profile, and so
we should naturally expect that preferentially populating halos with
high-concentration profiles should boost $\xi_{\rm gm}$ in the
one-halo regime.

The right-hand panel of Figure~\ref{fig:satcen} shows that this
expectation is born out: the galaxy-matter cross correlation gets a
$\sim10\%$ boost on $R \lesssim 400\kpc$ when $\abias=1$. Perhaps not
surprisingly, we can see that this effect is far more important for
satellite galaxies than for centrals. Referring to
Eq.~(\ref{eq:onehalotermlensing}), we can see that there are two
distinct effects responsible for this difference.  The first effect is
due to the different spatial distributions of centrals and
satellites. Because we model satellite galaxies to trace the
underlying dark matter potential, then by boosting satellite
occupations in high-concentration halos, both $\rho_{\rm m}(r)$ and $n_{\rm sat}(r)$ 
get a boost to the effective concentration parameter, $c_{\rm eff}.$ 
The convolution factor $\Xi_{\rm gm}$ in
the $\mean{\nsat}{\mvir}$ term therefore gets a boost that is quadratic in $c_{\rm eff}.$
Central galaxies, on the other hand, are assumed to sit at the halo
center, and so the $\Xi_{\rm gm}$ factor in the $\mean{\ncen}{\mvir}$
term is only boosted in linear proportion to $c_{\rm eff}.$

The second reason central and satellite assembly bias have distinct
contributions to one-halo lensing has to do with the combinatorics of
assembly bias.  As discussed in
\S~\ref{subsubsec:smallscaleclustering}, the halo mass range over
which assembly bias can influence central galaxy occupations is
restricted by the constraint that $0 < \langle\ncen\rangle <1;$ as a
result, central assembly bias for our fiducial $\mstar>10^{10.5}\msun$
sample ceases to be operative for $\mvir \gtrsim 5 \times 10^{12}
\msun$ (see Fig.~\ref{fig:occupations}). Hence, it is not possible for
central occupation assembly bias in this sample to have any effect at
all on scales $R \gtrsim \rvir(\mvir=5\times10^{12}\msun) \approx
350\kpc$.

Finally, the one-to-two-halo transition region on scales
$R\sim1-3\mpc$ exhibits a marked ``bump" feature, with contributions
from both centrals and satellites.  As first pointed out in
\citet{sunayama_etal15}, this characteristic scale-dependent signature
of assembly bias is associated with so-called ``splashback" material
that is physically bound to the halos of massive groups and clusters
and is congregating at the point of first apocentric passage
\citep{adhikari_etal14, diemer_kravtsov14, more_etal15}.  In fact,
this signature can also be seen in the dashed, yellow curve in the
left-hand panel of Figure~\ref{fig:satcen}, but it is not visible in
either the black or dot-dashed magenta curves because it is swamped by
the $\langle\nsat^{2}\rangle$ boost discussed in
\S\ref{subsubsec:smallscaleclustering}.

%------------------------------------------------------------------
\subsection{The Strength of Assembly Bias}
\label{subsec:strengthrange}

A practical conclusion from the results presented in
Fig.~\ref{fig:satcen} is that galaxy assembly bias can potentially be
a very significant effect compared to the standards of contemporary
and future measurements of galaxy clustering and lensing. Even in this
simple model, the shift in large-scale clustering of galaxies can be
as large as $\sim 10\%$. Moreover, the effect is strongly scale
dependent and assembly bias can cause a change in small-scale
clustering strength of nearly a factor of $\sim 2$. Likewise, the
galaxy-mass cross correlation can plausibly be shifted by $\sim 10\%$
or more by assembly bias and that shift may have non-trivial scale
dependence.

In \S~\ref{subsubsec:gglensing} we restricted attention to the case
where assembly bias in centrals and satellites is of maximum strength
and of the same positive sign.  In this section we study how
variations in both the strength and sign of $\abias$ manifest in
galaxy clustering and lensing.  Various assembly bias strengths
$\Abias$ are depicted in Fig.~\ref{fig:clustering}.  In these
examples, both the central galaxy and satellite galaxies are assigned
the same values of $\Abias$ and in all cases the baseline HOD is, once
more, the model described in \S~\ref{subsec:baselineHOD}.

Perhaps even more intriguing than the dynamic range of assembly bias
effects on large and small scales is the fact that the small-scale
galaxy-galaxy clustering shift is {\em always positive} while the
large-scale shift in the galaxy correlation function induced by
assembly bias can be of either sign.  The large-scale galaxy
clustering can be thought of as a weighted average of the halo bias
(cf., the term in brackets in Eq.~[\ref{eq:twohaloterm}]). Shifting
galaxies away from weakly-clustered host halos toward
strongly-clustered host halos increases the overall galaxy clustering
signal and vice versa. Therefore, assembly bias can induce either an
increment or a decrement on large-scale clustering. Contrarily, on
small scales, as discussed in \S~\ref{subsubsec:smallscaleclustering},
the signal is dominated by satellite-satellite pairs and shifting a
fixed total number of satellite galaxies into fewer host halos {\em
  always} increases $\langle\nsat^{2}\rangle$.

Figure~\ref{fig:clustering_opposite_sign} is analogous to
Figure~\ref{fig:clustering}, except that in
Fig.~\ref{fig:clustering_opposite_sign} the central and satellite
assembly biases are of opposite sign (e.g., $ \Abias^{\rm
  cen}=-\Abias^{\rm sat}$).  Comparing the left and right panels in
Fig.~\ref{fig:clustering} and Fig.~\ref{fig:clustering_opposite_sign},
it is evident that the physical principles discussed in
\S~\ref{subsec:basicfeatures} result in qualitatively different
relative effects on large- and small-scales between $\xigg$ and
$\xigm.$ Assembly bias can induce numerous changes to two-point
statistics and these shifts can be of a wide range of amplitudes and
scale dependences. However, as a consequence of this, simultaneous
measurements of both galaxy-galaxy and galaxy-matter clustering over a
wide range of length scales may, in principle, enable one to determine
both the sign and amplitude of assembly bias.  Further degeneracy
breaking may be possible through the distinct manifestations of the
``bump" feature at $R \sim1-2\mpc $ in galaxy clustering and
galaxy-galaxy lensing.  We defer further discussion of these points to
\S~\ref{sec:discussion}.

%-----------------------------------------------------------------------------------
\subsection{Dependence Upon Baseline HOD}
\label{subsec:dependencebaseline}

%%%%%%%%%%%%%%%%%%%%%%%%% FIGURE %%%%%%%%%%%%%%%%%%%%%%%%%%%%%
\begin{figure*}
\begin{center}
\includegraphics[width=8.3cm]{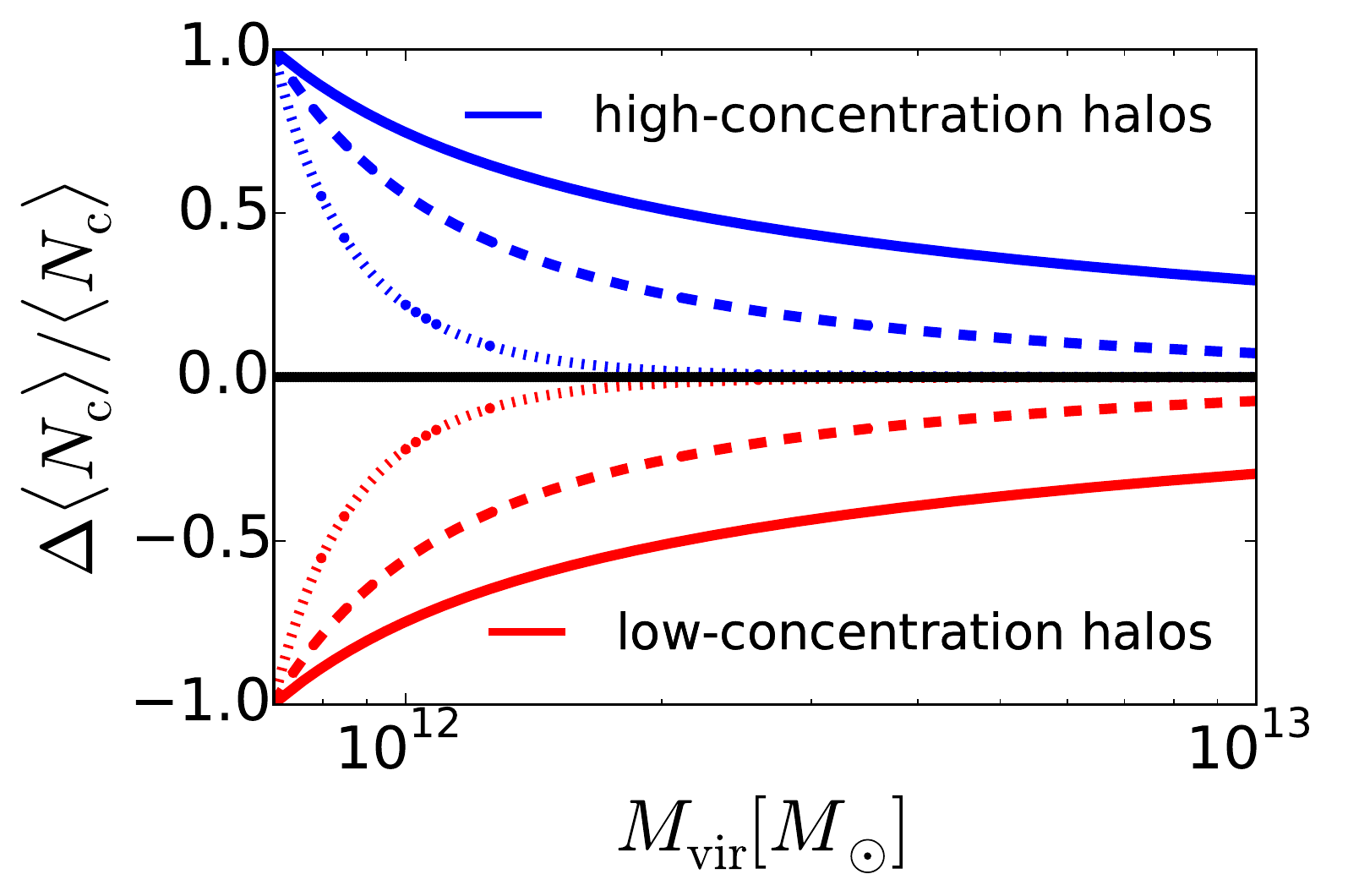}
\includegraphics[width=8.3cm]{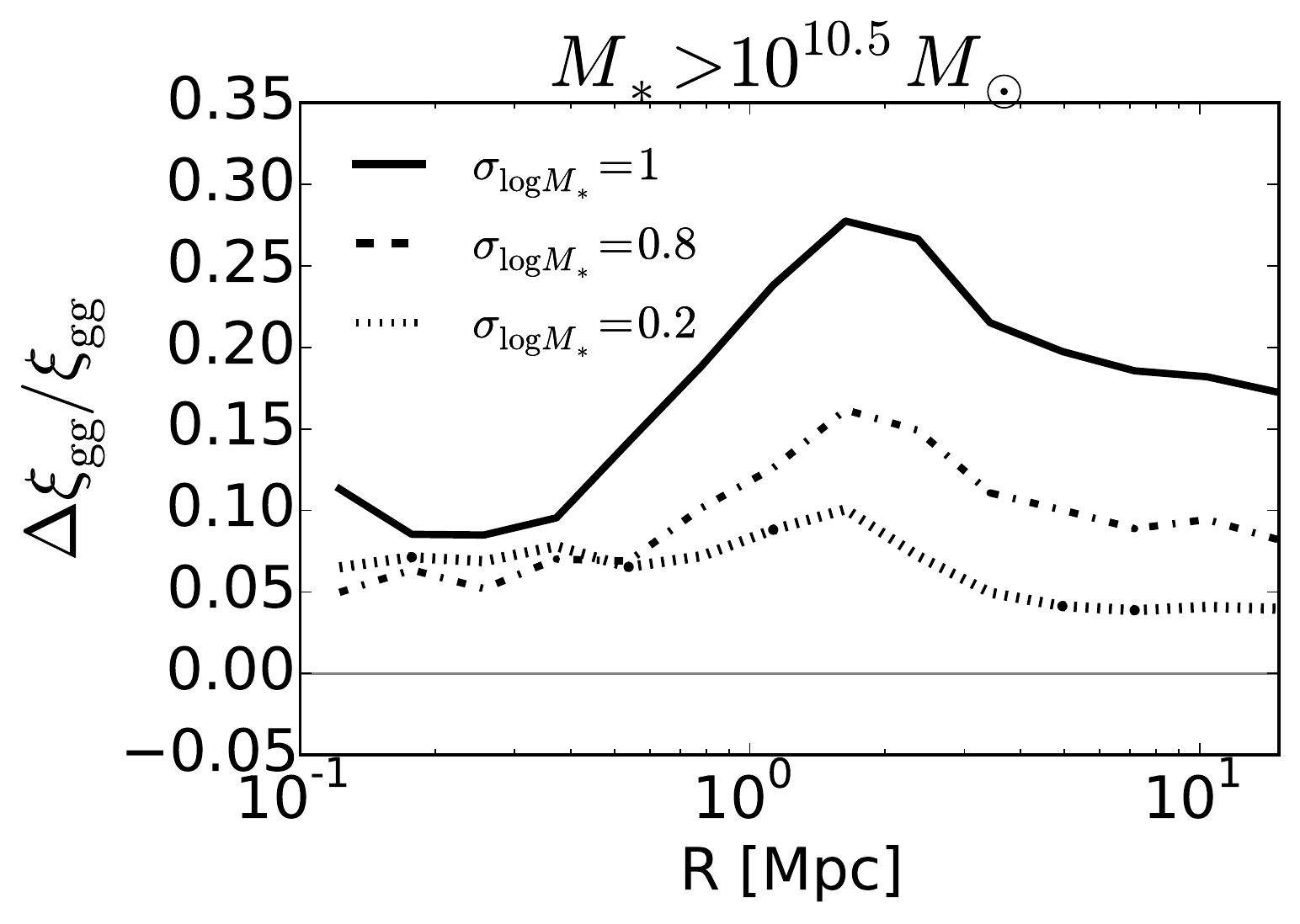}
\caption{ {\bf The role of scatter in the stellar-to-halo mass
    relation}.  The {\em left} panel shows the fractional difference
  in the central galaxy halo occupation in our simple model of galaxy
  assembly bias for various values of $\sigmalogm$.  Larger values of
  the scatter increase the dynamic range over which $0 <
  \mean{\ncen}{\mhalo} < 1,$ creating a longer baseline in halo mass
  for assembly bias to be operative.  In these figures, the assembly
  bias strength is set as $\Abias^{\rm cen}=1$ and $\Abias^{\rm
    sat}=0.2.$ The {\em right} panel exhibits the deviations in galaxy
  clustering at various values of $\sigmalogm$.  }
\label{fig:hod_dependence}
\end{center}
\end{figure*}
%%%%%%%%%%%%%%%%%%%%%%%%%%%%%%%%%%%%%%%%%%%%%%%%%%%%%%%%%%%%%%%%%%%%

The degree to which assembly bias alters galaxy clustering statistics
can be quite sensitive to the underlying baseline, mass-only, HOD of
the galaxy population under consideration. In particular, the impact
of assembly bias on $\xigg$ is quite sensitive to the steepness of the
transition from $\mean{\ncen}{\mhalo}_{\rm std}=0$ at low host masses
to $\mean{\ncen}{\mhalo}_{\rm std}=1$ at high host masses. This
steepness is controlled by the level of stochasticity in the central
galaxy stellar mass at fixed halo mass, parameterized in our baseline
model by $\sigmalogm$ in Eq.~(\ref{eq:ncendef}).

This sensitivity to the underlying mass-only HOD can be understood
quite simply. First, in typical samples, central galaxies constitute
$\sim 70\% - 90\%$ of the sample, so the behavior of centrals is of
primary importance\footnote{This is not necessarily true and could be
  violated if the galaxy selection function is tailored to favor
  satellites.}.  If $\mean{\ncen}{\mhalo}_{\rm std}=0$, then there are
no galaxies at all to be apportioned to specific halos according to a
secondary property other than halo mass.  If
$\mean{\ncen}{\mhalo}_{\rm std} = 1$, then all halos of that mass
contain a central galaxy and again, there is only one way in which
these galaxies can be apportioned among the halos of fixed mass. The
flexibility to apportion central galaxies to halos according to a
secondary property other than mass is available only when
$\mean{\ncen}{\mhalo}_{\rm std}>0$ {\em and}
$\mean{\ncen}{\mhalo}_{\rm std}<1$.  The greater the range of host
halo masses over which this condition is met, the larger the fraction
of the sample that is subject to assembly bias. Therefore, increasing
$\sigmalogm$ increases the potential importance of assembly bias,
particularly on large scales.

The importance of the baseline mass-only HOD, in particular the
parameter $\sigmalogm$, is shown in
Figure~\ref{fig:hod_dependence}. In Fig.~\ref{fig:hod_dependence}, and
in all subsequent figures, we choose to represent assembly bias with
$\Abias^{\rm cen}=1$ and $\Abias^{\rm sat}=0.2$ so as to depress the
strong small-scale influence of satellite assembly bias relative to
the more mild large-scale influence of central galaxy assembly
bias. These parameters designate our ``fiducial" model of assembly
bias for this and forthcoming comparisons.

Notice in Fig.~\ref{fig:hod_dependence} that the large-scale
clustering of the sample can be enhanced by more than $15-20\%$ for
large values of $\sigmalogm$, whereas assembly bias effects are
relatively small ($\sim 5\%$ or less) for $\sigmalogm \lesssim 0.2$.
This fact is especially interesting in light of the findings presented
in \citet{zentner_etal14}: in HOD analyses of galaxy samples in which
significant levels of assembly bias are erroneously ignored, one may
infer erroneously small values of the $\sigmalogm$ parameter
($\sigmalogm \lesssim 0.2$) when the true underlying values are
actually large (e.g., $\sigmalogm \gtrsim 0.8$). Therefore, performing
a standard HOD analysis on a sample and concluding that $\sigmalogm <
0.2$ is not sufficient to render assembly bias effects to be on the
order of a few percent or less \citep[see][for detailed discussion of this
  point]{zentner_etal14}.

%-----------------------------------------------------------------------------------
\subsection{Dependence Upon Secondary Halo Property}
\label{subsec:secondary_property}

In the previous subsections, we illustrated assembly bias using halo
concentration, $c$, at fixed halo mass as the secondary halo property
used to condition the HOD; however, halo clustering is now known to
depend upon a number of halo properties and it is not clear which, if
any, secondary halo property should be most closely related to galaxy
properties. Numerous quantities may be sensible to explore in this
context and in this section we demonstrate the induced galaxy assembly
bias upon conditioning the HOD on a small subset of these host halo
properties.

Measures of halo formation time are sensible halo properties on which
to condition the HOD, particularly because it is not unreasonable to
suspect that the formation history of a halo may be related in some
way to the formation histories of the galaxies that the halo contains.
We examine two properties associated with the formation histories of
halos. First, we explore assembly bias induced by conditioning the HOD
on the half-mass formation time of the host halo, $a_{1/2}$. As this
name suggests, $a_{1/2}$ is the scale-factor at which the mass of the
host halo's main progrenitor first exceeded half the mass of the host
halo at $z=0$. Second, we condition HODs by the accretion rate of the
halos, $\dd \mhalo/\dd t$, defined as the change in virial mass of the
main progenitor halo over the past dynamical time $\tau_{\rm dyn}
\approx 2$ Gyr.

\begin{figure*}
\begin{center}
\includegraphics[width=8.3cm]{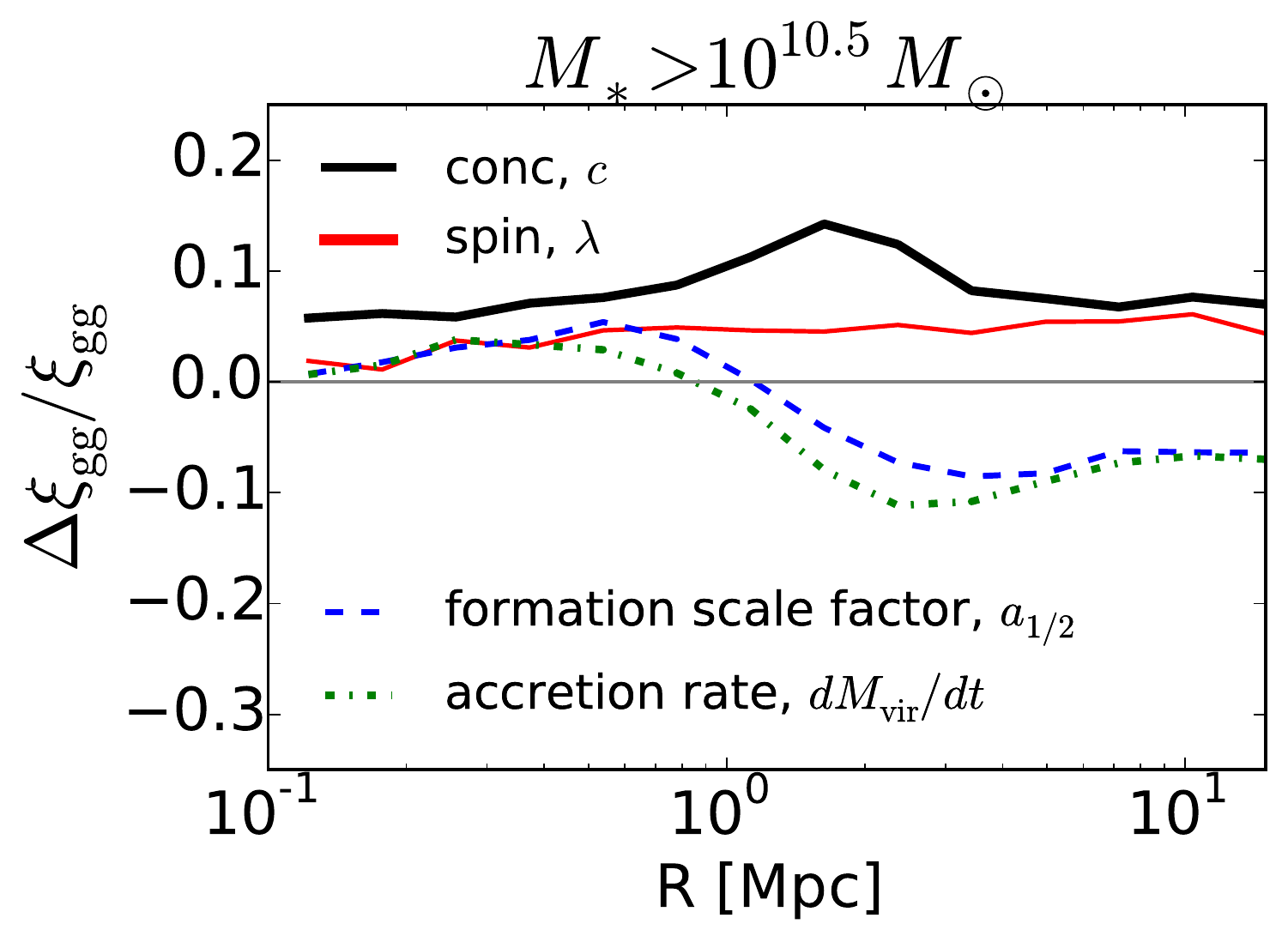}
\includegraphics[width=8.3cm]{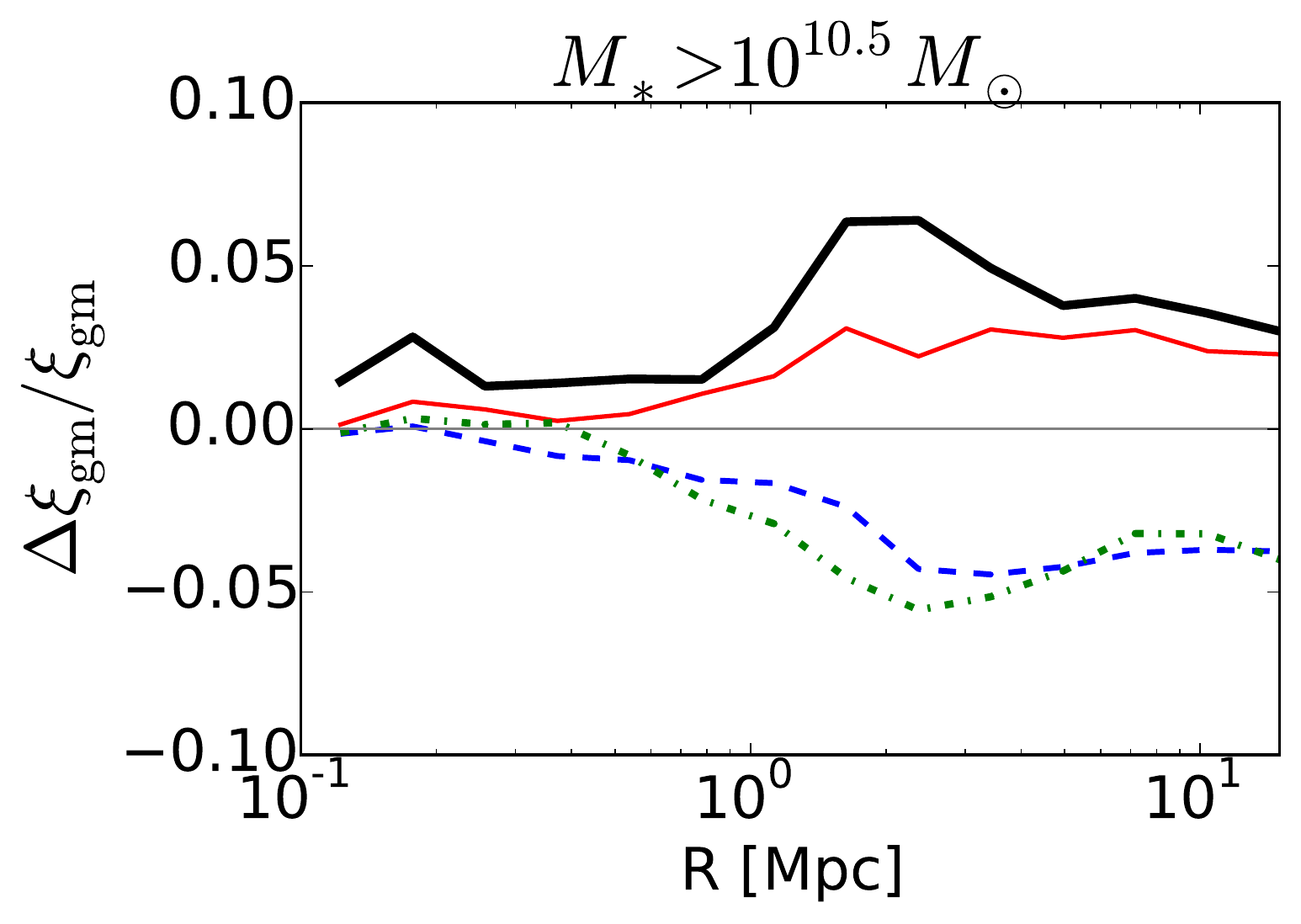}
\caption{ {\bf The role of the secondary halo property driving
    assembly bias. }  As in Fig.~\ref{fig:hod_dependence}, we set the
  assembly bias strength parameters as $\Abias^{\rm cen}=1$ and
  $\Abias^{\rm sat}=0.2;$ here our baseline HOD is for a
  $\mstar>10^{10.5}\msun$ threshold sample with $\sigmalogm=0.4.$ Our
  sign convention is such that when $\abias>0,$ assembly bias boosts
  the occupation of halos with above average values of the secondary
  halo property.  We choose different secondary halo properties to
  modulate galaxy occupation statistics as indicated in the legend.  }
\label{fig:sec_haloprop}
\end{center}
\end{figure*}

The assembly bias induced by these measures of halo formation history
are shown as the green, dotted (for $a_{1/2}$) and magenta dot-dashed
(for $\dd \mhalo/\dd t$) curves in Fig.~\ref{fig:sec_haloprop}. In
both cases, assembly bias reduces the large-scale clustering strength
of galaxies. We remind the reader of our sign convention: when
$\abias>0,$ assembly bias boosts the occupation of halos with above
average values of the secondary halo property at fixed mass. Formation
time is anti-correlated with concentration (higher concentrations lead
to earlier formation times and thus smaller values of either $a_{1/2}$
or $\dd \mhalo/\dd t$), so this result is qualitatively in accord with
naive expectations.

%%%%%%%%%%%%%%%%%%%%%%%%% FIGURE %%%%%%%%%%%%%%%%%%%%%%%%%%%%%

%%%%%%%%%%%%%%%%%%%%%%%%%%%%%%%%%%%%%%%%%%%%%%%%%%%%%%%%%%%%%%%%%%%%

Notice that on small scales ($r \lesssim 1~\mathrm{Mpc}$), assembly
bias induced by $a_{1/2}$ and $\dd \mhalo/\dd t$ leads to an
enhancement in galaxy clustering rather than a decrement.  This is
primarily driven by the $\langle\nsat^2\rangle$ boost effect described
in \S~\ref{subsec:basicfeatures}: assembly bias of this kind always
packs satellite galaxies into fewer host halos, enhancing small-scale
clustering.
 
However, small-scale clustering is also sensitive to a secondary
effect related to the spatial profiles.  When the sign of $\abias$ is
positive, more galaxies are packed into halos with above-average
values for the secondary property. For our fiducial case where the
secondary property is halo concentration, $x=c,$ this results in an
{\em additional} boost to the two-point function because the effective
spatial distribution of satellites and dark matter is boosted (this
manifests itself in the $\Xi(r)$ term in Eqs.~[\ref{eq:onehaloterm}]
and~[\ref{eq:onehalotermlensing}]).  Because $a_{1/2}$ and $\dd
\mhalo/\dd t$ are anti-correlated with $c,$ this spatial profile
effect partially counteracts the $\langle\nsat^2\rangle$ boost
described above, and so the weaker one-halo enhancement seen in these
models relative to our fiducial $c-$based model is expected on these
grounds.  Of course, the relations between halo formation time
measures and concentration themselves exhibit significant scatter, so
other factors may also contribute to this distinction.

As a final example, we demonstrate assembly bias induced by
conditioning the HOD on the angular momentum of the host halo. To be
specific, we condition the HOD on the halo spin parameter,
\begin{equation}
\lambda = \frac{J\, \sqrt{\vert E\vert}}{G\, \mhalo^{5/2}}, 
\end{equation}
where $J$ is the total angular momentum of the halo and $E$ is the
total energy of the halo in a system defined such that the potential
energy at infinity is zero (and therefore $E$ is negative because
halos are bound objects). The spin parameter is the ratio of the
rotational kinetic energy of the system to the total kinetic energy of
the halo and serves as a measure of the importance of rotation, as
opposed to random motion, in supporting the system against
gravitational collapse. The red, thin, solid lines in
Fig.~\ref{fig:sec_haloprop} show the galaxy assembly bias imposed by
conditioning the HOD on $\lambda$ in our simple two-population
model.

As can easily be seen, although halo spin is correlated with
halo formation and halo concentration, the large-scale galaxy assembly
bias is considerably milder upon conditioning on $\lambda$ than it is
upon conditioning on $c$, $a_{1/2}$, or $\dd \mhalo/\dd
t$. Consequently, the scale-dependence of the assembly bias induced by
selecting on halo spin differs from the previous cases and is markedly
weaker.

%-----------------------------------------------------------------------------------
\subsection{Mass and Redshift Dependence of Assembly Bias}
\label{subsec:mass_redshift}

In the previous sections, we explored assembly bias for a specific HOD
model designed to approximate a sample of galaxies selected on a
stellar mass threshold of $M_* > 10^{10.5}\, \Msun$ at $z \approx
0$. The strength of the assembly bias of halos varies with both halo
mass and redshift and so it is instructive to examine examples of
various mass thresholds at various epochs.

\subsubsection{Stellar Mass Dependence}
\label{subsubsec:smdep}

In the left panel of Figure~\ref{fig:smzdep} we use our fiducial model
to study how the impact of assembly bias on galaxy clustering depends
on the stellar mass threshold of the sample.  Assembly bias of dark
matter halos generically weakens with increasing halo mass
\citep{wechsler06}, and in our HOD model $\mstar$ increases
monotonically with $\mhalo.$ Thus at fixed assembly bias parameter
strength $\abias,$ we should expect the impact of assembly bias on
$\xigg$ on large scales to become weaker in galaxy samples with higher
stellar mass thresholds.

While the red curve in the left panel of Figure~\ref{fig:smzdep} shows
that this is true for the $\mstar>10^{11.25}\Msun$ sample, the
clustering of the $\mstar>10^{10.75}\Msun$ sample appears to be
slightly {\em more} sensitive to assembly bias relative to the
$\mstar>10^{10.25}\Msun$ sample. We attribute this surprising result
to insufficient resolution of the Bolshoi simulation.  As shown
in Figure 2 of \citet{sunayama_etal15}, Bolshoi exhibits non-monotonic
behavior in the variation of halo assembly bias strength with halo
mass for $\mvir\lesssim10^{11.7}\msun.$ Naively, this may be
surprising since this corresponds to a halo with more than $3000$
particles, vastly exceeding the industry convention of deeming halos
with several hundred particles to be adequately resolved. However, we
remind the reader that for models in which assembly bias is governed
by concentration (or in the case of \citet{sunayama_etal15}, $\vmax$),
one must resolve the halo's internal structure, not just the halo
itself. For halos with $\mvir\approx10^{11.7}\msun,$ the mean
concentration is roughly ten. For such a concentration, the mass
inside the scale radius is over ten times less than the total halo
mass. Resolving the scale radius of such halos should therefore result
in a tenfold increase in the simulation resolution requirements. For
the $\mstar>10^{10.25}\Msun$ sample, $\sim10\%$ of the centrals in the
sample reside in halos with $\mvir<10^{11.7}\msun,$ whereas for the
$\mstar>10^{10.75}\Msun$ sample this fraction is less than $0.1\%.$ We
consider these estimates to be highly suggestive that Bolshoi suffers
from resolution effects for this particular science target, though we
leave a proper numerical resolution study as a subject for future
investigation.

The $\mstar-$dependence of small-scale clustering
displays complex behavior that can again be understood using the
physical considerations of \S\ref{subsec:basicfeatures}. The
small-scale assembly bias of the $\mstar>10^{11.25}\Msun$ sample
begins to overtake the effect seen in the other two samples on scales
$R\lesssim400\kpc,$ which corresponds to the virial radius of a halo of mass
$\mvir\approx5\times10^{12}\msun.$ For the two lower mass thresholds,
$\langle\ncen\rangle=1$ at this mass, whereas for the
$\mstar>10^{11.25}\Msun$ sample, $\langle\ncen\rangle\approx0.25.$
Thus in the highest threshold sample, $\xigg(R\approx400\kpc)$
actually receives a significant contribution from central-central and
central-satellite pairs in halos where assembly bias is significant,
whereas the statistics of HOD conservation prohibits the lower
$\mstar-$threshold samples from receiving such a contribution. This 
boosts the small-scale clustering effect of
the $\mstar>10^{11.25}\Msun$ sample relative to the other two.

%------- redshift
\subsubsection{Redshift Dependence}
\label{subsubsec:zdep}

\begin{figure*}
\begin{center}
\includegraphics[width=8.3cm]{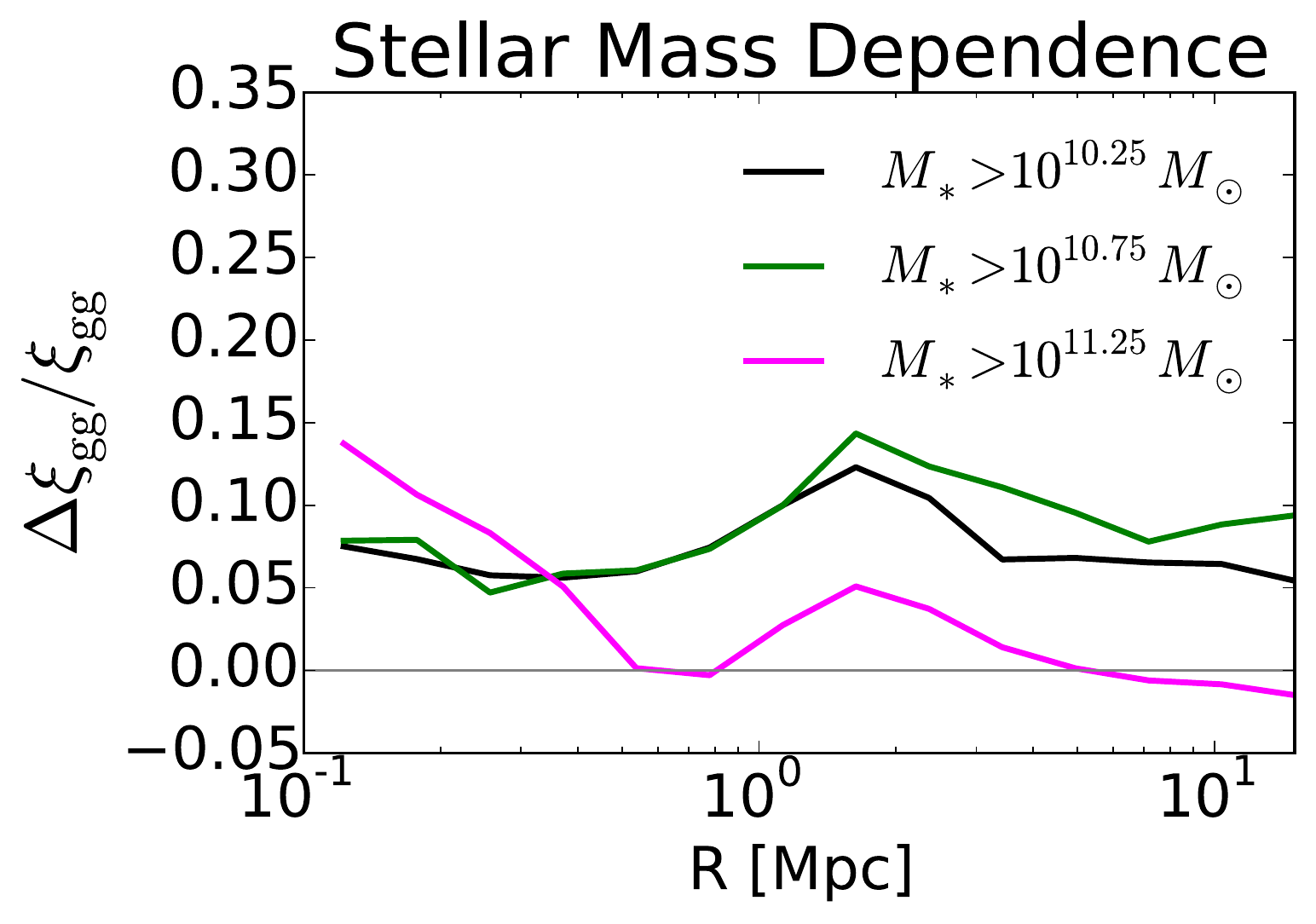}
\includegraphics[width=8.3cm]{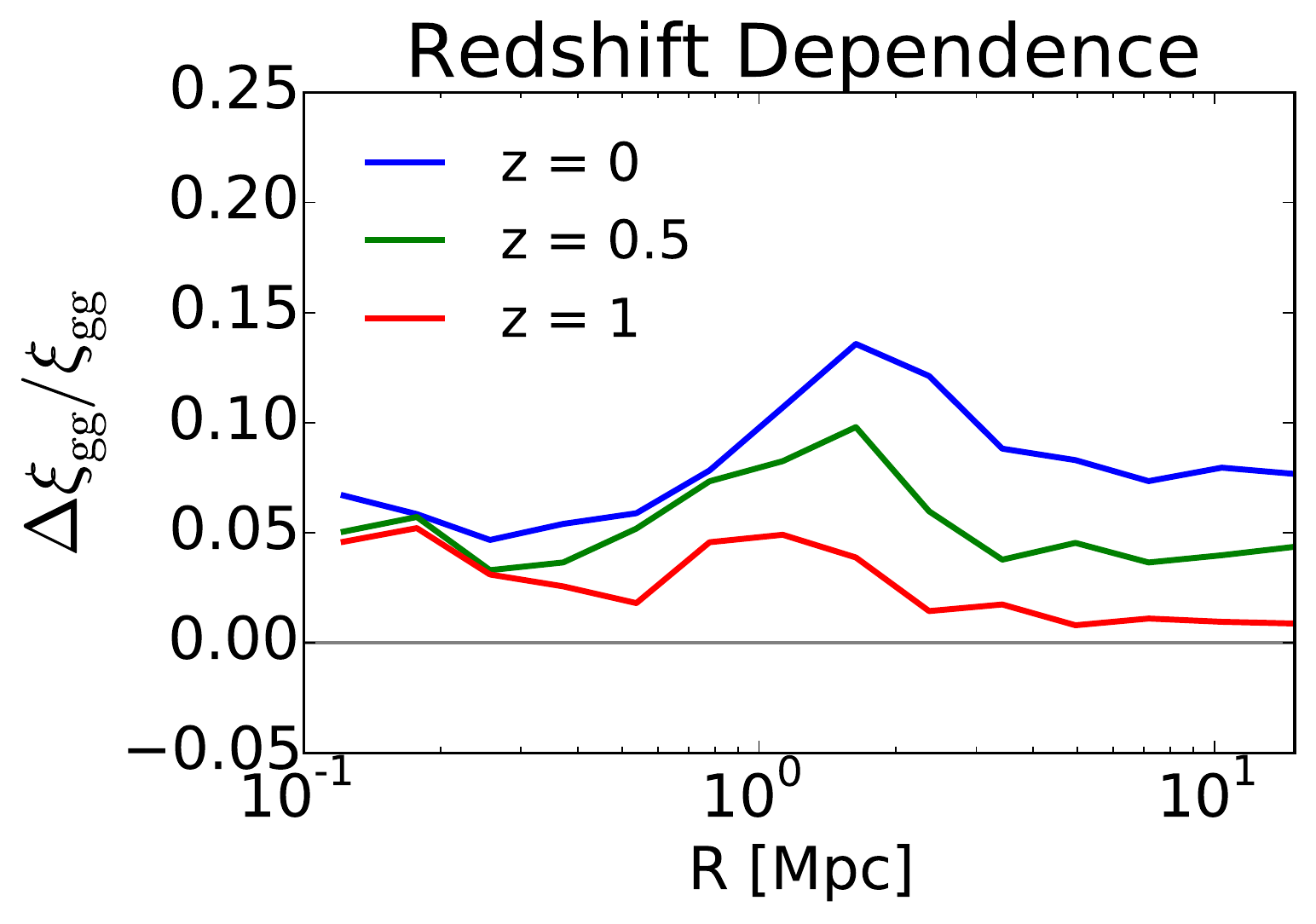}
\caption{ {\bf $\mstar-$ and $z-$dependence of assembly bias.} All
  curves in both panels pertain to our fiducial assembly bias model,
  in which $\abias^{\rm cen} = 1,$ $\abias^{\rm sat} = 0.2,$ the
  secondary halo property is concentration and $\mstarscatter=0.4.$ In
  the {\em left panel}, we show the fractional effect on $\xigg$ for
  the three different $\mstar-$thresholds indicated in the legend; in
  our baseline HOD model, the $\mstar$ threshold-dependence derives
  from the $\mstar-\mhalo$ relation (see Eq.~[\ref{eq:ncendef}]).
  Over this stellar mass range, the impact of assembly bias generally
  weakens as the $\mstar$ threshold increases; the apparent
  non-monotonic behavior seen by comparing the green and black curves
  in the left panel is a numerical resolution artifact, as discussed
  in the text.  In the {\em right panel} we show how a fixed level of
  assembly bias has an impact on galaxy clustering with a strength
  that varies with redshift. In our model $z-$dependence derives from
  the evolution of the stellar-to-halo-mass relation (see
  Eq.~[\ref{eq:zdepsmhmparams}]).  For $L^{\ast}-$type galaxy samples
  defined by a fixed stellar mass (or luminosity) threshold, the
  impact of assembly bias on galaxy clustering becomes less important
  at higher redshift, a generic result.  }
\label{fig:smzdep}
\end{center}
\end{figure*}

The right panel of Figure~\ref{fig:smzdep} shows that at fixed
strength $\abias,$ the impact of assembly bias on galaxy clustering
weakens for galaxy samples at higher redshift. This effect is
straightforward to understand. Over the range $z \lesssim 1,$ there is
very little evolution in the stellar-to-halo-mass relation
\citep[e.g.,][]{behroozi13}. For example, in the \citet{leauthaud11a}
HOD model, the mean halo mass for a central galaxy of
$\mstar\approx10^{10.5}\msun$ is $\mvir\approx10^{11.9}\msun$ at $z=0$
and $\mvir\approx10^{12.1}\msun$ at $z=1.$ However, it is not the {\em
  absolute} mass $\mvir$ that sets the strength of halo assembly bias,
but rather the mass relative to the collapse mass $\mcoll.$ At $z=0,$
the collapse mass $\mcoll\approx10^{12.7}\msun,$ but by redshift
$z=1$, the collapse mass has declined to $\sim3 \times 10^{11}\,
\Msun.$ So by comparing $\xigg$ for samples at different redshift at
fixed $\mstar,$ we are effectively comparing $\xigg$ of halos with
quite different values of $\mvir/\mcoll$ \citep[see][for an
  alternative discussion of this point in the context of galactic
  conformity]{hearin15}. This effect is purely monotonic, and indeed
we can see the impact of assembly bias is considerably weaker at $z=1$
in $\mstar>10^{10.5}\Msun$ samples relative to the present day.

%%%%%%%%%%%%%%%%%%%%%%%%% FIGURE %%%%%%%%%%%%%%%%%%%%%%%%%%%%%

%%%%%%%%%%%%%%%%%%%%%%%%%%%%%%%%%%%%%%%%%%%%%%%%%%%%%%%%%%%%%%%%%%%%

%----------------------------------------------------------------------------------------
\section{Discussion}
\label{sec:discussion}

In this paper, we have described a new analytical formalism, the
decorated HOD, for encoding the effects of assembly bias on the
galaxy-halo connection.  For the sake of definiteness, we have chosen
to focus on a simple toy model implementation of the decorated HOD in
which halos of a given mass are divided into two sub-populations based
on some secondary halo property. We have used this simple
implementation to enumerate the primary set of effects that assembly
bias can have on two-point clustering of stellar-mass threshold galaxy
samples. In \S\ref{subsec:hodgenerality}, we describe how the scope of
the decorated HOD framework is far broader than the toy model
implementation in this work, and in \S\ref{subsec:previouswork} we
compare the decorated HOD to previous attempts to encode assembly bias
effects into galaxy-halo modeling. We conclude in
\S\ref{subsec:cosmology} with a discussion of the relevance of this
work to the precision-cosmology program.

\subsection{Generality of the Decorated HOD Framework}
\label{subsec:hodgenerality}

In \S\ref{sub:basic}, we wrote the mean number of galaxies in a halo
of mass $\mhalo$ and some secondary property $x$ as
$$\mean{\ngal}{\mhalo; x}\equiv \mean{\ngal}{\mhalo} + \delta
\ngal(\mhalo; x).$$ For any given ``mass-only" HOD model describing
$\mean{\ngal}{\mhalo},$ we can construct new ``decorated" versions of
such a model. We further suggested that in order to isolate the
effects of assembly bias, it may be useful to assume that the
decorated HOD satisfies ``HOD conservation," which we defined as
$P_{\rm dec}(\ngal \vert \mhalo) = P_{\rm std}(\ngal \vert \mhalo)$,
i.e., the marginalized probability function of the decorated HOD
reduces to that of the baseline model. In practice, the HOD is used most 
often to model two-point statistics for which only the first two moments of 
the HODs are needed. As we describe below, the
concepts of HOD decoration and conservation have many potential applications.

\subsubsection{Decoration of arbitrary galaxy properties}

The decorated HOD framework is not limited to applications concerning
the prediction of $\avg{\ngal}.$ As stated above, the notion of ``HOD
conservation" implies a mathematical condition on the decoration
function $\delta\ngal(\mhalo; x)$ such that the value of
$\mean{\ngal}{\mhalo}$ of the baseline model is left unchanged by the
decoration.\footnote{For brevity we focus exclusively on the
  first-order moments here, but everything extends trivially to higher
  orders as well.}  {\em This concept applies to any one-point
  function in the galaxy-halo connection}, $\mean{f}{\mhalo},$ not
just $\mean{\ngal}{\mhalo}.$

For example, if $\mean{f}{\mhalo} = F^{\rm sat}_{\rm red}(\mhalo)$ is
the red fraction of satellites as a function of halo mass, then the
decorated HOD can be used to introduce assembly bias in the colors of
satellites. In this case, we ``decorate the red fraction," instead of
$\avg{\ngal}$, and our first-moment conservation equation 
becomes $$ 0 = \int\, \delta F^{\rm sat}_{\rm red}(\mhalo, x)\,
P(x\vert\mhalo) \dd x. $$ By modeling the clustering and lensing of
red galaxies as function of luminosity or stellar mass, such an
approach can provide valuable insight as to which halo properties are
most strongly correlated with satellite quenching.

We list a few further examples below to demonstrate the power of this
generalization:
\ben

\item Let $x=t_{\rm lmm}$ be the time since the halo's last major
  merger merger and let $\mean{f}{\mhalo}$ express the average
  morphology of central galaxies (for example, let $f$ correspond to
  the Sersic index $n$ that best fits the galaxy's surface brightness
  profile). By fitting measurements of $n-$dependent clustering with
  an HOD decorated according to $t_{\rm lmm}$, one can statistically
  test the hypothesis that galaxy morphology is physically connected
  to major mergers.
  
\item Let $x=\Delta\mvir/\Delta\tau$ be halo mass accretion rate,
  defined over some timescale $\tau,$ and let $\mean{f}{\mhalo}$ be some
  baseline model for the duty cycles of quasars. By fitting such a
  decorated HOD to the two-point functions of quasar samples and
  finding the time $\tau$ over which the correlation is strongest, it
  becomes possible to statistically quantify the timescale over which
  the cosmic supply of fresh gas impacts the quasar duty cycle.
  
\item Let $x=\ncen(\mhalo)$ be the number of central galaxies in a
  halo (i.e., zero or one), and let $\mean{f}{\mhalo} =
  \mean{\nsat}{\mhalo}.$ This is the variation on the decorated HOD
  alluded to in \S\ref{subsec:censat} that allows one to explore
  intermediate cases between the two extreme assumptions for how
  $\mean{\ncen\nsat}{\mvir}$ is computed in ordinary HOD models.
  
\een

We conclude this section by noting that the {\tt Halotools} code base
already supports all of the above generality. As described in
the code documentation {\tt http://halotools.readthedocs.org}, the 
{\tt HeavisideAssembias} orthogonal mix-in class can be used to decorate
{\em any} one-point function with the step-function style assembly
bias employed in the present paper. There is also freedom to explore
$\abias=\abias(\mhalo),$ as well as $\mhalo-$dependence in how the
halos are split into sub-populations, i.e., $P_1= P_1(\mhalo)$. This level of generality is
made possible through the use of a python decorator, from which the
decorated HOD derives its name. We will explore many of these
interesting avenues for extending the HOD approach in forthcoming
papers.

%---------------------------------------------------------------------------------------------------------
\subsection{Previous Formulations}
\label{subsec:previouswork}

\subsubsection{Early HOD-style models}
\label{subsubsec:earlyhods}

The general approach taken in our work is most closely related to
\citet{wechsler06}.  In that paper, the authors considered
generalizing the halo model to predict dark matter and/or galaxy
clustering given that the clustering of halos is both mass- and
density-profile dependent. Our mathematical formulation of the
decorated HOD builds naturally upon this early work by introducing the
concept of ``HOD conservation" into the framework outlined in Section
4.4 of \citet{wechsler06}.

Another early formulation of assembly-biased HOD-style models appears
in \citet{tinker08b}.  There the authors introduced a dependence of
the first occupation moment of central galaxies $\avg{\ncen}$ on
$\delta_5,$ the number density of dark matter particles smoothed with a 
spherical tophat window of radius $5\, \mpc.$ In this model,
$\mean{\ncen}{\mhalo;\delta_5}\neq\mean{\ncen}{\mhalo},$ and the level
at which the equality is violated is controlled by additional
parameters. The baseline HOD model explored in \citet{tinker08b}
included a parameter $M_{\rm min},$ which can intuitively-but-roughly
be thought of as the minimum mass required for a halo to host a
central galaxy above a given brightness. In \citet{tinker08b}, the
authors implemented assembly bias by manually changing the $M_{\rm
  min}$ parameter in high- and low-density regions such that the
overall number density of galaxies is held fixed.
 
The model presented in \citet{marin_etal11} has much in common with
the \citet{tinker08b} model. In \citet{marin_etal11}, the authors
partition the dark matter density field into three disjoint
categories, {\em voids, filaments} and {\em nodes}, and allow the
values of all the HOD parameters to separately vary in each of these
regions.

These early formulations of assembly bias are conceptually quite
different from the decorated HOD. For example, altering the value of
$M_{\rm min}$ as a function of large-scale dark matter density
introduces an explicit covariance between the HOD parameter $M_{\rm
  min}$ and the additional parameters encoding the assembly bias. The
advantage of our formulation is that one is free to change the
assembly bias parameters $\abias$ while fixing {\em both} the number
density $\bar{n}_{\rm g}$ {\em and} all parameters of the baseline
model. Decorated HODs have the distinct advantage that the new
assembly bias parameters are orthogonal to those that
describe the standard baseline model.

\subsubsection{Abundance matching-style models}
\label{subsubsec:shammodels}

Subhalo abundance matching
\citep[SHAM;][]{kravtsov04a,vale_ostriker04} is a class of galaxy-halo
models that differs from HOD-style models in several respects. In
SHAM, functional forms for $\mean{\ncen}{\mhalo}$ and
$\mean{\nsat}{\mhalo}$ are never explicitly chosen. Instead, there is
an assumed monotonic correspondence between stellar mass $\mstar$ and
some (sub)halo property $x.$ In the absence of scatter between
$\mstar$ and $x,$ there exists a unique mapping $\mean{\mstar}{x}$
that yields the correct one-point function, which in this example is
the stellar mass function $\phi(\mstar)$. By associating host halos
with central galaxies and subhalos with satellites, for any stellar
mass threshold the functions $\mean{\ncen}{\mhalo}$ and
$\mean{\nsat}{\mhalo}$ are determined by this unique mapping.

In the most quantitatively successful SHAM models, the property $x$ is
chosen to be some measure of the subhalo circular velocity $\vmax$
\citep[e.g.,][]{conroy06,reddick12,hearin_etal12b}. As pointed out in
\citet{zentner_etal14}, this choice results in significant levels of
assembly bias in the following sense:
$\mean{\ncen}{\mvir, \vmax}\neq\mean{\ncen}{\mvir}$ and
$\mean{\nsat}{\mvir, \vmax}\neq\mean{\nsat}{\mvir}.$ In an interesting
recent advance of the abundance matching formalism,
\citet{lehmann_etal15} explored SHAM models in which the subhalo
property $x$ varies between $\mvir$ and $\vmax$ via a
continuously-valued parameter $\alpha.$ They fit the newly introduced
parameter $\alpha$ to SDSS clustering measurements at $z\approx0,$ and
their best-fit model significantly prefers ``$\vmax-$like" subhalo
properties over ``pure-$\mvir$" properties. That is, low-redshift
clustering measurements are better described by SHAM-style models in
which $\avg{\nsat}$ and $\avg{\ncen}$ do not depend on virial mass
$\mvir$ alone \citep[see also][for closely related work]{mao_etal15}.

The model presented in this paper implements assembly bias in a very
different manner from these and other SHAM-style models of
volume-limited galaxy samples. First of all,
there are the usual differences between traditional SHAM and HOD
implementations: the use of subhalos vs. analytical descriptions of
the spatial distribution of satellite galaxies, the explicit
vs. implicit parameterization of the occupation moments, etc. Beyond
these differences, the above SHAM-style models depend exclusively on a
{\em single} halo property at a time, whereas the decorated HOD
permits exploration of the galaxy-halo connection upon two (or more) 
halo properties simultaneously. Of equal importance, 
the decorated HOD provides fully independent
control over the level of assembly bias and the ``$\mvir-$only"
occupation moments. By contrast, there is no such freedom in SHAM. As
a specific example, in the \citet{lehmann_etal15} model, the level of
assembly bias in $\avg{\ncen}$ is controlled by the $\alpha$
parameter, but varying this parameter {\em also} changes the
$\mvir-$dependence of the satellite fraction.

\subsubsection{Models of galaxy color}

The ``age-matching" model introduced in \citet{HW13a} makes
predictions for the color dependence of the galaxy-halo
connection. Age matching is a particular implementation of a class of
models called {\em conditional abundance matching}
\citep[CAM,][]{hearin_etal13b}. As described in
\S\ref{subsubsec:shammodels}, abundance matching models use the
one-point function $\phi(\mstar)$ as an input that determines the
mapping $\mean{\mstar}{x}$ between $\mstar$ and some halo property
$x.$ Analogously, conditional abundance matching models use the
conditional one-point functions $\phi(g-r\vert\mstar)$ to determine
the continuously varying probability distribution $P(g-r\vert x, y)$
of $g-r$ colors as a function of two subhalo properties $x$ and $y.$
These models are formulated such that at fixed values of the primary
halo property $x,$ the galaxy property $g-r$ and the subhalo property
$y$ are in monotonic correspondence. Motivated by \cite{wechsler02}
and \cite{bullock01}, the age matching implementations of CAM that
have been most successful \citep{HW13a, hearin_etal13b,watson_etal14,
  yamamoto_etal15, saito_etal15} choose the primary halo property
$x=V_{\rm peak}$ and the secondary halo property $y=z_{\rm starve},$ a
proxy for the age of the halo \citep[see also][]{masaki13,
  kulier_ostriker15}.

CAM models can be thought of as one way to generalize the decorated
HOD framework to a continuously variable galaxy property.  As shown in
Campbell et al. (2015, in prep), the strength of the correlation
between, for example, $g-r$ and $z_{\rm starve}$ can be smoothly
varied, with age matching representing the extreme, maximum correlation
strength. In such variations, the conditional one-point functions are
held exactly fixed in an analogous fashion to how the first moments of
decorated HOD models are held fixed as the $\abias$ parameter is
varied.

The model presented in \citet{paranjape15} is another example of an
HOD-style model implementing assembly bias. In the color dependence of
this model, one first specifies a model for the red fraction of
galaxies \citep[separately for centrals and satellites, as in][]{skibba_sheth09}. Around this
baseline model, the red fraction varies according to halo
concentration in such a way that the overall red fraction is held
fixed. Intriguingly, \citet{paranjape15} find that their models prefer
relatively strong correlations between the red fraction and halo
concentration when comparing to SDSS measurements of ``1-halo
conformity": the tendency for red satellites to reside in groups with
a red central at fixed group mass.

The \citet{paranjape15} model is HOD-conserving and is therefore an
example of a decorated HOD model. As described in \citet{paranjape15},
this generalization of the HOD offers a promising means to understand
recent measurements of the galactic conformity signal \citep[see][and
  references therein for further details]{kauffmann_etal13,
  hearin_etal14, hearin15, kawinwanichakij_etal15}.

\subsection{Significance for Cosmology}
\label{subsec:cosmology}

In addition to galaxy formation, the effects of assembly bias can have
important consequences for cosmology.  HOD-style models are often
employed to model the galaxy--dark matter connection in studies that
aim to use relatively small-scale galaxy clustering and lensing
statistics to constrain cosmological parameters
\citep[e.g.][]{tinker05,vdBosch07,cacciato_etal09,leauthaud_etal12,cacciato_etal13,mandelbaum_etal13,more_etal13,villaescusa-navarro_etal14,more_etal15}.
These methods assume that assembly bias is either not present in the
observed universe or that it is present only at levels that do not
hinder cosmological parameter inference. As assembly bias can
compromise the inferred galaxy--dark matter connection
\citep{zentner_etal14}, it may also induce systematic errors in
inferred cosmological parameters and impair tests of general
relativity \citep{hearin15b}.  The degree to which such effects may
threaten the program of using galaxy clustering on quasilinear and
nonlinear scales to constrain cosmology has not yet been
quantified. We are actively pursuing this line of research.

In this work, we have studied a new class of models that may be used
to incorporate assembly bias into HOD-style analyses of galaxy
clustering and lensing statistics. Utilizing these models in
cosmological analyses will render any inferred parameters more robust
against systematic errors induced by assembly bias. However, the new
parameters of assembly bias models may be degenerate with cosmological
parameters, and to the degree to which they are degenerate this will
degrade the statistical constraints on inferred cosmological
parameters. Assessing the degeneracy between cosmological parameters
and models of assembly bias is another subject of our ongoing
collaborative work.

\section{Summary}
\label{sec:summary}

We conclude the paper with a summary of our primary findings. 

\ben
\item We introduce the {\em decorated HOD}, a new class of models for the
  galaxy-halo connection designed to account for assembly bias.
\item Using a simple two-population decorated HOD, we exhaustively
  enumerate the litany of signatures that assembly bias imprints on
  the clustering and lensing of $\mstar-$threshold galaxy samples. For the
  clustering of SDSS Main Galaxy Sample-type galaxies, the effects can
  be as large as a factor of two on $200$ kpc scales, and up to
  $\sim15\%$ in the linear regime. For lensing, the effects are
  limited to the $\sim10-15\%$ level on all scales.
\item For galaxy samples selected by a fixed stellar mass threshold,
  the impact of assembly bias on clustering and lensing generally
  weakens with redshift.
\item The scale dependence of assembly bias is complex. We advocate
  that flexible analysis techniques such as those provided by the
  open-source {\tt Halotools} package will be necessary in order for
  the precision-cosmology program to proceed into the quasilinear and
  nonlinear regime.
\een

We refer readers to the repository stored at {\tt https://github.com/aphearin/decorated-hod-paper}, 
which contains an annotated IPython Notebook that can be used to reproduce our figures, 
as well as a frozen copy of the exact version of the {\tt Halotools} code base we used to generate our results. 

\section*{Acknowledgments}
The work of ARZ was funded by the U.S. National Science Foundation
under grant AST 1517563 and by the Pittsburgh Particle physics,
Astrophysics, and Cosmology Center (Pitt PACC) at the University of
Pittsburgh.  The work of FvdB was funded by the U.S. National Science
Foundation under grant AST 1516962.  A portion of this work was also
supported by the National Science Foundation under grant PHYS-1066293
and the hospitality of the Aspen Center for Physics.  Support for EJT
was provided by NASA through Hubble Fellowship grants \#51316.01
awarded by the Space Telescope Science Institute, which is operated by
the Association of Universities for Research in Astronomy, Inc., for
NASA, under contract NAS 5-26555.  
APH is funded by the Yale Center for Astronomy \& Astrophysics, 
and thanks Doug Watson for useful comments 
and Link Wray for the generous, sublime boogie of {\em I'm So Glad, I'm So Proud}.

\bibliography{./abthy.bib}

\begin{thebibliography}{}

\bibitem[\protect\citeauthoryear{{Adhikari}, {Dalal} \&
  {Chamberlain}}{{Adhikari} et~al.}{2014}]{adhikari_etal14}
{Adhikari} S.,  {Dalal} N.,    {Chamberlain} R.~T.,  2014, JCAP, 11, 19

\bibitem[\protect\citeauthoryear{{Astropy Collaboration}, {Robitaille},
  {Tollerud}, {Greenfield}, {Droettboom}, {Bray}, {Aldcroft} et~al.,}{{Astropy
  Collaboration} et~al.}{2013}]{astropy}
{Astropy Collaboration} {Robitaille} T.~P.,  {Tollerud} E.~J.,  {Greenfield}
  P.,  {Droettboom} M.,  {Bray} E.,  {Aldcroft} T.,    et~al., 2013, AAP, 558,
  A33

\bibitem[\protect\citeauthoryear{Behnel, Bradshaw, Citro, Dalcin, Seljebotn \&
  Smith}{Behnel et~al.}{2011}]{cython}
Behnel S.,  Bradshaw R.,  Citro C.,  Dalcin L.,  Seljebotn D.,    Smith K.,
  2011, Computing in Science Engineering, 13, 31

\bibitem[\protect\citeauthoryear{{Behroozi}, {Conroy} \& {Wechsler}}{{Behroozi}
  et~al.}{2010}]{behroozi10}
{Behroozi} P.~S.,  {Conroy} C.,    {Wechsler} R.~H.,  2010, \apj, 717, 379

\bibitem[\protect\citeauthoryear{{Behroozi}, {Wechsler} \& {Conroy}}{{Behroozi}
  et~al.}{2013}]{behroozi13}
{Behroozi} P.~S.,  {Wechsler} R.~H.,    {Conroy} C.,  2013, \apj, 770, 57

\bibitem[\protect\citeauthoryear{{Behroozi}, {Wechsler} \& {Wu}}{{Behroozi}
  et~al.}{2011}]{behroozi_rockstar11}
{Behroozi} P.~S.,  {Wechsler} R.~H.,    {Wu} H.-Y.,  2011, ArXiv:1110.4372

\bibitem[\protect\citeauthoryear{{Behroozi}, {Wechsler}, {Wu}, {Busha},
  {Klypin} \& {Primack}}{{Behroozi} et~al.}{2013}]{behroozi_trees13}
{Behroozi} P.~S.,  {Wechsler} R.~H.,  {Wu} H.-Y.,  {Busha} M.~T.,  {Klypin}
  A.~A.,    {Primack} J.~R.,  2013, \apj, 763, 18

\bibitem[\protect\citeauthoryear{{Berlind} \& {Weinberg}}{{Berlind} \&
  {Weinberg}}{2002}]{berlind02}
{Berlind} A.~A.,  {Weinberg} D.~H.,  2002, \apj, 575, 587

\bibitem[\protect\citeauthoryear{{Blanton} \& {Berlind}}{{Blanton} \&
  {Berlind}}{2007}]{blanton_berlind07}
{Blanton} M.~R.,  {Berlind} A.~A.,  2007, \apj, 664, 791

\bibitem[\protect\citeauthoryear{{Blanton}, {Eisenstein}, {Hogg} \&
  {Zehavi}}{{Blanton} et~al.}{2006}]{blanton06}
{Blanton} M.~R.,  {Eisenstein} D.,  {Hogg} D.~W.,    {Zehavi} I.,  2006, \apj,
  645, 977

\bibitem[\protect\citeauthoryear{{Bray} et~al.,}{{Bray}
  et~al.}{2016}]{bray_etal15}
{Bray} A.~D.,  et~al., 2016, \mnras, 455, 185

\bibitem[\protect\citeauthoryear{{Bullock}, {Kolatt}, {Sigad}, {Somerville},
  {Kravtsov}, {Klypin}, {Primack} \& {Dekel}}{{Bullock}
  et~al.}{2001}]{bullock01}
{Bullock} J.~S.,  {Kolatt} T.~S.,  {Sigad} Y.,  {Somerville} R.~S.,  {Kravtsov}
  A.~V.,  {Klypin} A.~A.,  {Primack} J.~R.,    {Dekel} A.,  2001, \mnras, 321,
  559

\bibitem[\protect\citeauthoryear{{Cacciato}, {van den Bosch}, {More}, {Li},
  {Mo} \& {Yang}}{{Cacciato} et~al.}{2009}]{cacciato_etal09}
{Cacciato} M.,  {van den Bosch} F.~C.,  {More} S.,  {Li} R.,  {Mo} H.~J.,
  {Yang} X.,  2009, \mnras, 394, 929

\bibitem[\protect\citeauthoryear{{Cacciato}, {van den Bosch}, {More}, {Mo} \&
  {Yang}}{{Cacciato} et~al.}{2013}]{cacciato_etal13}
{Cacciato} M.,  {van den Bosch} F.~C.,  {More} S.,  {Mo} H.,    {Yang} X.,
  2013, \mnras, 430, 767

\bibitem[\protect\citeauthoryear{{Collister} \& {Lahav}}{{Collister} \&
  {Lahav}}{2005}]{collister05}
{Collister} A.~A.,  {Lahav} O.,  2005, \mnras, 361, 415

\bibitem[\protect\citeauthoryear{{Conroy} \& {Wechsler}}{{Conroy} \&
  {Wechsler}}{2009}]{conroy_wechsler09}
{Conroy} C.,  {Wechsler} R.~H.,  2009, \apj, 696, 620

\bibitem[\protect\citeauthoryear{{Conroy}, {Wechsler} \& {Kravtsov}}{{Conroy}
  et~al.}{2006}]{conroy06}
{Conroy} C.,  {Wechsler} R.~H.,    {Kravtsov} A.~V.,  2006, \apj, 647, 201

\bibitem[\protect\citeauthoryear{{Cooray} \& {Sheth}}{{Cooray} \&
  {Sheth}}{2002}]{cooray02}
{Cooray} A.,  {Sheth} R.,  2002, \physrep, 372, 1

\bibitem[\protect\citeauthoryear{{Croton}, {Gao} \& {White}}{{Croton}
  et~al.}{2007}]{croton_etal07}
{Croton} D.~J.,  {Gao} L.,    {White} S.~D.~M.,  2007, \mnras, 374, 1303

\bibitem[\protect\citeauthoryear{{Dalal}, {White}, {Bond} \&
  {Shirokov}}{{Dalal} et~al.}{2008}]{dalal_etal08}
{Dalal} N.,  {White} M.,  {Bond} J.~R.,    {Shirokov} A.,  2008, \apj, 687, 12

\bibitem[\protect\citeauthoryear{{Diemer} \& {Kravtsov}}{{Diemer} \&
  {Kravtsov}}{2014}]{diemer_kravtsov14}
{Diemer} B.,  {Kravtsov} A.~V.,  2014, \apj, 789, 1

\bibitem[\protect\citeauthoryear{{Feldmann} \& {Mayer}}{{Feldmann} \&
  {Mayer}}{2015}]{feldmann_mayer14}
{Feldmann} R.,  {Mayer} L.,  2015, \mnras, 446, 1939

\bibitem[\protect\citeauthoryear{{Gao}, {Springel} \& {White}}{{Gao}
  et~al.}{2005}]{gao_etal05}
{Gao} L.,  {Springel} V.,    {White} S.~D.~M.,  2005, \mnras, 363, L66

\bibitem[\protect\citeauthoryear{{Gao} \& {White}}{{Gao} \&
  {White}}{2007}]{gao_white07}
{Gao} L.,  {White} S.~D.~M.,  2007, \mnras, 377, L5

\bibitem[\protect\citeauthoryear{{Gil-Mar{\'{\i}}n}, {Jimenez} \&
  {Verde}}{{Gil-Mar{\'{\i}}n} et~al.}{2011}]{marin_etal11}
{Gil-Mar{\'{\i}}n} H.,  {Jimenez} R.,    {Verde} L.,  2011, \mnras, 414, 1207

\bibitem[\protect\citeauthoryear{{Gottloeber} \& {Klypin}}{{Gottloeber} \&
  {Klypin}}{2008}]{gottloeber_klypin08}
{Gottloeber} S.,  {Klypin} A.,  2008, ArXiv:0803.4343

\bibitem[\protect\citeauthoryear{{Guo}, {Zheng}, {Zehavi}, {Xu}, {Eisenstein},
  {Weinberg}, {Bahcall}, {Berlind}, {Comparat}, {McBride}, {Ross}, {Schneider},
  {Skibba}, {Swanson}, {Tinker}, {Tojeiro} \& {Wake}}{{Guo}
  et~al.}{2014}]{guo_etal14}
{Guo} H.,  {Zheng} Z.,  {Zehavi} I.,  {Xu} H.,  {Eisenstein} D.~J.,  {Weinberg}
  D.~H.,  {Bahcall} N.~A.,  {Berlind} A.~A.,  {Comparat} J.,  {McBride} C.~K.,
  {Ross} A.~J.,  {Schneider} D.~P.,  {Skibba} R.~A.,  {Swanson} M.~E.~C.,
  {Tinker} J.~L.,  {Tojeiro} R.,    {Wake} D.~A.,  2014, ArXiv:1401.3009

\bibitem[\protect\citeauthoryear{{Guo}, {White}, {Boylan-Kolchin}, {De Lucia},
  {Kauffmann}, {Lemson}, {Li}, {Springel} \& {Weinmann}}{{Guo}
  et~al.}{2011}]{guo_etal11b}
{Guo} Q.,  {White} S.,  {Boylan-Kolchin} M.,  {De Lucia} G.,  {Kauffmann} G.,
  {Lemson} G.,  {Li} C.,  {Springel} V.,    {Weinmann} S.,  2011, \mnras, 413,
  101

\bibitem[\protect\citeauthoryear{{Hearin}}{{Hearin}}{2015}]{hearin15b}
{Hearin} A.~P.,  2015, \mnras, 451, L45

\bibitem[\protect\citeauthoryear{{Hearin}, {Behroozi} \& {van den
  Bosch}}{{Hearin} et~al.}{2015}]{hearin15}
{Hearin} A.~P.,  {Behroozi} P.~S.,    {van den Bosch} F.~C.,  2015,
  ArXiv:1504.05578

\bibitem[\protect\citeauthoryear{{Hearin} \& {Watson}}{{Hearin} \&
  {Watson}}{2013}]{HW13a}
{Hearin} A.~P.,  {Watson} D.~F.,  2013, ArXiv:1304.5557

\bibitem[\protect\citeauthoryear{{Hearin}, {Watson}, {Becker}, {Reyes},
  {Berlind} \& {Zentner}}{{Hearin} et~al.}{2013}]{hearin_etal13b}
{Hearin} A.~P.,  {Watson} D.~F.,  {Becker} M.~R.,  {Reyes} R.,  {Berlind}
  A.~A.,    {Zentner} A.~R.,  2013, ArXiv:1310.6747

\bibitem[\protect\citeauthoryear{{Hearin}, {Watson} \& {van den
  Bosch}}{{Hearin} et~al.}{2014}]{hearin_etal14}
{Hearin} A.~P.,  {Watson} D.~F.,    {van den Bosch} F.~C.,  2014,
  ArXiv:1404.6524

\bibitem[\protect\citeauthoryear{{Hearin}, {Zentner}, {Berlind} \&
  {Newman}}{{Hearin} et~al.}{2012}]{hearin_etal12b}
{Hearin} A.~P.,  {Zentner} A.~R.,  {Berlind} A.~A.,    {Newman} J.~A.,  2012,
  ArXiv:1210.4927

\bibitem[\protect\citeauthoryear{{Heitmann}, {Higdon}, {White}, {Habib},
  {Williams}, {Lawrence} \& {Wagner}}{{Heitmann}
  et~al.}{2009}]{heitmann_etal08}
{Heitmann} K.,  {Higdon} D.,  {White} M.,  {Habib} S.,  {Williams} B.~J.,
  {Lawrence} E.,    {Wagner} C.,  2009, \apj, 705, 156

\bibitem[\protect\citeauthoryear{{Heitmann}, {White}, {Wagner}, {Habib} \&
  {Higdon}}{{Heitmann} et~al.}{2010}]{heitmann_etal10}
{Heitmann} K.,  {White} M.,  {Wagner} C.,  {Habib} S.,    {Higdon} D.,  2010,
  \apj, 715, 104

\bibitem[\protect\citeauthoryear{{Jiang} \& {van den Bosch}}{{Jiang} \& {van
  den Bosch}}{2015}]{Jiang_vdB15}
{Jiang} F.,  {van den Bosch} F.~C.,  2015, \mnras, 453, 3575

\bibitem[\protect\citeauthoryear{{Kauffmann}, {Li}, {Zhang} \&
  {Weinmann}}{{Kauffmann} et~al.}{2013}]{kauffmann_etal13}
{Kauffmann} G.,  {Li} C.,  {Zhang} W.,    {Weinmann} S.,  2013, \mnras, 430,
  1447

\bibitem[\protect\citeauthoryear{{Kauffmann}, {White}, {Heckman}, {M{\'e}nard},
  {Brinchmann}, {Charlot}, {Tremonti} \& {Brinkmann}}{{Kauffmann}
  et~al.}{2004}]{kauffmann_etal04}
{Kauffmann} G.,  {White} S.~D.~M.,  {Heckman} T.~M.,  {M{\'e}nard} B.,
  {Brinchmann} J.,  {Charlot} S.,  {Tremonti} C.,    {Brinkmann} J.,  2004,
  \mnras, 353, 713

\bibitem[\protect\citeauthoryear{{Kawinwanichakij} et~al.,}{{Kawinwanichakij}
  et~al.}{2015}]{kawinwanichakij_etal15}
{Kawinwanichakij} L.,  et~al., 2015, ArXiv:1511.02862

\bibitem[\protect\citeauthoryear{{Klypin}, {Trujillo-Gomez} \&
  {Primack}}{{Klypin} et~al.}{2011}]{bolshoi_11}
{Klypin} A.~A.,  {Trujillo-Gomez} S.,    {Primack} J.,  2011, \apj, 740, 102

\bibitem[\protect\citeauthoryear{{Kravtsov}, {Berlind}, {Wechsler}, {Klypin},
  {Gottl{\"o}ber}, {Allgood} \& {Primack}}{{Kravtsov}
  et~al.}{2004}]{kravtsov04a}
{Kravtsov} A.~V.,  {Berlind} A.~A.,  {Wechsler} R.~H.,  {Klypin} A.~A.,
  {Gottl{\"o}ber} S.,  {Allgood} B.,    {Primack} J.~R.,  2004, \apj, 609, 35

\bibitem[\protect\citeauthoryear{{Kravtsov}, {Klypin} \& {Khokhlov}}{{Kravtsov}
  et~al.}{1997}]{kravtsov_eta97}
{Kravtsov} A.~V.,  {Klypin} A.~A.,    {Khokhlov} A.~M.,  1997, \apjs, 111, 73

\bibitem[\protect\citeauthoryear{{Kulier} \& {Ostriker}}{{Kulier} \&
  {Ostriker}}{2015}]{kulier_ostriker15}
{Kulier} A.,  {Ostriker} J.~P.,  2015, \mnras, 452, 4013

\bibitem[\protect\citeauthoryear{{Lacerna} \& {Padilla}}{{Lacerna} \&
  {Padilla}}{2011}]{lacerna11}
{Lacerna} I.,  {Padilla} N.,  2011, \mnras, 412, 1283

\bibitem[\protect\citeauthoryear{{Landy} \& {Szalay}}{{Landy} \&
  {Szalay}}{1993}]{landyszalay93}
{Landy} S.~D.,  {Szalay} A.~S.,  1993, \apj, 412, 64

\bibitem[\protect\citeauthoryear{{Leauthaud} et~al.,}{{Leauthaud}
  et~al.}{2011}]{leauthaud11a}
{Leauthaud} A.,  et~al., 2011, ArXiv:1104.0928

\bibitem[\protect\citeauthoryear{{Leauthaud} et~al.,}{{Leauthaud}
  et~al.}{2012}]{leauthaud_etal12}
{Leauthaud} A.,  et~al., 2012, \apj, 744, 159

\bibitem[\protect\citeauthoryear{{Leauthaud}, {Tinker}, {Behroozi}, {Busha} \&
  {Wechsler}}{{Leauthaud} et~al.}{2011}]{leauthaud11b}
{Leauthaud} A.,  {Tinker} J.,  {Behroozi} P.~S.,  {Busha} M.~T.,    {Wechsler}
  R.~H.,  2011, \apj, 738, 45

\bibitem[\protect\citeauthoryear{{Lehmann}, {Mao}, {Becker}, {Skillman} \&
  {Wechsler}}{{Lehmann} et~al.}{2015}]{lehmann_etal15}
{Lehmann} B.~V.,  {Mao} Y.-Y.,  {Becker} M.~R.,  {Skillman} S.~W.,
  {Wechsler} R.~H.,  2015, ArXiv:1510.05651

\bibitem[\protect\citeauthoryear{{Lin}, {Mandelbaum}, {Huang}, {Huang},
  {Dalal}, {Diemer}, {Jian} \& {Kravtsov}}{{Lin}
  et~al.}{2015}]{lin_mandelbaum_etal15}
{Lin} Y.-T.,  {Mandelbaum} R.,  {Huang} Y.-H.,  {Huang} H.-J.,  {Dalal} N.,
  {Diemer} B.,  {Jian} H.-Y.,    {Kravtsov} A.,  2015, ArXiv:1504.07632

\bibitem[\protect\citeauthoryear{{Ma} \& {Fry}}{{Ma} \& {Fry}}{2000}]{mafry00}
{Ma} C.-P.,  {Fry} J.~N.,  2000, \apj, 543, 503

\bibitem[\protect\citeauthoryear{{Mandelbaum} et~al.,}{{Mandelbaum}
  et~al.}{2005}]{mandelbaum05}
{Mandelbaum} R.,  et~al., 2005, \mnras, 361, 1287

\bibitem[\protect\citeauthoryear{{Mandelbaum}, {Slosar}, {Baldauf}, {Seljak},
  {Hirata}, {Nakajima}, {Reyes} \& {Smith}}{{Mandelbaum}
  et~al.}{2013}]{mandelbaum_etal13}
{Mandelbaum} R.,  {Slosar} A.,  {Baldauf} T.,  {Seljak} U.,  {Hirata} C.~M.,
  {Nakajima} R.,  {Reyes} R.,    {Smith} R.~E.,  2013, \mnras, 432, 1544

\bibitem[\protect\citeauthoryear{{Mao}, {Williamson} \& {Wechsler}}{{Mao}
  et~al.}{2015}]{mao_etal15}
{Mao} Y.-Y.,  {Williamson} M.,    {Wechsler} R.~H.,  2015, \apj, 810, 21

\bibitem[\protect\citeauthoryear{{Masaki}, {Lin} \& {Yoshida}}{{Masaki}
  et~al.}{2013}]{masaki13}
{Masaki} S.,  {Lin} Y.-T.,    {Yoshida} N.,  2013, ArXiv:1301.1217

\bibitem[\protect\citeauthoryear{{Miyatake}, {More}, {Takada}, {Spergel},
  {Mandelbaum}, {Rykoff} \& {Rozo}}{{Miyatake} et~al.}{2015}]{miyatake_etal15}
{Miyatake} H.,  {More} S.,  {Takada} M.,  {Spergel} D.~N.,  {Mandelbaum} R.,
  {Rykoff} E.~S.,    {Rozo} E.,  2015, ArXiv:1506.06135

\bibitem[\protect\citeauthoryear{{Mo}, {van den Bosch} \& {White}}{{Mo}
  et~al.}{2010}]{mo_vdb_white10}
{Mo} H.,  {van den Bosch} F.~C.,    {White} S.,  2010, {Galaxy Formation and
  Evolution}.
Cambridge University Press, Cambridge, UK

\bibitem[\protect\citeauthoryear{{Mo}, {Yang}, {van den Bosch} \& {Jing}}{{Mo}
  et~al.}{2004}]{mo_etal04}
{Mo} H.~J.,  {Yang} X.,  {van den Bosch} F.~C.,    {Jing} Y.~P.,  2004, \mnras,
  349, 205

\bibitem[\protect\citeauthoryear{{More}, {Diemer} \& {Kravtsov}}{{More}
  et~al.}{2015}]{more_etal15}
{More} S.,  {Diemer} B.,    {Kravtsov} A.~V.,  2015, \apj, 810, 36

\bibitem[\protect\citeauthoryear{{More}, {van den Bosch}, {Cacciato}, {More},
  {Mo} \& {Yang}}{{More} et~al.}{2013}]{more_etal13}
{More} S.,  {van den Bosch} F.~C.,  {Cacciato} M.,  {More} A.,  {Mo} H.,
  {Yang} X.,  2013, \mnras, 430, 747

\bibitem[\protect\citeauthoryear{{Moster}, {Naab} \& {White}}{{Moster}
  et~al.}{2013}]{moster13}
{Moster} B.~P.,  {Naab} T.,    {White} S.~D.~M.,  2013, \mnras, 428, 3121

\bibitem[\protect\citeauthoryear{{Navarro}, {Frenk} \& {White}}{{Navarro}
  et~al.}{1997}]{nfw97}
{Navarro} J.~F.,  {Frenk} C.~S.,    {White} S.~D.~M.,  1997, \apj, 490, 493

\bibitem[\protect\citeauthoryear{{Paranjape}, {Kova{\v c}}, {Hartley} \&
  {Pahwa}}{{Paranjape} et~al.}{2015}]{paranjape15}
{Paranjape} A.,  {Kova{\v c}} K.,  {Hartley} W.~G.,    {Pahwa} I.,  2015,
  \mnras, 454, 3030

\bibitem[\protect\citeauthoryear{{Porciani}, {Magliocchetti} \&
  {Norberg}}{{Porciani} et~al.}{2004}]{porciani04}
{Porciani} C.,  {Magliocchetti} M.,    {Norberg} P.,  2004, \mnras, 355, 1010

\bibitem[\protect\citeauthoryear{{Porciani} \& {Norberg}}{{Porciani} \&
  {Norberg}}{2006}]{porciani06}
{Porciani} C.,  {Norberg} P.,  2006, \mnras, 371, 1824

\bibitem[\protect\citeauthoryear{{Purcell}, {Bullock} \& {Zentner}}{{Purcell}
  et~al.}{2007}]{purcell_etal07}
{Purcell} C.~W.,  {Bullock} J.~S.,    {Zentner} A.~R.,  2007, \apj, 666, 20

\bibitem[\protect\citeauthoryear{{Reddick}, {Wechsler}, {Tinker} \&
  {Behroozi}}{{Reddick} et~al.}{2012}]{reddick12}
{Reddick} R.~M.,  {Wechsler} R.~H.,  {Tinker} J.~L.,    {Behroozi} P.~S.,
  2012, ArXiv:1207.2160

\bibitem[\protect\citeauthoryear{{Reid}, {Seo}, {Leauthaud}, {Tinker} \&
  {White}}{{Reid} et~al.}{2014}]{reid_etal14}
{Reid} B.~A.,  {Seo} H.-J.,  {Leauthaud} A.,  {Tinker} J.~L.,    {White} M.,
  2014, \mnras, 444, 476

\bibitem[\protect\citeauthoryear{{Riebe}, {Partl}, {Enke}, {Forero-Romero},
  {Gottloeber}, {Klypin}, {Lemson}, {Prada}, {Primack}, {Steinmetz} \&
  {Turchaninov}}{{Riebe} et~al.}{2011}]{riebe_etal11}
{Riebe} K.,  {Partl} A.~M.,  {Enke} H.,  {Forero-Romero} J.,  {Gottloeber} S.,
  {Klypin} A.,  {Lemson} G.,  {Prada} F.,  {Primack} J.~R.,  {Steinmetz} M.,
  {Turchaninov} V.,  2011, ArXiv:1109.0003

\bibitem[\protect\citeauthoryear{{Rodriguez-Puebla}, {Avila-Reese}, {Firmani}
  \& {Colin}}{{Rodriguez-Puebla} et~al.}{2011}]{rod_puebla11}
{Rodriguez-Puebla} A.,  {Avila-Reese} V.,  {Firmani} C.,    {Colin} P.,  2011,
  ArXiv:1103.4151

\bibitem[\protect\citeauthoryear{{Ross} \& {Brunner}}{{Ross} \&
  {Brunner}}{2009}]{ross_brunner09}
{Ross} A.~J.,  {Brunner} R.~J.,  2009, \mnras, 399, 878

\bibitem[\protect\citeauthoryear{{Saito}, {Leauthaud}, {Hearin}, {Bundy},
  {Zentner}, {Behroozi}, {Reid}, {Sinha}, {Coupon}, {Tinker}, {White} \&
  {Schneider}}{{Saito} et~al.}{2015}]{saito_etal15}
{Saito} S.,  {Leauthaud} A.,  {Hearin} A.~P.,  {Bundy} K.,  {Zentner} A.~R.,
  {Behroozi} P.~S.,  {Reid} B.~A.,  {Sinha} M.,  {Coupon} J.,  {Tinker} J.~L.,
  {White} M.,    {Schneider} D.~P.,  2015, ArXiv:1509.00482

\bibitem[\protect\citeauthoryear{{Scoccimarro}, {Sheth}, {Hui} \&
  {Jain}}{{Scoccimarro} et~al.}{2001}]{scoccimarro01a}
{Scoccimarro} R.,  {Sheth} R.~K.,  {Hui} L.,    {Jain} B.,  2001, \apj, 546, 20

\bibitem[\protect\citeauthoryear{{Seljak}}{{Seljak}}{2000}]{seljak00}
{Seljak} U.,  2000, \mnras, 318, 203

\bibitem[\protect\citeauthoryear{{Seljak}, {Makarov}, {Mandelbaum}, {Hirata},
  {Padmanabhan}, {McDonald}, {Blanton}, {Tegmark}, {Bahcall} \&
  {Brinkmann}}{{Seljak} et~al.}{2005}]{seljak05}
{Seljak} U.,  {Makarov} A.,  {Mandelbaum} R.,  {Hirata} C.~M.,  {Padmanabhan}
  N.,  {McDonald} P.,  {Blanton} M.~R.,  {Tegmark} M.,  {Bahcall} N.~A.,
  {Brinkmann} J.,  2005, PRD, 71, 043511

\bibitem[\protect\citeauthoryear{{Skibba} \& {Sheth}}{{Skibba} \&
  {Sheth}}{2009}]{skibba_sheth09}
{Skibba} R.~A.,  {Sheth} R.~K.,  2009, \mnras, 392, 1080

\bibitem[\protect\citeauthoryear{{Sunayama}, {Hearin}, {Padmanabhan} \&
  {Leauthaud}}{{Sunayama} et~al.}{2015}]{sunayama_etal15}
{Sunayama} T.,  {Hearin} A.~P.,  {Padmanabhan} N.,    {Leauthaud} A.,  2015,
  ArXiv:1509.06417

\bibitem[\protect\citeauthoryear{{Tinker}, {Kravtsov}, {Klypin}, {Abazajian},
  {Warren}, {Yepes}, {Gottl{\"o}ber} \& {Holz}}{{Tinker}
  et~al.}{2008}]{tinker08a}
{Tinker} J.,  {Kravtsov} A.~V.,  {Klypin} A.,  {Abazajian} K.,  {Warren} M.,
  {Yepes} G.,  {Gottl{\"o}ber} S.,    {Holz} D.~E.,  2008, \apj, 688, 709

\bibitem[\protect\citeauthoryear{{Tinker}, {Conroy}, {Norberg}, {Patiri},
  {Weinberg} \& {Warren}}{{Tinker} et~al.}{2008}]{tinker08b}
{Tinker} J.~L.,  {Conroy} C.,  {Norberg} P.,  {Patiri} S.~G.,  {Weinberg}
  D.~H.,    {Warren} M.~S.,  2008, \apj, 686, 53

\bibitem[\protect\citeauthoryear{{Tinker}, {Leauthaud}, {Bundy}, {George},
  {Behroozi}, {Massey}, {Rhodes} \& {Wechsler}}{{Tinker}
  et~al.}{2013}]{tinker_etal13}
{Tinker} J.~L.,  {Leauthaud} A.,  {Bundy} K.,  {George} M.~R.,  {Behroozi} P.,
  {Massey} R.,  {Rhodes} J.,    {Wechsler} R.,  2013, ArXiv:1308.2974

\bibitem[\protect\citeauthoryear{{Tinker}, {Robertson}, {Kravtsov}, {Klypin},
  {Warren}, {Yepes} \& {Gottl{\"o}ber}}{{Tinker} et~al.}{2010}]{tinker10}
{Tinker} J.~L.,  {Robertson} B.~E.,  {Kravtsov} A.~V.,  {Klypin} A.,  {Warren}
  M.~S.,  {Yepes} G.,    {Gottl{\"o}ber} S.,  2010, \apj, 724, 878

\bibitem[\protect\citeauthoryear{{Tinker}, {Weinberg}, {Zheng} \&
  {Zehavi}}{{Tinker} et~al.}{2005}]{tinker05}
{Tinker} J.~L.,  {Weinberg} D.~H.,  {Zheng} Z.,    {Zehavi} I.,  2005, \apj,
  631, 41

\bibitem[\protect\citeauthoryear{{Vale} \& {Ostriker}}{{Vale} \&
  {Ostriker}}{2004}]{vale_ostriker04}
{Vale} A.,  {Ostriker} J.~P.,  2004, \mnras, 353, 189

\bibitem[\protect\citeauthoryear{{van den Bosch}, {Mo} \& {Yang}}{{van den
  Bosch} et~al.}{2003}]{vdBosch03b}
{van den Bosch} F.~C.,  {Mo} H.~J.,    {Yang} X.,  2003, \mnras, 345, 923

\bibitem[\protect\citeauthoryear{{van den Bosch}, {More}, {Cacciato}, {Mo} \&
  {Yang}}{{van den Bosch} et~al.}{2013}]{vdBosch13}
{van den Bosch} F.~C.,  {More} S.,  {Cacciato} M.,  {Mo} H.,    {Yang} X.,
  2013, \mnras, 430, 725

\bibitem[\protect\citeauthoryear{{van den Bosch}, {Tormen} \& {Giocoli}}{{van
  den Bosch} et~al.}{2005}]{vdBosch05}
{van den Bosch} F.~C.,  {Tormen} G.,    {Giocoli} C.,  2005, \mnras, 359, 1029

\bibitem[\protect\citeauthoryear{{van den Bosch}, {Yang} \& {Mo}}{{van den
  Bosch} et~al.}{2003}]{vdBosch03a}
{van den Bosch} F.~C.,  {Yang} X.,    {Mo} H.~J.,  2003, \mnras, 340, 771

\bibitem[\protect\citeauthoryear{{van den Bosch}, {Yang}, {Mo}, {Weinmann},
  {Macci{\`o}}, {More}, {Cacciato}, {Skibba} \& {Kang}}{{van den Bosch}
  et~al.}{2007}]{vdBosch07}
{van den Bosch} F.~C.,  {Yang} X.,  {Mo} H.~J.,  {Weinmann} S.~M.,
  {Macci{\`o}} A.~V.,  {More} S.,  {Cacciato} M.,  {Skibba} R.,    {Kang} X.,
  2007, \mnras, 376, 841

\bibitem[\protect\citeauthoryear{{Villaescusa-Navarro}, {Marulli}, {Viel},
  {Branchini}, {Castorina}, {Sefusatti} \& {Saito}}{{Villaescusa-Navarro}
  et~al.}{2014}]{villaescusa-navarro_etal14}
{Villaescusa-Navarro} F.,  {Marulli} F.,  {Viel} M.,  {Branchini} E.,
  {Castorina} E.,  {Sefusatti} E.,    {Saito} S.,  2014, \jcap, 3, 11

\bibitem[\protect\citeauthoryear{{Wake}, {Sheth}, {Nichol}, {Baugh},
  {Bland-Hawthorn}, {Colless}, {Couch}, {Croom}, {de Propris}, {Drinkwater},
  {Edge}, {Loveday}, {Lam}, {Pimbblet}, {Roseboom}, {Ross}, {Schneider},
  {Shanks} \& {Sharp}}{{Wake} et~al.}{2008}]{wake_etal08}
{Wake} D.~A.,  {Sheth} R.~K.,  {Nichol} R.~C.,  {Baugh} C.~M.,
  {Bland-Hawthorn} J.,  {Colless} M.,  {Couch} W.~J.,  {Croom} S.~M.,  {de
  Propris} R.,  {Drinkwater} M.~J.,  {Edge} A.~C.,  {Loveday} J.,  {Lam} T.~Y.,
   {Pimbblet} K.~A.,  {Roseboom} I.~G.,  {Ross} N.~P.,  {Schneider} D.~P.,
  {Shanks} T.,    {Sharp} R.~G.,  2008, \mnras, 387, 1045

\bibitem[\protect\citeauthoryear{{Wake}, {Whitaker}, {Labb{\'e}}, {van Dokkum},
  {Franx}, {Quadri}, {Brammer}, {Kriek}, {Lundgren}, {Marchesini} \&
  {Muzzin}}{{Wake} et~al.}{2011}]{wake_etal11}
{Wake} D.~A.,  {Whitaker} K.~E.,  {Labb{\'e}} I.,  {van Dokkum} P.~G.,  {Franx}
  M.,  {Quadri} R.,  {Brammer} G.,  {Kriek} M.,  {Lundgren} B.~F.,
  {Marchesini} D.,    {Muzzin} A.,  2011, \apj, 728, 46

\bibitem[\protect\citeauthoryear{{Wang}, {Weinmann}, {De Lucia} \&
  {Yang}}{{Wang} et~al.}{2013}]{wang_etal13}
{Wang} L.,  {Weinmann} S.~M.,  {De Lucia} G.,    {Yang} X.,  2013, \mnras, 433,
  515

\bibitem[\protect\citeauthoryear{{Wang}, {Yang}, {Mo}, {van den Bosch},
  {Weinmann} \& {Chu}}{{Wang} et~al.}{2008}]{wang_etal08}
{Wang} Y.,  {Yang} X.,  {Mo} H.~J.,  {van den Bosch} F.~C.,  {Weinmann} S.~M.,
    {Chu} Y.,  2008, \apj, 687, 919

\bibitem[\protect\citeauthoryear{{Watson}, {Berlind} \& {Zentner}}{{Watson}
  et~al.}{2011}]{watson_powerlaw11}
{Watson} D.~F.,  {Berlind} A.~A.,    {Zentner} A.~R.,  2011, \apj, 738, 22

\bibitem[\protect\citeauthoryear{{Watson}, {Hearin}, {Berlind}, {Becker},
  {Behroozi}, {Skibba}, {Reyes}, {Zentner} \& {van den Bosch}}{{Watson}
  et~al.}{2015}]{watson_etal14}
{Watson} D.~F.,  {Hearin} A.~P.,  {Berlind} A.~A.,  {Becker} M.~R.,  {Behroozi}
  P.~S.,  {Skibba} R.~A.,  {Reyes} R.,  {Zentner} A.~R.,    {van den Bosch}
  F.~C.,  2015, \mnras, 446, 651

\bibitem[\protect\citeauthoryear{{Wechsler}, {Bullock}, {Primack}, {Kravtsov}
  \& {Dekel}}{{Wechsler} et~al.}{2002}]{wechsler02}
{Wechsler} R.~H.,  {Bullock} J.~S.,  {Primack} J.~R.,  {Kravtsov} A.~V.,
  {Dekel} A.,  2002, \apj, 568, 52

\bibitem[\protect\citeauthoryear{{Wechsler}, {Zentner}, {Bullock}, {Kravtsov}
  \& {Allgood}}{{Wechsler} et~al.}{2006}]{wechsler06}
{Wechsler} R.~H.,  {Zentner} A.~R.,  {Bullock} J.~S.,  {Kravtsov} A.~V.,
  {Allgood} B.,  2006, \apj, 652, 71

\bibitem[\protect\citeauthoryear{{Weinmann}, {van den Bosch}, {Yang} \&
  {Mo}}{{Weinmann} et~al.}{2006}]{weinmann06b}
{Weinmann} S.~M.,  {van den Bosch} F.~C.,  {Yang} X.,    {Mo} H.~J.,  2006,
  \mnras, 366, 2

\bibitem[\protect\citeauthoryear{{Yamamoto}, {Masaki} \& {Hikage}}{{Yamamoto}
  et~al.}{2015}]{yamamoto_etal15}
{Yamamoto} M.,  {Masaki} S.,    {Hikage} C.,  2015, ArXiv:1503.03973

\bibitem[\protect\citeauthoryear{{Yang}, {Mo} \& {van den Bosch}}{{Yang}
  et~al.}{2003}]{yang03}
{Yang} X.,  {Mo} H.~J.,    {van den Bosch} F.~C.,  2003, \mnras, 339, 1057

\bibitem[\protect\citeauthoryear{{Yang}, {Mo} \& {van den Bosch}}{{Yang}
  et~al.}{2006}]{yang_etal06a}
{Yang} X.,  {Mo} H.~J.,    {van den Bosch} F.~C.,  2006, \apjl, 638, L55

\bibitem[\protect\citeauthoryear{{Yang}, {Mo} \& {van den Bosch}}{{Yang}
  et~al.}{2009a}]{yang09a}
{Yang} X.,  {Mo} H.~J.,    {van den Bosch} F.~C.,  2009a, \apj, 695, 900

\bibitem[\protect\citeauthoryear{{Yang}, {Mo} \& {van den Bosch}}{{Yang}
  et~al.}{2009b}]{yang09b}
{Yang} X.,  {Mo} H.~J.,    {van den Bosch} F.~C.,  2009b, \apj, 693, 830

\bibitem[\protect\citeauthoryear{{Yang}, {Mo}, {van den Bosch}, {Zhang} \&
  {Han}}{{Yang} et~al.}{2012}]{yang12}
{Yang} X.,  {Mo} H.~J.,  {van den Bosch} F.~C.,  {Zhang} Y.,    {Han} J.,
  2012, \apj, 752, 41

\bibitem[\protect\citeauthoryear{{Yang}, {Mo}, {Zhang} \& {van den
  Bosch}}{{Yang} et~al.}{2011}]{yang11a}
{Yang} X.,  {Mo} H.~J.,  {Zhang} Y.,    {van den Bosch} F.~C.,  2011, \apj,
  741, 13

\bibitem[\protect\citeauthoryear{{Yoo}, {Tinker}, {Weinberg}, {Zheng}, {Katz}
  \& {Dav{\'e}}}{{Yoo} et~al.}{2006}]{yoo06}
{Yoo} J.,  {Tinker} J.~L.,  {Weinberg} D.~H.,  {Zheng} Z.,  {Katz} N.,
  {Dav{\'e}} R.,  2006, \apj, 652, 26

\bibitem[\protect\citeauthoryear{{Zehavi} et~al.,}{{Zehavi}
  et~al.}{2005}]{zehavi05a}
{Zehavi} I.,  et~al., 2005, \apj, 630, 1

\bibitem[\protect\citeauthoryear{{Zehavi} et~al.,}{{Zehavi}
  et~al.}{2011}]{zehavi11}
{Zehavi} I.,  et~al., 2011, \apj, 736, 59

\bibitem[\protect\citeauthoryear{{Zentner}}{{Zentner}}{2007}]{zentner07}
{Zentner} A.~R.,  2007, International Journal of Modern Physics D, 16, 763

\bibitem[\protect\citeauthoryear{{Zentner}, {Berlind}, {Bullock}, {Kravtsov} \&
  {Wechsler}}{{Zentner} et~al.}{2005}]{zentner05}
{Zentner} A.~R.,  {Berlind} A.~A.,  {Bullock} J.~S.,  {Kravtsov} A.~V.,
  {Wechsler} R.~H.,  2005, \apj, 624, 505

\bibitem[\protect\citeauthoryear{{Zentner}, {Hearin} \& {van den
  Bosch}}{{Zentner} et~al.}{2014}]{zentner_etal14}
{Zentner} A.~R.,  {Hearin} A.~P.,    {van den Bosch} F.~C.,  2014, \mnras, 443,
  3044

\bibitem[\protect\citeauthoryear{{Zheng}, {Coil} \& {Zehavi}}{{Zheng}
  et~al.}{2007}]{Zheng07}
{Zheng} Z.,  {Coil} A.~L.,    {Zehavi} I.,  2007, \apj, 667, 760

\bibitem[\protect\citeauthoryear{{Zhu}, {Zheng}, {Lin}, {Jing}, {Kang} \&
  {Gao}}{{Zhu} et~al.}{2006}]{zhu_etal06}
{Zhu} G.,  {Zheng} Z.,  {Lin} W.~P.,  {Jing} Y.~P.,  {Kang} X.,    {Gao} L.,
  2006, \apjl, 639, L5

\bibitem[\protect\citeauthoryear{{Zu} \& {Mandelbaum}}{{Zu} \&
  {Mandelbaum}}{2015}]{zu_mandelbaum15b}
{Zu} Y.,  {Mandelbaum} R.,  2015, ArXiv:1509.06758

\end{thebibliography}

\end{document}